\documentclass[preprint,aps,floatfix]{revtex4}
\usepackage{lmodern}
\usepackage{lmodern}
\usepackage[latin9]{inputenc}
\usepackage{bm}
\usepackage{amsmath}
\usepackage{amssymb}
\usepackage{graphicx}
\usepackage{esint}

\makeatletter

\providecommand{\tabularnewline}{\\}

\@ifundefined{textcolor}{}
{%
 \definecolor{BLACK}{gray}{0}
 \definecolor{WHITE}{gray}{1}
 \definecolor{RED}{rgb}{1,0,0}
 \definecolor{GREEN}{rgb}{0,1,0}
 \definecolor{BLUE}{rgb}{0,0,1}
 \definecolor{CYAN}{cmyk}{1,0,0,0}
 \definecolor{MAGENTA}{cmyk}{0,1,0,0}
 \definecolor{YELLOW}{cmyk}{0,0,1,0}
}

\usepackage{textcomp}
\usepackage{bm}

\@ifundefined{textcolor}{}{%
 \definecolor{BLACK}{gray}{0}
 \definecolor{WHITE}{gray}{1}
 \definecolor{RED}{rgb}{1,0,0}
 \definecolor{GREEN}{rgb}{0,1,0}
 \definecolor{BLUE}{rgb}{0,0,1}
 \definecolor{CYAN}{cmyk}{1,0,0,0}
 \definecolor{MAGENTA}{cmyk}{0,1,0,0}
 \definecolor{YELLOW}{cmyk}{0,0,1,0}
}
\newcommand{\beq}{\begin{equation}}
\newcommand{\eeq}{\end{equation}}
\newcommand{\bea}{\begin{eqnarray}}
\newcommand{\eea}{\end{eqnarray}}
\newcommand{\be}{\begin{equation}}
\newcommand{\ee}{\end{equation}}

\newcommand{\Ut}[1]{\tilde{U}}
\newcommand{\Utt}[1]{\tilde{\tilde{U}}}

\makeatother

\begin{document}

\title{Magnetism, superconductivity, and spontaneous orbital order in iron-based
superconductors: who comes first and why?}

\author{Andrey V. Chubukov$^{1}$, M. Khodas $^{2,3}$, and Rafael M. Fernandes$^{1}$}
\affiliation{$^{1}$ School of Physics and Astronomy, University of Minnesota,
Minneapolis, MN 55455, USA \\
 $^{2}$ Racah Institute of Physics, The Hebrew University, Jerusalem 91904, Israel\\
 $^{3}$ Department of Physics and Astronomy, University of Iowa, Iowa City, IA 52242, USA
 }

\begin{abstract}
Magnetism and nematic order are the two non-superconducting orders
observed in iron-based superconductors. To elucidate the interplay
between them and ultimately unveil the pairing mechanism, several
models have been investigated. In models with quenched orbital degrees
of freedom, magnetic fluctuations promote stripe magnetism which induces
orbital order. In models with quenched spin degrees of freedom, charge
fluctuations promote spontaneous orbital order which induces stripe
magnetism. Here we develop an unbiased approach, in which we treat
magnetic and orbital fluctuations on equal footing. Key to our approach
is the inclusion of the orbital character of the low-energy electronic
states into renormalization group analysis. Our results show that
in  systems with large Fermi energies,
 such as BaFe$_{2}$As$_{2}$, LaFeAsO, and NaFeAs, orbital order
is induced by stripe magnetism. However, in systems with small Fermi
energies, such as FeSe, the system develops a spontaneous orbital
order, while magnetic order does not develop. Our results provide
a unifying description of different iron-based materials.
\end{abstract}
\maketitle
\newpage{}

\textbf{{Introduction.}}~~~~ The interplay between magnetism
and orbital order and how the two affect superconductivity are the
most interesting, yet, most controversial aspects of the physics of
iron-based superconducting materials (FeSCs). Both orbital and magnetic
fluctuations have been proposed as the glue that binds electrons together
for superconductivity, yielding different pairing states \cite{kontani,mazin,kuroki,kemper,chubukov,yamaze,kotliar_AFM,ashwin}.
However, which of the two degrees of freedom, orbital or spin, is
the driving force, is a hotly debated topic \cite{fcs,dagotto,kruger,ku,leni,phillips,singh,brydon,fernandes_12,stanev,moreo,gastiasoro,cfs_15,chu_r}.

The proponents of either orbital or magnetic fluctuations put forward
models in which the unwanted degree of freedom is quenched. In a class
of models where spin degrees of freedom are quenched~\cite{yamaze,efremov,ashwin},
density fluctuations with opposite signs on the Fe $d_{xz}$ and $d_{yz}$
orbitals are enhanced as the temperature is lowered. Consequently,
below a temperature $T_{s}$ the occupation of the $d_{xz}$ and $d_{yz}$
orbitals becomes unequal, breaking the tetragonal symmetry of the system
and triggering a structural transition. This orbital order can either
be homogeneous (ferro-orbital order) or with a lattice wavevector
(antiferro-orbital order (AFO)). In the band basis, ferro-orbital
order is a Pomeranchuk-type (POM) order in the $d-$wave charge channel
\cite{i_paul}. Such an order has been extensively studied in recent
years in the context of quantum criticality~\cite{kiv_led,ashwin}.
Orbital fluctuations can mediate superconductivity (SC) and favor
a sign-preserving $s^{++}$ SC order~\cite{kontani,ashwin,yamaze}.

In models where
 orbital degrees of freedom are quenched, orbital order is a spin-off
of stripe spin-density-wave (SDW) magnetism. Stripe SDW order breaks
the tetragonal symmetry between the $x$ and $y$ directions in addition
to breaking the spin-rotational symmetry \cite{kivelson}. It has
been shown~\cite{fcs,fernandes_12} that the breaking of the discrete
tetragonal symmetry occurs prior to the breaking of the continuous
spin-rotational symmetry, via the development of a composite Ising-nematic
order. By symmetry arguments, this order induces orbital order \cite{fs}.
Magnetic fluctuations that drive Ising-nematic order also favor a
sign-changing $s^{+-}$ SC~\cite{mazin,kuroki,chubukov,fcs,khodas}.

Each set of models uses approximations which
have been strongly questioned. Orbital models assume attractive local
(Hubbard) intra-pocket interaction, in variance with first-principle
calculations~\cite{first_principles}. Magnetic models either
 assume a priori that superconductivity is magnetically mediated~\cite{kemper},
or treat superconductivity and magnetism on equal footing, but neglect
the orbital content of low-energy excitations~\cite{cee,maiti}.
In reality,
 however, magnetic and orbital degrees of freedom are coupled and
affect each other \cite{fanfarillo,christensen}.

\begin{figure}[h]
\centering{}\includegraphics[width=0.6\columnwidth]{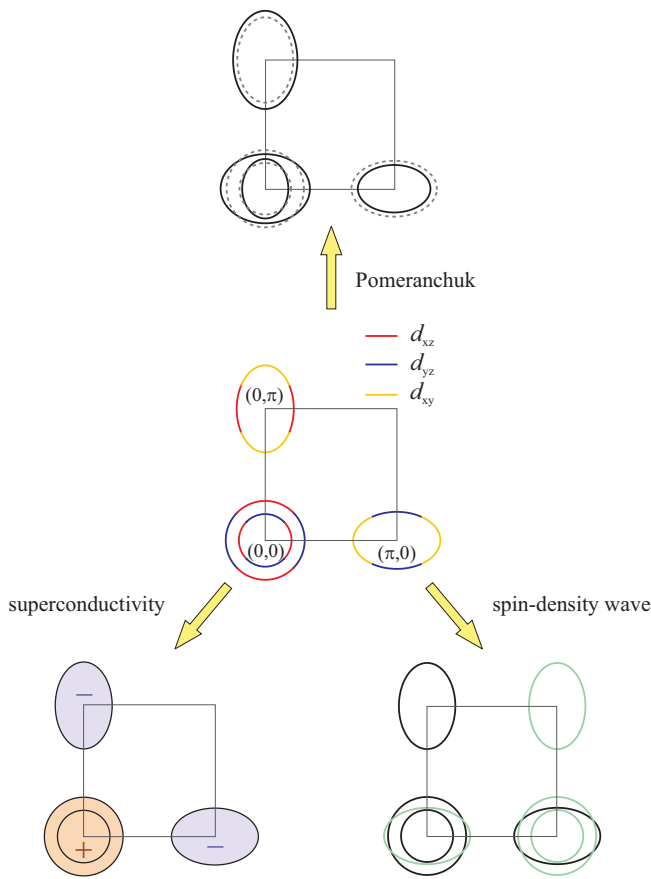} \protect\caption{\textbf{Low-energy states and potential instabilities.}
The orbital content of the 2D Fermi surface of the Fe-based superconductors
is plotted together with the changes in the fermionic excitations
promoted by one of three electronic instabilities -- $s^{+-}$ superconductivity,
stripe SDW magnetism, and nematicity (breaking of $C_{4}$ lattice
rotational symmetry), which necessary gives rise to orbital order.
The low-energy excitations live near hole-pockets centered at the
$\Gamma$ point ($k_{x}=k_{y}=0$), and near electron pockets centered
at $(0,\pi)$ $(\pi,0)$ in 1Fe Brillouin zone. Excitations near
the
hole pockets are made out of $d_{xz}$ and $d_{yz}$ orbitals, while
the ones near the electron pockets are made out of $d_{xz}$ and $d_{xy}$
($d_{yz}$ and $d_{xy}$) orbitals. (Refs.\protect\cite{kuroki,kemper}).
In some systems, there exists a third hole pocket (not shown) centered
at $(\pi,\pi)$ and made out of the $d_{xy}$ orbital. $s^{+-}$ superconductivity
gaps out low-energy excitations, and the superconducting order parameter
changes sign between hole and electron pockets. Stripe SDW magnetism
with momentum $(0,\pi)$ or $(\pi,0)$ (shown) mixes hole and electron
states by band-folding and split hole and electron pockets into even
smaller sub-pockets. Orbital order elongates the two hole pockets
in opposite directions and makes one electron pocket larger and the
other one smaller. \label{fig:schematic} }
\end{figure}

In this work we treat magnetism, superconductivity, and orbital order
on equal footing. We use the renormalization group (RG) technique,
which is the most unbiased way to analyze how different interaction
channels affect each other and what is the leading (and the subleading)
instability in the system~\cite{cee,maiti,rice,honerkamp,thomalle,lee}.
We list potential instabilities in Fig. \ref{fig:schematic} and show
how each reconstructs the fermionic states. We consider a model with
repulsive intra-pocket interaction,  like in earlier studies of the interplay between magnetism and
superconductivity.
However, in distinction to earlier works~\cite{fernandes_12,cee,maiti}
we explicitly include into consideration the orbital composition of
the low-energy electronic states. This allows us to consider fluctuations
in the orbital channel on equal footing with fluctuations in the magnetic
and superconducting channels. We assume that there is a substantial
energy range of metallic behavior and do not discuss Hund metallic
behavior~\cite{kotliar} and orbitally-selective Mottness~\cite{medici}.

The two key questions we address are: (i) How one can get an attraction
in the orbital channel
out of purely repulsive bare interactions? (ii) If the orbital channel
is attractive, can orbital order develop, upon lowering the temperature,
prior to magnetism and superconductivity?

We show that the outcome depends on whether the leading instability
develops at a temperature $T_{\mathrm{ins}}$ smaller or larger than
the Fermi energy $E_{F}$. When $T_{\mathrm{ins}}<E_{F}$, the orbital
composition of the low-energy excitations does not play a crucial
role and the system develops either SDW and Ising-nematic order or
$s^{+-}$ SC order. This is the case for most iron-based systems.
However, when $T_{\mathrm{ins}}>E_{F}$, which is the case of FeSe,
the orbital composition is crucial, and the $d$-wave POM instability
occurs prior to the SDW and SC instabilities, giving rise to a spontaneous
orbital order. This orbital order, however, is not the consequence
of a strong attraction in the POM channel at the bare (mean-field)
level. Instead, the instability in the POM channel is induced and
pushed to a higher temperature by magnetic fluctuations, which take
advantage of the repulsive electronic interactions of the system.
At the same time, this magnetically-driven Pomeranchuk instability
is very different from the Ising-nematic instability, because it is
not a vestige of a stripe magnetic order.

The interplay between SDW, SC, and POM orders in FeSCs has been earlier
analyzed numerically using functional RG (fRG) approach~\cite{honerkamp,thomalle,lee},
and the POM channel was found to be a distant third, after SDW and
SC. This conclusion, however, follows from comparative analysis of
the running couplings in different channels. We argue that the analysis
of the couplings is insufficient, and to analyze which channel becomes
unstable first, one has to compare the corresponding susceptibilities.
This is how POM channel comes ahead of SC and SDW channels. We use
analytical parquet RG (pRG), which allows us to analyze the flow not
only of the couplings, but also of susceptibilities.

\textbf{The Model}~~~~ We depart from the actual underlying 2D
microscopic model in the orbital basis. The kinetic energy is given
by the hopping terms involving all five Fe-orbitals (direct and via
pnictigen/chalcogen sites) and the potential energy describes onsite
interactions between the Fe-orbitals. These interactions include intra-orbital
and inter-orbital Hubbard and Hund terms (Refs. \cite{kuroki,kemper,kotliar}).
We convert from orbital into band basis and obtain the corresponding
band model. The orbital composition of the excitations does not show
up in the kinetic part of the Hamiltonian in the band basis, but it
imposes angular dependencies on the four-fermion interaction terms.
As we will show, the terms with different angular dependencies flow
differently under pRG.

The fermionic structure in the band basis contains hole and electron
pockets. Two hole pockets are centered at the $\Gamma$ point ($k_{x}=k_{y}=0$)
and are constructed out of $d_{xz}$ and $d_{yz}$ orbitals (Fig.
\ref{fig:schematic}). In some materials there exists another hole
pocket, centered at $(\pi,\pi)$ in the 1-Fe zone and made fully out
of $d_{xy}$ orbital~\cite{fifth_pocket}. This pocket will not play
a role in our analysis and we neglect it. The Fourier components of
the $d_{xz}$ and $d_{yz}$ operators with momenta $k$ near $\Gamma$
are related to $c_{k}$ and $d_{k}$ operators describing excitations
near the two $\Gamma-$centered hole pockets by a rotation~\cite{Cvetkovic2013}
\begin{equation}
d_{xz,k}=\cos{\theta}_{k}c_{k}+{\sin\theta}_{k}d_{k},~~d_{yz,k}=\cos{\theta}_{k}d_{k}-\sin{\theta}_{k}c_{k},~~\label{a_2}
\end{equation}
The rotation angle $\theta_{k}$ coincides with the angle along the
hole Fermi surface if the hole pockets can be approximated as circular,
which we assume to be the case. The extension to a more general Fermi
surface geometry complicates the formulas but does not
introduce new physics. The kinetic energy in the band basis is ${\cal H}_{2,h}=\sum_{k}\epsilon_{c,k}c_{k}^{\dagger}c_{k}+\epsilon_{d,k}d_{k}^{\dagger}d_{k}$,
where $\epsilon_{c,k}=\mu-k^{2}/(2m_{c})$ and $\epsilon_{d,k}=\mu-k^{2}/(2m_{d})$,
with $k$ near the $\Gamma$ point. The two dispersions are are not
identical when $m_{c}\neq m_{d}$, but are degenerate by symmetry
at $k=0$ in the absence of spin-orbit coupling~\cite{Cvetkovic2013}.
The degeneracy implies that both $\Gamma$-centered hole pockets
\textit{must be present simultaneously already in the minimal model}.

The two electron pockets are centered at $Q_{1}=(0,\pi)$ and $Q_{2}=(\pi,0)$
in the 1-Fe Brillouin zone (Fig. \ref{fig:schematic}). The kinetic
energy of the fermions near the electron pockets is ${\cal H}_{2,e}=\sum_{k}\epsilon_{f_{1},k}f_{1,k}^{\dagger}f_{1,k}+\epsilon_{f_{2},k}f_{2,k}^{\dagger}f_{2,k}$,
where $\epsilon_{f_{1},k}=\epsilon_{0}+k_{x}^{2}/(2m_{x})+k_{y}^{2}/(2m_{y})-\mu$
and $\epsilon_{f_{2},k}=\epsilon_{0}+k_{x}^{2}/(2m_{y})+k_{y}^{2}/(2m_{x})-\mu$,
with $k$ measured with respect to $Q_{i}$ for $f_{i,k}\equiv f_{i,k+Q_{i}}$.
The two electron pockets are related by $C_{4}$ symmetry and transform
into each other under a $\pi/2$ rotation. The band fermions $f_{1,k+Q_{1}}$
and $f_{2,k+Q_{2}}$ are linear combinations of $d_{xz}/d_{xy}$ and
$d_{yz}/d_{xy}$ orbitals, respectively~\cite{kuroki,kemper}, and
the relative amplitude of the spectral weights depends on system parameters.

The interactions between low-energy fermions are Hubbard and Hund
terms expressed via corresponding band operators. Although there are
only four interactions at the bare level ($U,U',J,J'$), the number
of topologically distinct invariant combinations of 4-fermion terms
is much higher and equals 30 for a generic 4-band model in the absence
of spin-orbit coupling~\cite{Cvetkovic2013}. The bare values of
all 30 couplings are expressed
in terms of $U,U',J,J'$, but under pRG the couplings flow to different
values. To make the problem analytically treatable, we neglect the
$d_{xy}$ spectral weight on the electron pockets, i.e. we identify
the excitations near the $(0,\pi)$ ($(\pi,0)$) pocket with the $d_{xz}$
($d_{yz}$) orbital, $f_{1,k+Q_{1}}=d_{xz,k+Q_{1}}$ and $f_{2,k+Q_{2}}=d_{yz,k+Q_{2}}$.
This approximation reduces the number of couplings to manageable 14.
As a verification, we considered the opposite case, when we kept only
the $d_{xy}$ spectral weight on the two electron pockets. We obtained
the same results as with pure $d_{xz}$ ($d_{yz}$) pockets. This
gives us confidence that the approximation we make does not change
the physics.

The 14 different interaction parameters are the prefactors for 14
combinations of the original $152$ interaction terms in the band
basis (96 involving $c$ and $d$ fermions, 8 involving $f_{1}$ and
$f_{2}$ fermions, and 48 cross-terms), combined using the symmetry
condition that under rotation by $\pi/2$, $c_{k}\to-d_{k}$, $d_{k}\to c_{k}$,
and $f_{1}\to f_{2}$. We present the full form of the interaction
term ${\cal H}_{4}$ in the Supplementary Material (SM), and here
show a representative
set from each combination:
\begin{eqnarray}
 &  & {\cal H}_{4}=\sum_{k_{i}}c_{k_{1},\alpha}^{\dagger}f_{1;k_{2},\beta}^{\dagger}f_{1;k_{3},\beta}c_{k_{4},\alpha}\left[U_{1}\cos{\theta_{\mathbf{k}_{1}}}\cos{\theta_{\mathbf{k}_{4}}}+{\bar{U}}_{1}\sin{\theta_{\mathbf{k}_{1}}}\sin{\theta_{\mathbf{k}_{4}}}\right]\nonumber \\
 &  & +\sum_{k_{i}}c_{k_{1},\alpha}^{\dagger}f_{1;k_{2},\beta}^{\dagger}c_{k_{3},\beta}f_{1;k_{4},\alpha}\left[U_{2}\cos{\theta_{\mathbf{k}_{1}}}\cos{\theta_{\mathbf{k}_{3}}}+{\bar{U}}_{2}\sin{\theta_{\mathbf{k}_{1}}}\sin{\theta_{\mathbf{k}_{3}}}\right]\nonumber \\
 &  & +\sum_{k_{i}}c_{k_{1},\alpha}^{\dagger}c_{k_{2},\beta}^{\dagger}f_{1;k_{3},\beta}f_{1;k_{4},\alpha}\left[\frac{U_{3}}{2}\cos{\theta_{\mathbf{k}_{1}}}\cos{\theta_{\mathbf{k}_{2}}}+\frac{{\bar{U}}_{3}}{2}\sin{\theta_{\mathbf{k}_{1}}}\sin{\theta_{\mathbf{k}_{2}}}\right]\nonumber \\
 &  & +\sum_{k_{i}}c_{k_{1},\alpha}^{\dagger}c_{k_{2},\beta}^{\dagger}c_{k_{3},\beta}c_{k_{4},\alpha}\left[\frac{U_{4}}{2}\cos{\theta_{\mathbf{k}_{1}}}\cos{\theta_{\mathbf{k}_{2}}}\cos{\theta_{\mathbf{k}_{3}}}\cos{\theta_{\mathbf{k}_{4}}}+\frac{{\bar{U}}_{4}}{2}\cos{\theta_{\mathbf{k}_{1}}}\cos{\theta_{\mathbf{k}_{2}}}\sin{\theta_{\mathbf{k}_{3}}}\sin{\theta_{\mathbf{k}_{4}}}\right]\nonumber \\
 &  & +\sum_{k_{i}}c_{k_{1},\alpha}^{\dagger}c_{k_{2},\beta}^{\dagger}c_{k_{3},\beta}c_{k_{4},\alpha}\left[{\tilde{U}}_{4}\cos{\theta_{\mathbf{k}_{1}}}\sin{\theta_{\mathbf{k}_{2}}}\sin{\theta_{\mathbf{k}_{3}}}\cos{\theta_{\mathbf{k}_{4}}}+{\tilde{\tilde{U}}}_{4}\cos{\theta_{\mathbf{k}_{1}}}\sin{\theta_{\mathbf{k}_{2}}}\cos{\theta_{\mathbf{k}_{3}}}\sin{\theta_{\mathbf{k}_{4}}}\right]\nonumber \\
 &  & +\sum_{k_{i}}\frac{U_{5}}{2}f_{1;k_{1},\alpha}^{\dagger}f_{1;k_{2},\beta}^{\dagger}f_{1;k_{3},\beta}f_{1;k_{4},\alpha}+\frac{{\bar{U}}_{5}}{2}f_{1;k_{1},\alpha}^{\dagger}f_{1;k_{2},\beta}^{\dagger}f_{2;k_{3},\beta}f_{2;k_{4},\alpha}\nonumber \\
 &  & +\sum_{k_{i}}{\tilde{U}}_{5}f_{1;k_{1},\alpha}^{\dagger}f_{2;k_{2},\beta}^{\dagger}f_{2;k_{3},\beta}f_{1;k_{4},\alpha}+{\tilde{\tilde{U}}}_{5}f_{1;k_{1},\alpha}^{\dagger}f_{2;k_{2},\beta}^{\dagger}f_{1;k_{3},\beta}f_{2;k_{4},\alpha}+\dots\label{ya_1}
\end{eqnarray}
where $\dots$ stand for other terms in each of the 14 combinations
in (\ref{ya_1}). Out of the 14 interactions, 4 are density-density,
exchange and pair-hopping terms for fermions near the two hole pockets
($U_{4},{\bar{U}}_{4},\tilde{U}_{4},\tilde{\tilde{U}}_{4}$), another
4 are
analogous interactions for fermions near the two electron pockets
($U_{5},{\bar{U}}_{5},\tilde{U}_{5},\tilde{\tilde{U}}_{5}$), and
6 involve fermions near both hole and electron pockets ($U_{1},{\bar{U}}_{1},U_{2},{\bar{U}}_{2},U_{3},{\bar{U}}_{3})$.
The bare values of these 14 couplings are $U_{1}=U_{2}=U_{3}=U_{4}=U_{5}=U$,
${\bar{U}}_{1}={\tilde{U}}_{4}={\tilde{U}}_{5}=U'$, ${\bar{U}}_{2}={\tilde{\tilde{U}}}_{4}={\tilde{\tilde{U}}}_{5}=J$,
${\bar{U}}_{3}={\bar{U}}_{4}={\bar{U}}_{5}=J'$.

\textbf{RG equations}~~~~ In the mean-field approach the bare
values of these 14 couplings are used to compute susceptibilities
in SDW, SC, POM and other channels. A simple analysis shows that in
mean-field, SDW wins over SC and orbital order. However, the mean-field
approach is strongly questionable because it effectively isolates
each electronic channel, neglecting their interplay and mutual feedback.
To overcome this limitation, here we implement a pRG approach and
calculate how the couplings and the susceptibilities in different
channels evolve as high-energy degrees of freedom are integrated out.
In this approach, each dimensionless coupling $u_{i}=(A_{i}/4\pi)U_{i}$,
where $A_{i}$ are combinations of effective masses, acquires a dependence
on the running energy/temperature scale $E$ via $L=\log{W/E}$, where
$W$ is of the order of the bandwidth.

The derivation of the one-loop RG equations is tedious but straightforward.
We present the details and the full equations in the SM and here list
the 14 pRG equations in the approximation $m_{c}=m_{d}=m_{h},m_{x}=m_{y}=m_{e}$:
\begin{align}
\dot{u}_{1} & =u_{1}^{2}+u_{3}^{2}/C^{2},~~\dot{\bar{u}}_{1}=\bar{u}_{1}^{2}+\bar{u}_{3}^{2}/C^{2},~~\dot{u}_{2}=2u_{1}u_{2}-2u_{2}^{2},~~\dot{\bar{u}}_{2}=2\bar{u}_{1}\bar{u}_{2}-2\bar{u}_{2}^{2}\notag\label{RG_13}\\
\dot{u}_{3} & =-u_{3}u_{4}-\bar{u}_{3}\bar{u}_{4}+4u_{3}u_{1}-u_{5}u_{3}-u_{8}\bar{u}_{3}-2u_{2}u_{3},~\dot{\bar{u}}_{3}=-\bar{u}_{3}u_{4}-u_{3}\bar{u}_{4}+4\bar{u}_{3}\bar{u}_{1}-u_{5}\bar{u}_{3}-u_{8}u_{3}-2\bar{u}_{2}\bar{u}_{3}\notag\\
\dot{u}_{4} & =-u_{4}^{2}-\bar{u}_{4}^{2}-u_{3}^{2}-\bar{u}_{3}^{2},~~\dot{\bar{u}}_{4}=-2u_{4}\bar{u}_{4}-2u_{3}\bar{u}_{3}~~\dot{u}_{5}=-u_{5}^{2}-u_{8}^{2}-u_{3}^{2}-\bar{u}_{3}^{2},~~\dot{{\bar{u}}}_{5}=-2u_{5}{\bar{u}}_{5}-2u_{3}\bar{u}_{3}\notag\\
\dot{\tilde{u}}_{4} & =-(\tilde{u}_{4}^{2}+\tilde{\tilde{u}}_{4}^{2}),~~\dot{\tilde{\tilde{u}}}_{4}=-2\tilde{u}_{4}\tilde{\tilde{u}}_{4},~~\dot{{\tilde{u}}}_{5}=-({\tilde{u}}_{5}^{2}+{\tilde{\tilde{u}}}_{5}^{2}),~~\dot{{\tilde{\tilde{u}}}}_{5}=-2{\tilde{u}}_{5}{\tilde{\tilde{u}}}_{5}
\end{align}
where $C=(m_{e}+m_{h})/(2\sqrt{m_{e}m_{h}})$.

One can immediately verify that the running couplings flow to different
values under the pRG, and these new values \textit{cannot} be re-expressed
just in terms of running $U,U',J,J'$. As a consequence, the model
with only local interactions does not survive under renormalization
and longer-range interactions emerge in the process of pRG flow. The
minimal model with 14 couplings includes all symmetry-allowed interactions
within
a plaquette of four Fe atoms. We present the corresponding Hamiltonian
in the SM.

\begin{figure}[h]
\centering{}\includegraphics[width=0.7\columnwidth]{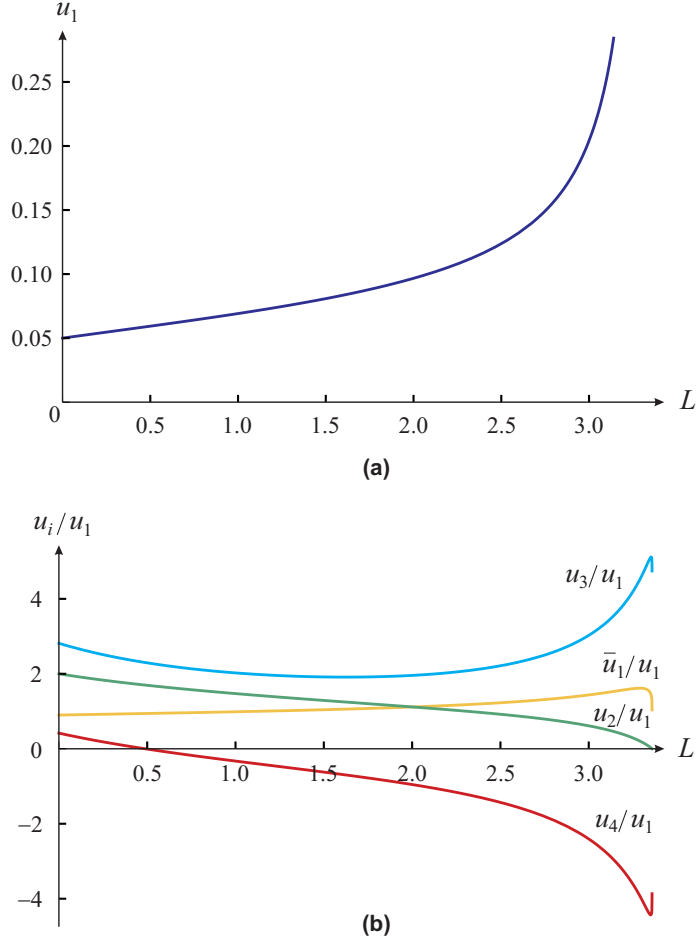} \protect\caption{\textbf{The pRG flow of the couplings.}
 Panel (a) -- $u_{1}(L)$, where $L=\log{W/E}$ is
the RG scale, $W$ is the bandwidth, and $E$ is the running energy
(temperature) at which one probes the system. The flow of other couplings
is similar. The couplings $u_{1}-u_{5}$ and ${\bar{u}}_{1}-{\bar{u}}_{5}$
all diverge as $1/(L_{0}-L)$ when $L$ approaches $L_{0}$, whose
value depends on the initial conditions. The couplings ${\tilde{u}}_{4},{\tilde{\tilde{u}}}_{4},{\tilde{u}}_{5},{\tilde{\tilde{u}}}_{5}$
tend to zero at $L\to L_{0}$. Panel (b) -- flow of the ratios of
the couplings. All ratios tend to fixed finite values as $L$ approaches
$L_{0}$: $\bar{u}_{1}=u_{1}$, $u_{3}={\bar{u}}_{3}=4.7u_{1}$, $u_{4}=u_{5}={\bar{u}}_{4}={\bar{u}}_{5}=-3.8u_{1}$
(see Eq. \protect\ref{a_4}). The ratios $u_{2}/u_{1}$ and ${\bar{u}}_{2}/u_{1}$
tend to zero as $L$ approaches $L_{0}$. The initial values used
were ${\bar{u}}_{1}/u_{1}=0.9$, $u_{2}/u_{1}=2$, $u_{3}/u_{1}=2.8$,
$u_{4}/u_{1}=0.4$. In both panels we set $C=(m_{e}+m_{h})/(2\sqrt{m_{e}m_{h}})=1.1$
for definiteness.
For the model with  electron pockets the fixed trajectory is the same, but the system approaches it much faster (see SM).
 \label{fig:exponents} }
\end{figure}

\textbf{RG flow}~~~~ The analysis of Eq. (\ref{RG_13}) readily
reveals that the last four RG equations decouple from the other ten,
and that ${\tilde{u}}_{4,5}$ and ${\tilde{\tilde{u}}}_{4,5}$ flow
to zero under pRG. The remaining ten pRG equations are all coupled
and have to be solved self-consistently. For $U'=J=J'=0$, the bare
values of the couplings ${\bar{u}}_{i}$ ($i=1-5$) are zero, and
a straightforward analysis of Eq. (\ref{RG_13}) shows that they remain
zero under pRG. This leads to the same system behavior as found in
previous studies~\cite{maiti}. However, the solution with ${\bar{u}}_{i}=0$
becomes unstable already for arbitrarily small $U',J$ and $J'$,
i.e. for arbitrarily small bare ${\bar{u}}_{i}$. We have analyzed
the pRG equations for non-zero bare ${\bar{u}}_{i}$ and found that
the system flows towards a single stable fixed trajectory, along which
$u_{i}$ and ${\bar{u}}_{i}$ become equivalent. This implies that
the terms ${\bar{u}}_{i}$, which were originally of order $U'$ or
even $J$, grow under pRG and eventually become comparable to $u_{i}$,
which were originally of order $U$. In other words, the initial hierarchy
of interactions disappears under the pRG flow towards the fixed trajectory~\cite{comm}.
We show the RG flow in Fig. \ref{fig:exponents}.

Along the stable fixed trajectory, the ratios between various couplings
become pure numbers: $u_{2}=\gamma_{2}u_{1},~u_{3}={\bar{u}}_{3}=\gamma_{3}u_{1},~u_{4}={\bar{u}}_{4}=\gamma_{4}u_{1},~u_{5}={\bar{u}}_{5}=\gamma_{5}u_{1}$.
Solving (\ref{RG_13}) for $u_{1}$ and $\gamma_{i}$ we obtain
\begin{eqnarray}
 &  & u_{1}=\frac{a}{(L_{0}-L)},~a=1/(8C^{2}+4\sqrt{1-C^{2}+4C^{4}}),~\gamma_{2}={\bar{\gamma}}_{2}=0,\nonumber \\
 &  & \gamma_{3}=C\sqrt{8C^{2}-1+4\sqrt{1-C^{2}+4C^{4}}},~\gamma_{4}=\gamma_{5}=1-2C^{2}-\sqrt{1-C^{2}+4C^{4}}\label{a_4}
\end{eqnarray}
For $C=1$, which corresponds to perfect nesting, we have $\gamma_{3}=\sqrt{15}$
and $\gamma_{4}=-3$. The couplings diverge at the logarithmic scale
$L=L_{0}=O(W/U)\gg1$, whose exact value depends on $U,U'$ and $J$.
Note that $\gamma_{4}=\gamma_{5}$ is negative for arbitrary $C$,
hence the couplings $u_{4}$ and $u_{5}$ necessary change sign under
the pRG and become negative along the fixed trajectory. We emphasize
that the RG equations are valid up to $L_{F}=\log{W/E_{F}}$, where
$E_{F}$ is the \textit{largest} of the Fermi energies. For $E<E_{F}$,
particle-particle and particle-hole channels no longer ``talk''
to each other and the flow equation is different (see below).

\textbf{Competition between channels}~~~~ We now use the results
for the pRG flow to find which of the many electronic channels becomes
unstable first upon lowering the running energy $E$, which from physics
perspective is equivalent to lowering the temperature $T$. For this
we introduce infinitesimally small vertices $\Gamma_{0,i}$ for the
coupling between fermions and order parameters in different channels
($i$ = SDW, SC, POM, or AFO), and identify the combinations of the
couplings $U^{i}$ which renormalize $\Gamma_{i}^{(0)}$ into $\Gamma_{i}=\Gamma_{i}^{(0)}(1+U^{i}\Pi_{i}+...)$,
where $\Pi_{i}$ are the corresponding polarization bubbles. We present
the details in SM and list $U^{i}$ in Table \ref{tab:couplings}.

Earlier pRG and fRG studies assumed that the channel with the largest
$U^{i}$ along the fixed trajectory wins. We argue that this procedure
is incomplete, and to compare different channels one actually needs
to obtain and solve another set of pRG equations for $\Gamma_{i}$,
then compute the corresponding susceptibilities, find which ones diverge,
and compare the exponents. The leading instability will be in the
channel in which the exponent is the largest. This procedure has been
applied to the one-band Hubbard model~\cite{metzner} and bi-layer
graphene~\cite{oskar}, but has not yet been applied to FeSCs. The
advantage of using analytical pRG in this procedure is that the RG
equations for $\Gamma_{i}$ and for the susceptibilities can be obtained
in a straightforward way.

The analysis of the susceptibilities is different for the SC/SDW channels
and the POM channel. For the SC and SDW channels, $\Pi_{i}$ is logarithmic,
and, to logarithmic accuracy,
\begin{equation}
\chi_{\mathrm{SDW}}(L)\propto\int_{L}dL'\Gamma_{\mathrm{SDW}}^{2}(L'),~~\chi_{\mathrm{SC}}=\int_{L}dL'\left(\Gamma_{\mathrm{SC}}^{s+-}\right)^{2}(L')
,\label{a_6}
\end{equation}
where $\Gamma_{\mathrm{SDW}}(L')$ and $\Gamma_{\mathrm{SC}}^{s+-}(L')$
are the fully renormalized SDW and SC vertices obtained from the solutions
of pRG equations $\dot{\Gamma}_{i}\propto\Gamma_{i}U^{i}$. We derive
these equations in the SM and present them here for the couplings
along the fixed trajectory:
\begin{equation}
\dot{\Gamma}_{\mathrm{SDW}}=\Gamma_{\mathrm{SDW}}u_{1}\left(1+\frac{\gamma_{3}}{C}\right),~~\dot{\Gamma}_{\mathrm{SC}}=\Gamma_{\mathrm{SC}}u_{1}\left(2\gamma_{3}+2|\gamma_{4}|\right),
\label{RG_13_1}
\end{equation}
where $\gamma_{3,4}$ are given by (\ref{a_4}). Solving these two
equations and substituting the results into (\ref{a_6}) we obtain
\begin{equation}
\chi_{\mathrm{SDW}}(L)\propto\frac{1}{(L_{0}-L)^{\alpha_{\mathrm{SDW}}}},~~\chi_{\mathrm{SC}}(L)\propto\frac{1}{(L_{0}-L)^{\alpha_{\mathrm{SC}}}}\label{a_7}
\end{equation}
with the exponents
\begin{equation}
\alpha_{\mathrm{SDW}}=2\frac{1+\gamma_{3}/C}{1+\gamma_{3}^{2}/C^{2}}-1,~~\alpha_{\mathrm{SC}}=4\frac{|\gamma_{4}|+\gamma_{3}}{1+\gamma_{3}^{2}/C^{2}}-1\, .
\label{a_8}
\end{equation}

In Fig. \ref{fig:exponents}a we plot $\alpha_{i}$ as a function
of $C=(m_{e}+m_{h})/(2\sqrt{m_{e}m_{h}})\geq 1$. We see that for all
values of $C$, $1>\alpha_{\mathrm{SC}}>0$, while $\alpha_{\mathrm{SDW}}<0$.
This implies that that only SC order develops. SDW order does not
develop, despite that at the bare level SDW channel was the only attractive
channel. We show the behavior of the susceptibilities in SDW and SC
($s^{+-}$) channels
 along the fixed trajectory in Fig. \ref{fig:exponents}b.

The phenomenon
 in which SDW interaction pushes up superconductivity but by itself
gets cut by the feedback effect from the rising superconducting fluctuations
had already been found in earlier pRG and fRG studies of multi-band
FeSCs~\cite{thomalle,maiti} as well as in pRG and fRG analysis of
doped graphene~\cite{rahul,thomalle_2}. In our case this effect
is additionally enhanced because $\alpha_{\mathrm{SC}}$ contains
contributions from $u_{1}$, $u_{3}$ and ${\bar{u}}_{1}$, ${\bar{u}}_{3}$,
hence the factor of 2 in front of $\gamma-$dependent term in (\ref{a_8}),
while $\alpha_{\mathrm{SDW}}$ contains contributions only from $u_{3}$
and $u_{4}$, but not ${\bar{u}}_{3}$ and ${\bar{u}}_{4}$.

We now turn to the POM channel. Here the situation is different because
the particle-hole polarization bubble at energy $E$ is determined
by fermions with energies of order $E$. As a result, within one-loop,
s-wave ($s^{+-}$) and d-wave Pomeranchuk susceptibilities obey algebraic
rather than differential equations (see SM) and behave as
\begin{equation}
\chi_{\mathrm{POM}}^{s}\propto\frac{1}{1-u_{1}(4C+|\gamma_{4}|)}=\frac{1}{L_{P_{s}}-L},~~\chi_{\mathrm{POM}}^{d}\propto\frac{1}{1-u_{1}|\gamma_{4}|}=\frac{1}{L_{P_{d}}-L}\, .
\end{equation}
For both susceptibilities, the exponent $\alpha_{\mathrm{POM}}=1$
is larger than $\alpha_{\mathrm{SC}}<1$. Furthermore, for all values
of $C$, $L_{P_{s}}$ are smaller than $L_{0}$ (for $C=1$, $L_{P_{s}}=L_{0}-7/16,L_{P_{d}}=L_{0}-3/8$).
As a result, within one-loop pRG, \textit{the first instability upon
lowering the temperature actually occurs in the Pomeranchuk channel}.
We show the behavior of the susceptibilities in SDW, SC ($s^{+-}$)
and POM channels in Fig. \ref{fig:exponents}b.

 Note that at $L=L_{P_{s,d}}$, $u_{1}\sim 1$, and the corrections
to one-loop pRG may become relevant. Still, the comparison of the
susceptibilities clearly favors the POM channel over SC and SDW channels.
Also, number-wise, for $L=L_{P_{d}}$ and $C=1$,
 $u_{1}=(1/16)/(L_{0}-L)=1/6$, which is still a small number.

Of the two Pomeranchuk susceptibilities, the larger one is in the
$s^{+-}$ ($A_{1g}$) channel. An order of this kind splits the chemical
potentials on hole and electron pockets, but conserves the total number
of carriers. Because this does not correspond to a true symmetry breaking,
the divergence of $\chi_{\mathrm{POM}}^{s}$ must be softened by terms
beyond RG, such as the fermionic self-energy~\cite{lara}. Yet, the
relative chemical potential shift $\mu_{h}-\mu_{e}$ must be enhanced
near the temperature at which $\chi_{\mathrm{POM}}^{s}$ diverges
within the RG. Interestingly, the analysis of ARPES data for several
FeSCs did find~\cite{dhaka} some evidence for temperature-dependent
$\mu_{h}-\mu_{e}$.

The true Pomeranchuk instability is in the $d$-wave ($B_{1g}$) channel,
signaled by the divergence of $\chi_{\mathrm{POM}}^{d}$. This instability
implies that the mean-values of $\Delta_{1h}=\sum_{k}\left\langle c_{k}^{\dagger}c_{k}-d_{k}^{\dagger}d_{k}\right\rangle \cos{2\theta_{k}}$,
$\Delta_{2h}=\sum_{k}\left\langle c_{k}^{\dagger}d_{k}+d_{k}^{\dagger}c_{k}\right\rangle \sin{2\theta_{k}}$,
and $\Delta_{e}=\sum_{k}\left\langle f_{1,k+Q_{1}}^{\dagger}f_{1,k+Q_{1}}\right\rangle =-\sum_{k}\left\langle f_{2,k+Q_{2}}^{\dagger}f_{2,k+Q_{2}}\right\rangle $
become non-zero. The solution of the set of coupled equations for
$\Delta_{1h},\Delta_{2h}$, and $\Delta_{e}$ at $L=L_{P_{d}}$ yields
$\Delta_{1h}=\Delta_{2h}=4\Delta_{e}$ (see SM). Converting these
results to the orbital basis, we find that $\left\langle d_{xz}^{\dagger}d_{xz}\right\rangle -\left\langle d_{yz}^{\dagger}d_{yz}\right\rangle $
becomes non-zero, while the cross term $\left\langle d_{xz}^{\dagger}d_{yz}+d_{yz}^{\dagger}d_{xz}\right\rangle $
remains zero. This corresponds precisely to ferro-orbital order. We
emphasize that the origin of this ferro-orbital order is not just
an attraction in the POM channel, as proposed by other works. In our
case the bare interaction well may be repulsive (when $U+J>2U'$,
see above), yet the POM channel becomes attractive in the process
of pRG flow and eventually wins over SC and SDW. The attraction in
POM chanel is driven by the coupling to magnetic fluctuations, and
in this respect the pRG scenario or orbital ordering falls into the
orbit of ``magnetic scenarios''.

Therefore, the full one-loop pRG analysis shows that the system first
develops a ferro-orbital order at $T_{s}$ and then becomes a superconductor
at a lower $T_{c}$. SDW order does not develop. This sequence of
transitions is fully consistent with that in FeSe. In other FeSCs,
however, the system does develop SDW order at $T_{N}$ at small dopings,
and the nematic transition line follows $T_{N}$, suggesting that
nematic order is a vestige of the SDW order.

To understand this difference between FeSe and other FeSCs, we note
that in our analysis we assumed that the pRG flow reaches the fixed
trajectory at $L=L_{P_{d}}\approx L_{0}$, before the pRG analysis
breaks down at an energy comparable to the largest $E_{F}$ in the
system, i.e.,
at $L=L_{F}$. This holds when $L_{0}<L_{F}$, i.e., when \textit{all}
Fermi energies are small. If $L_{F}<L_{0}$,
the pRG flow runs up only to $L=L_{F}$, and at larger $L$ the particle-hole
and particle-particle channels decouple from each other. As a result,
the divergence of the Pomeranchuk susceptibility is cut and this channel
no longer competes with SC/SDW. Also,
because the SC and the SDW channels do not mix below $E_{F}$, each
develops independently in a mean-field fashion with the couplings
taken at $L=L_{F}$ (Ref~\cite{chubukov,maiti}). If $L_{F}$ is
small enough, these values are close to the bare ones and the system
develops SDW order (and Ising-nematic order above it, if SDW order
is a stripe). When doping gets larger (and nesting gets weaker), SDW
channel becomes less singular and SC order develops first. This behavior
is consistent with the one observed in most FeSCs, for which the largest
$E_{F}\sim100$ meV well exceeds $T_{N},\: T_{c}\sim10$ meV (Ref.
\cite{largeEF}). In FeSe, on the other hand, all $E_{F}\leq10$ meV
and are comparable to $T_{s}\sim7$ meV~ \cite{FSFeSe}.

\textbf{Summary}~~~ In this paper we employed the analytical pRG
technique to analyze the interplay between SDW, SC, and orbital POM
order in Fe-based superconducting materials. We computed the exponents
for susceptibilities in SDW, SC, and POM channels and found that in
FeSe, where all Fermi energies are small, the system develops a spontaneous
ferro-orbital order, followed by $s^{+-}$ superconductivity, while
SDW order does not develop. In systems in which at least one of
the pockets has $E_{F}$ large, as in LaFeAsO, BaFe$_{2}$As$_{2}$,
and NaFeAs, orbital order does not develop. Instead, SDW and SC orders
compete with each other, with SC winning at higher doping and SDW
winning at smaller doping. In this situation, nematic order is associated
with stripe SDW. Our work provides an appealing unified microscopic
description of the behavior of different families of FeSCs.

\textbf{Acknowledgments}~~~We thank G. Blumberg, A. Boehmer, I. Fisher, P. Hirschfeld, C. Honerkamp, I. Eremin,
S. Kivelson, H. Kontani, I. Mazin, C. Meingast, R. Thomale, V. K. Thorsm\o lle, O. Vafek,  R. Valenti, and Y. Wang for useful discussions.
This work was supported by the Office of Basic Energy Sciences, U.S. Department of Energy, under awards DE-FG02-ER46900 (AVC) and DE-SC0012336
(RMF). MK is supported by the Israel Science Foundation, Grant No. 1287/15 and NSF DMR-1506668.

\newpage{}

\begin{table}[h]
\centering %
\begin{tabular}{|c|c|c|c|}
\hline
SDW  & CDW-r  & CDW-i  & \tabularnewline
\hline
\hline
$u_{1}+u_{3}/C$  & $u_{1}-u_{3}/C-2u_{2}$  & $u_{1}-2u_{2}+u_{3}/C$  & \tabularnewline
\hline
\hline
SC $s^{+-}$  & POM $s$  & POM $d$  & \tabularnewline
\hline
\hline
$2(-u_{4}+u_{3})$  & $2(-u_{4}+4Cu_{1})$  & $-2u_{4}$ & \tabularnewline
\hline
\end{tabular}\protect\caption{The interactions in different channels along the stable fixed trajectory.
All interactions scale as $1/(L_{0}-L)$ and diverge at RG scale $L_{0}$.
We use these interactions to compute vertices and susceptibilities
in SDW channel, CDW channels with real and imaginary order parameters
(CDW-r and CDW-i), $s^{+-}$ superconducting channel, and $s-$wave
and $d-$wave Pomeranchuk channels (POM $s$ and POM $d$). \label{tab:couplings} }

\end{table}

\begin{figure}[h]
\centering{}\includegraphics[width=0.65\columnwidth]{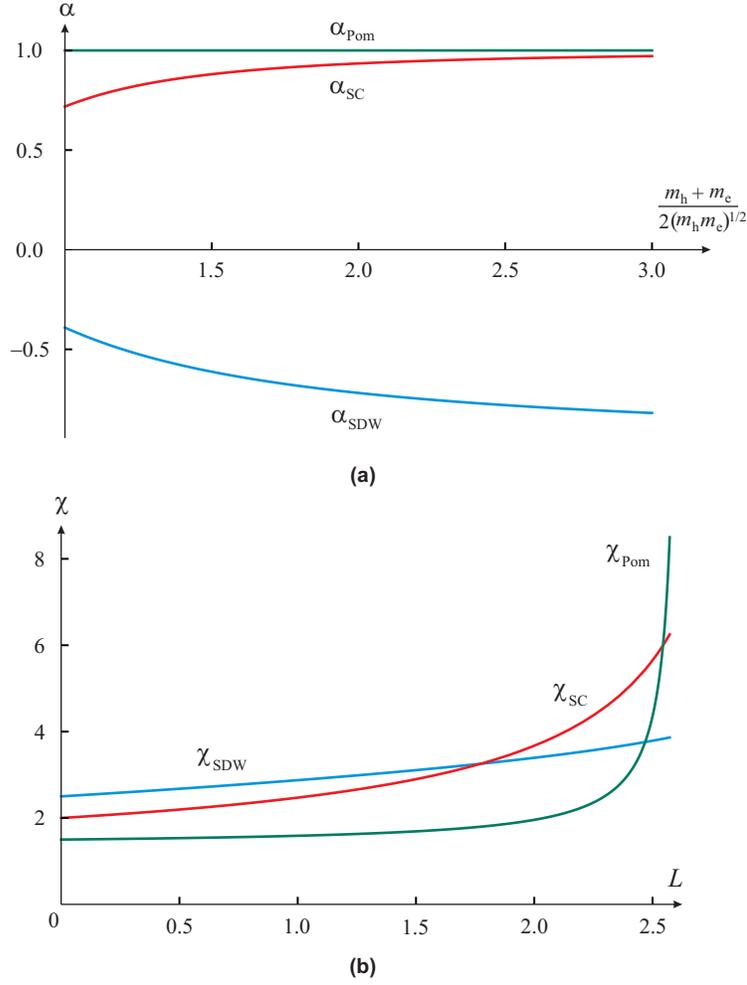}
\protect\caption{\textbf{The pRG flow of the susceptibilities.} (a) The exponents for
the susceptibilities $\chi_{i}\propto1/(L_{0}-L)_{i}^{\alpha}$ in
SDW,  $s^{+-}$ SC, and POM channels as functions of $C=(m_{e}+m_{h})/(2\sqrt{m_{e}m_{h}})$.
The largest exponent $\alpha_{\mathrm{POM}}=1$ is in the Pomeranchuk
channel. The exponent $\alpha_{\mathrm{SDW}}<0$, what implies that
within pRG $\chi_{\mathrm{SDW}}$ does not diverge. (b) The behavior
of susceptibilities in SDW,
$s^{+-}$ SC, and POM channels.  The Pomeranchuk susceptibility actually
diverges at $L=L_{P}<L_{0}$. As a result, the leading instability
upon lowering the temperature is towards d-wave orbital ordering.
$s^{+-}$ superconductivity develops at a smaller $T$, and SDW instability
does not develop. This holds when $L_{0}$ is smaller than $L_{F}=\log W/E_{F}$,
i.e., if the instability develops at an energy/temperature larger
than $E_{F}$. If $L_{F}<L_{0}$, the pRG flow runs up to $L=L_{F}$,
and at larger $L$ SDW and SC channel decouple and develop independent
on each other, while the Pomeranchuk channel gets frozen. In this
situation, the system first develops either SDW or SC order, depending
on the interplay between $L_{F}$ and $L_{0}$ and the degree of nesting.
\label{fig:exponents_1} }
\end{figure}

\pagebreak

\newpage

{}\textbf{\large{}Supplementary material for ``Magnetism,
superconductivity, and spontaneous orbital order in iron-based superconductors:
who comes first and why?''}{\large{} }\\
{\large \par}

\section{Model Hamiltonian}

We follow Ref.~\cite{Cvetkovic2013} to construct the low energy
Hamiltonian. We introduce low-energy spinor wave function $\psi_{\sigma}^{\dag}(\bm{q})=[f_{1,\sigma}^{\dag}(\bm{q}),f_{2,\sigma}^{\dag}(\bm{q}),d_{1,\sigma}^{\dag}(\bm{q}),d_{2,\sigma}^{\dag}(\bm{q})]$,
where the subscripts $\mu,\nu=1,2$ refer to the $xz$ and $yz$ orbital
content respectively. Below we use 1-Fe Brillouin zone and neglect
neglect processes with momentum transfer $(\pi,\pi)$, which may be
present due to the difference between the hopping via pnictogen/chalcogen
atoms above and below the iron layer. 

\subsection{Transformation from the orbital to the band basis}

The quadratic part of the Hamiltonian is expressed in terms of the
components of the spinor $\psi_{\sigma}^{\dag}(\bm{q})$ as follows,
\begin{align}
\mathcal{H}_{0}=\sum_{\bm{k},\alpha}\sum_{\mu,\nu=1,2}d_{\mu,\alpha}^{\dag}(\bm{k})\mathcal{H}_{\mu,\nu}^{\Gamma}(\bm{k})d_{\nu,\alpha}(\bm{k})+f_{\mu\alpha}^{\dag}(\bm{k})\mathcal{H}_{\mu,\nu}^{M}(\bm{k})f_{\nu,\alpha}(\bm{k})\,,\label{Hfree}
\end{align}
The effective Hamiltonian is specified by
\begin{align}
\mathcal{H}^{\Gamma}(\bm{k})=\begin{bmatrix}\epsilon_{\Gamma}+\frac{k^{2}}{2m_{\Gamma}}+ak^{2}\cos2\theta_{k} & ck^{2}\sin2\theta_{k}\\
ck^{2}\sin2\theta_{k} & \epsilon_{\Gamma}+\frac{k^{2}}{2m_{\Gamma}}-ak^{2}\cos2\theta_{k}
\end{bmatrix}\label{HGamma}
\end{align}
for holes, and by
\begin{align}
\mathcal{H}^{M}(\bm{k})=\begin{bmatrix}\epsilon_{M}+\frac{k^{2}}{2m_{M}}+bk^{2}\cos2\theta_{k} & 0\\
0 & \epsilon_{M}+\frac{k^{2}}{2m_{M}}-bk^{2}\cos2\theta_{k}
\end{bmatrix}\label{HM}
\end{align}
for electrons. In Eqs.~\eqref{HGamma} and \eqref{HM} we denote
$\theta_{k}=\arctan(k_{y}/k_{x})$, $\epsilon_{\Gamma,M}$, $1/m_{\Gamma,M}$,
$a$, $b$ and $c$ are parameters of the model which are determined
by the band structure calculations. In our approximation the electron's
Hamiltonian, \eqref{HM} is diagonal and $f_{1,2}$ are the actual
electron band operators. To simplify calculations, we set $a=c$ in
(\ref{HGamma}) in which case the two hole FSs are circular. The transformation
from orbital to band basis in Eq. (\ref{HGamma}) is just a rotation
\begin{align}
d_{1k\sigma} & =c_{k\sigma}\cos\theta_{k\sigma}+d_{k\sigma}\sin\theta_{k\sigma}\notag\label{band_G}\\
d_{2k\sigma} & =-c_{k\sigma}\sin\theta_{k\sigma}+d_{k\sigma}\cos\theta_{k\sigma}\,.
\end{align}
For $a\neq c$ the transformation to the band basis remains the same
as Eq.~\eqref{band_G}, with the rotation angle $\tilde{\theta}$
that is not identical to the angle $\theta$ formed by the vector
$\bm{k}$ with a given axis. In terms of the band operators, the kinetic
energy, Eq.~\eqref{Hfree} is diagonal,
\begin{align}
\mathcal{H}_{0}=\sum_{\bm{k},\alpha}\left[\epsilon_{c}(\bm{k})c_{k\sigma}^{\dag}c_{k\sigma}+\epsilon_{d}(\bm{k})d_{k\sigma}^{\dag}d_{k\sigma}+\epsilon_{1}(\bm{k})f_{1,k\sigma}^{\dag}f_{1,k\sigma}+\epsilon_{2}(\bm{k})f_{2,k\sigma}^{\dag}f_{2,k\sigma}\right],\label{Hfree1}
\end{align}
where we absorbed the constant terms into the chemical potential.
The band dispersions are
\begin{align}
\epsilon_{c,d}=-\frac{k^{2}}{2m_{c,d}}\,,\quad\epsilon_{1,2}(\bm{k})=\frac{k_{x}^{2}}{2m_{x,y}}+\frac{k_{y}^{2}}{2m_{x,y}}.\label{disp}
\end{align}
$m_{c,d}^{-1}=m_{\Gamma}^{-1}\pm2a$, and $m_{x,y}^{-1}=m_{M}^{-1}\pm2b$.

\subsection{Interaction Hamiltonian}

We depart from the local Hubbard-Hund interaction, in the notations
of Ref. \cite{kemper}
\begin{align}
H_{i}=U\sum_{i,\mu}n_{i,\mu\uparrow}n_{i,\mu\downarrow}+U'\sum_{i,\mu<\mu'}n_{i\mu}n_{i\mu'}+J\sum_{i,\mu'<\mu}\sum_{\sigma\sigma'}\psi_{i\mu\sigma}^{\dag}\psi_{i\mu'\sigma'}^{\dag}\psi_{i\mu\sigma'}\psi_{i\mu'\sigma}+J'\sum_{i,\mu'\neq\mu}\psi_{i\mu\uparrow}^{\dag}\psi_{i\mu\downarrow}^{\dag}\psi_{i\mu'\downarrow}\psi_{i\mu'\uparrow}\,,\label{interaction_K}
\end{align}
where the index $i$ enumerates the iron sites located at $\bm{R}_{i}$
and
\begin{align}
\psi_{\mu\sigma}(\bm{R}_{j})=\frac{1}{\sqrt{N}}\sum_{\bm{q}}\left[d_{\mu\sigma}(\bm{k})+f_{\mu\sigma}(\bm{k})e^{i\bm{Q}_{1(2)}\bm{R}_{j}}\right]e^{i\bm{k}\bm{R}_{j}}\label{site_d}
\end{align}
 is the annihilation operator of an electron at the iron site located
at $\bm{R}_{j}$ with spin $\sigma$ in the orbital state labeled
by $\mu$ ($\mu=1$ and $\mu=2$ refer to $xz$ and $yz$ orbitals respectively).
Further,
 $n_{i\mu\sigma}=\psi_{i\mu\sigma}^{\dag}\psi_{i\mu\sigma}$
is the density operator, $n_{i\mu}=n_{i\mu\uparrow}+n_{i\mu\downarrow}$,
and $N$ is the number of iron atoms. The Eq.~\eqref{interaction_K}
can be rewritten in an SU(2) invariant form as
\begin{align}
H_{i}=\frac{U}{2}\sum_{i,\mu}n_{i,\mu}n_{i,\mu}+\frac{U'}{2}\sum_{i,\mu\neq\mu'}n_{i\mu}n_{i\mu'}+\frac{J}{2}\sum_{i,\mu'\neq\mu}\sum_{\sigma\sigma'}\psi_{i\mu\sigma}^{\dag}\psi_{i\mu'\sigma'}^{\dag}\psi_{i\mu\sigma'}\psi_{i\mu'\sigma}+\frac{J'}{2}\sum_{i,\mu'\neq\mu}\psi_{i\mu\sigma}^{\dag}\psi_{i\mu\sigma'}^{\dag}\psi_{i\mu'\sigma'}\psi_{i\mu'\sigma}\,.\label{interaction_K_1}
\end{align}
Substituting Eq.~\eqref{site_d} into Eq.~\eqref{interaction_K_1}
we obtain
\begin{align}
H_{UJ}= & \frac{U}{2}\sum\nolimits'\left[(f_{1\sigma}^{\dag}f_{1\sigma}+d_{1\sigma}^{\dag}d_{1\sigma})^{2}+(f_{2\sigma}^{\dag}f_{2\sigma}+d_{2\sigma}^{\dag}d_{2\sigma})^{2}+(f_{1\sigma}^{\dag}d_{1\sigma}+d_{1\sigma}^{\dag}f_{1\sigma})^{2}+(f_{2\sigma}^{\dag}d_{2\sigma}+d_{2\sigma}^{\dag}f_{2\sigma})^{2}\right]\notag\label{H_T}\\
 & +U'\sum\nolimits'(f_{1\sigma}^{\dag}f_{1\sigma}+d_{1\sigma}^{\dag}d_{1\sigma})(f_{2\sigma'}^{\dag}f_{2\sigma'}+d_{2\sigma'}^{\dag}d_{2\sigma'})\notag\\
 & +J\sum\nolimits'(f_{1\sigma}^{\dag}f_{2\sigma}f_{2\sigma'}^{\dag}f_{1\sigma'}+d_{1\sigma}^{\dag}d_{2\sigma}d_{2\sigma'}^{\dag}d_{1\sigma'}+f_{1\sigma}^{\dag}d_{2\sigma}d_{2\sigma'}^{\dag}f_{1\sigma'}+d_{1\sigma}^{\dag}f_{2\sigma}f_{2\sigma'}^{\dag}d_{1\sigma'})\notag\\
 & +\frac{J'}{2}\sum\nolimits'(f_{1\sigma}^{\dag}f_{2\sigma}f_{1\sigma'}^{\dag}f_{2\sigma'}+d_{1\sigma}^{\dag}d_{2\sigma}d_{1\sigma'}^{\dag}d_{2\sigma'}+f_{1\sigma}^{\dag}d_{2\sigma}f_{1\sigma'}^{\dag}d_{2\sigma'}+d_{1\sigma}^{\dag}f_{2\sigma}d_{1\sigma'}^{\dag}f_{2\sigma'}+h.c.)
\end{align}
Here the momenta arguments of the fermion operators in each term,
$\bm{k}_{1}$, $\bm{k}_{2}$, $\bm{k}_{3}$, $\bm{k}_{4}$ are omitted
for clarity, and $\sum'$ stands for the summation over the spin indices,
$\sigma,\sigma'$ and over fermion momenta subject to $\bm{k}_{1}-\bm{k}_{2}+\bm{k}_{3}-\bm{k}_{4}=0$,
and also includes the normalization factor $1/N$.

\begin{figure}[h]
\begin{centering}
\includegraphics[width=0.6\columnwidth]{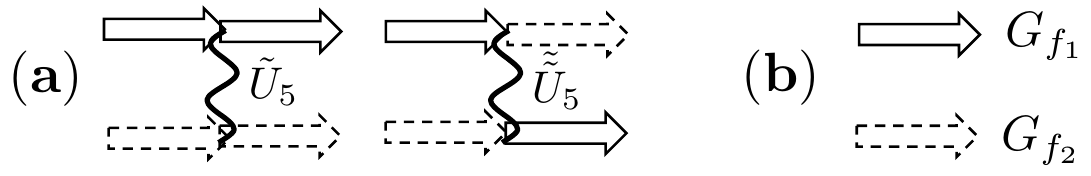} \protect\caption{(a) The subset of the interactions defined by Eq.~\eqref{H_int13}.
(b) The graphical representation of the electron propagators, introduced
in Eq.~\eqref{G_f}. \label{fig:amp1}}

\par\end{centering}

\centering{}
\end{figure}

\begin{figure}[h]
\begin{centering}
\includegraphics[width=0.85\columnwidth]{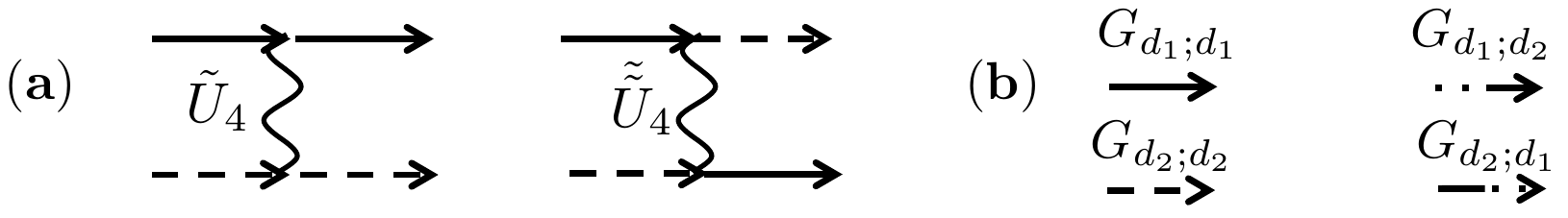} \protect\caption{(a) The subset of the interactions defined by Eq.~\eqref{H_int13}.
(b) The graphical representation of the hole propagators in the orbital
representation, introduced in Eqs.~\eqref{G_hb} and \eqref{G_orb}. \label{fig:amp2}}

\par\end{centering}

\centering{}
\end{figure}

\begin{figure}[h]
\begin{centering}
\includegraphics[width=0.6\columnwidth]{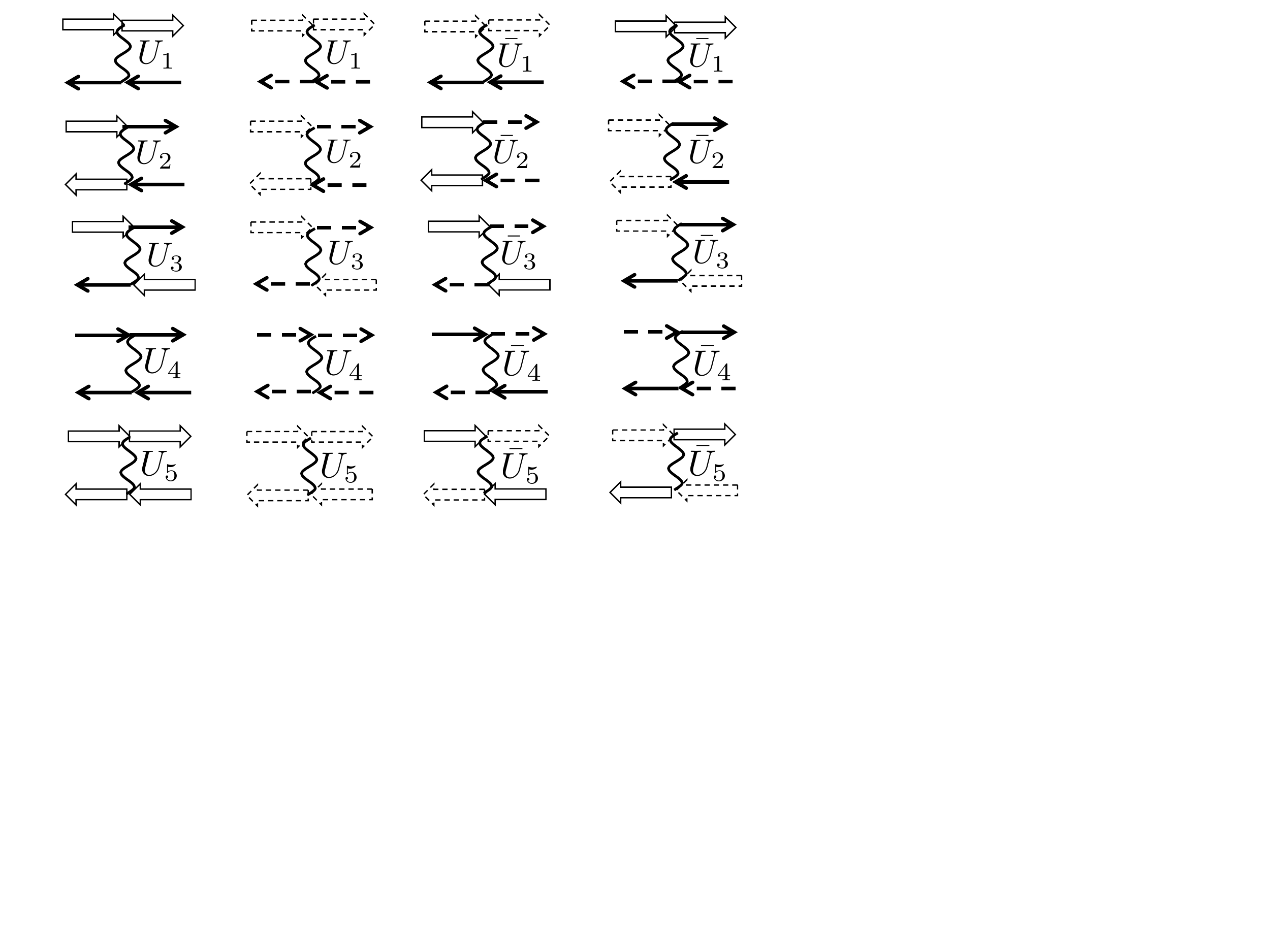} \protect\caption{The subset of the interactions defined by Eq.~\eqref{H_int13}. \label{fig:amp3}}

\par\end{centering}

\centering{}
\end{figure}

The initial observation, which sets the stage for the RG analysis
is that Eq.~\eqref{H_T} is not the most general one consistent with
the tetragonal symmetry. The most general interaction has the form
\begin{align}
H= & U_{1}\sum\nolimits'\left[f_{1\sigma}^{\dag}f_{1\sigma}d_{1\sigma'}^{\dag}d_{1\sigma'}+f_{2\sigma}^{\dag}f_{2\sigma}d_{2\sigma'}^{\dag}d_{2\sigma'}\right]+\bar{U}_{1}\sum\nolimits'\left[f_{2\sigma}^{\dag}f_{2\sigma}d_{1\sigma'}^{\dag}d_{1\sigma'}+f_{1\sigma}^{\dag}f_{1\sigma}d_{2\sigma'}^{\dag}d_{2\sigma'}\right]\notag\label{H_int13}\\
+ & U_{2}\sum\nolimits'\left[f_{1\sigma}^{\dag}d_{1\sigma}d_{1\sigma'}^{\dag}f_{1\sigma'}+f_{2\sigma}^{\dag}d_{2\sigma}d_{2\sigma'}^{\dag}f_{2\sigma'}\right]+\bar{U}_{2}\sum\nolimits'\left[f_{1\sigma}^{\dag}d_{2\sigma}d_{2\sigma'}^{\dag}f_{1\sigma'}+f_{2\sigma}^{\dag}d_{1\sigma}d_{1\sigma'}^{\dag}f_{2\sigma'}\right]\notag\\
+ & \frac{U_{3}}{2}\sum\nolimits'\left[f_{1\sigma}^{\dag}d_{1\sigma}f_{1\sigma'}^{\dag}d_{1\sigma'}+f_{2\sigma}^{\dag}d_{2\sigma}f_{2\sigma'}^{\dag}d_{2\sigma'}+h.c.\right]+\frac{\bar{U}_{3}}{2}\sum\nolimits'\left[f_{1\sigma}^{\dag}d_{2\sigma}f_{1\sigma'}^{\dag}d_{2\sigma'}+f_{2\sigma}^{\dag}d_{1\sigma}f_{2\sigma'}^{\dag}d_{1\sigma'}+h.c.\right]\notag\\
+ & \frac{U_{4}}{2}\sum\nolimits'\left[d_{1\sigma}^{\dag}d_{1\sigma}d_{1\sigma'}^{\dag}d_{1\sigma'}+d_{2\sigma}^{\dag}d_{2\sigma}d_{2\sigma'}^{\dag}d_{2\sigma'}\right]+\frac{\bar{U}_{4}}{2}\sum\nolimits'\left[d_{1\sigma}^{\dag}d_{2\sigma}d_{1\sigma'}^{\dag}d_{2\sigma'}+d_{2\sigma}^{\dag}d_{1\sigma}d_{2\sigma'}^{\dag}d_{1\sigma'}\right]\notag\\
+ & \tilde{U}_{4}\sum\nolimits'd_{1\sigma}^{\dag}d_{1\sigma}d_{2\sigma'}^{\dag}d_{2\sigma'}+\tilde{\tilde{U}}_{4}\sum\nolimits'd_{1\sigma}^{\dag}d_{2\sigma}d_{2\sigma'}^{\dag}d_{1\sigma'}\notag\\
+ & \frac{U_{5}}{2}\sum\nolimits'\left[f_{1\sigma}^{\dag}f_{1\sigma}f_{1\sigma'}^{\dag}f_{1\sigma'}+f_{2\sigma}^{\dag}f_{2\sigma}f_{2\sigma'}^{\dag}f_{2\sigma'}\right]+\frac{\bar{U}_{5}}{2}\sum\nolimits'\left[f_{1\sigma}^{\dag}f_{2\sigma}f_{1\sigma'}^{\dag}f_{2\sigma'}+f_{2\sigma}^{\dag}f_{1\sigma}f_{2\sigma'}^{\dag}f_{1\sigma'}\right]\notag\\
+ & \tilde{U}_{5}\sum\nolimits'f_{1\sigma}^{\dag}f_{1\sigma}f_{2\sigma'}^{\dag}f_{2\sigma'}+\tilde{\tilde{U}}_{5}\sum\nolimits'f_{1\sigma}^{\dag}f_{2\sigma}f_{2\sigma'}^{\dag}f_{1\sigma'}\,.
\end{align}
One can verify that each term in Eq.~\eqref{H_int13} obeys the tetragonal
symmetry separately. Eq. (\ref{H_int13}) contains 14 independent
coupling constants. We split these 14 couplings into a three subsets
that are not mixed with each other under the pRG flow. These three
subsets are presented graphically in Figs.~\ref{fig:amp1}, \ref{fig:amp2}
and \ref{fig:amp3}.

We show below that the amplitudes from different subsets do not mix under the pRG flow.The full three-orbital model, which includes $d_{xy}$ components on electron
pockets, contains 30 independent coupling constants \cite{Cvetkovic2013}.
The comparison between Eqs.~\eqref{H_T} and \eqref{H_int13} gives
the following relations,
\begin{align}
U_{1} & =U_{2}=U_{3}=U_{4}=U_{5}=U,\notag\label{Hubbard_relation}\\
\bar{U}_{1} & =\tilde{U}_{4}=\tilde{U}_{5}=U',\notag\\
\bar{U}_{2} & =\tilde{\tilde{U}}_{4}=\tilde{\tilde{U}}_{5}=J,\notag\\
\bar{U}_{3} & =\bar{U}_{4}=\bar{U}_{5}=J'\,.
\end{align}
These relations hold for bare couplings, but, as we will see, are
not preserved under the pRG flow. On the other hand RG flow does not
generate new couplings in addition to 14 in Eq. (\ref{H_int13}),
i.e., the model with 14 coupling is renormalizable.

The splitting between different couplings in Eq. (\ref{Hubbard_relation})
implies that RG flow generates non-local interactions. The information
extracted from the low-energy sector only is not sufficient to fully
specify which non-local interactions are generated, but the model
with 14 couplings can be constructed if one adds to local $U,U',J,J'$
also interactions of the same Hubbard and Hund type, but involving
fermions from different sites of each plaquette on a square lattice.
Thus 5 terms involving fermions from the same orbital $d_{xz}$ or
$d_{yz}$ ($U_{1}$, $U_{2}$, $U_{3}$, $U_{4}$, and $U_{5}$ terms)
appear with different couplings if we introduce, in addition to on-site
$U$, also the terms
\begin{eqnarray}
&&{\cal H}_{non-local}=\sum_{r}U_{a}d_{xz}^{\dagger}(r)d_{xz}(r)d_{xz}^{\dagger}(r+a_{y})d_{xz}(r+a_{y})+U_{b}d_{xz}^{\dagger}(r)d_{xz}(r)d_{xz}^{\dagger}(r)d_{xz}(r+a_{y})\nonumber \\
 &  & +U_{c}d_{xz}^{\dagger}(r)d_{xz}(r+a_{y})d_{xz}^{\dagger}(r+a_{x})d_{xz}(r+a_{x}+a_{y}) \nonumber \\
 && +U_{d}d_{xz}^{\dagger}(r)d_{xz}(r+a_{x})d_{xz}^{\dagger}(r+a_{y})d_{xz}(r+a_{x}+a_{y})+ h.c
 \label{chu_1}
\end{eqnarray}
and analogous (symmetry-related terms) for $d_{yz}$ orbital. In (\ref{chu_1})
$a_{x}$ and $a_{y}$ are the components of the lattice spacing ${\bf a}$.
The couplings $U_{i}$ ($i=1-5$) are now given by
\begin{eqnarray}
 &  & U_{1}=U+U_{a}-U_{b}-U_{c}-U_{d},~~U_{2}=U-U_{a}-U_{b}-U_{c}-U_{d},~~U_{3}=U-U_{a}+U_{b}+U_{c}-U_{d},\nonumber \\
 &  & U_{4}=U+U_{a}+U_{b}+U_{c}+U_{d},~~U_{5}=U+U_{a}-U_{b}+U_{c}+U_{d}
\end{eqnarray}
One can easily verify that the interactions within a given plaquette
involving fermions from different orbitals splits $U'$, $J$, and
$J'$ terms into subsets each consisting of three different interactions
(there are 5 terms in each subset, like in Eq. (\ref{chu_1}), but
there are only three non-equivalent combunations of different $U'_{i}$,
$J_{i}$ and $J'_{i}$.

\section{pRG equations and amplitudes}

We define the RG variable $L$ at energy/temperature scale $E$ as
$L=
\log\frac{W}{E}$, where $W$ is of order bandwidth. The variable $L$ increases starting
from $L=0$ at $E=W$.

The four-fermion interaction vertices in terms of band operators are
obtained by using Eq. (\ref{band_G}) and identifying $f_{1,2}$ with
the corresponding band operators. Each vertex involves two creation
and two annihilation fermionic operators either from one of two hole
pockets ($c_{k}$ and $d_{k}$) or from two electron pockets $f_{1,k}$
and $f_{2,k}$. The prefactors are the combinations of $\cos{\theta_{k}}$
and $\sin{\theta_{k}}$ from the transformation in Eq. (\ref{band_G}).
The total number of the interaction terms in the band basis is 152.
We verified that demonstrate that all the terms within each of the
14 combinations in Eq.~\eqref{H_int13} flow identically under pRG.
We show that the pRG equations split into three groups which remain
separate under pRG flow. The first group includes interactions ${\tilde{U}}_{5}$
and ${\tilde{\tilde{U}}}_{5}$. The second group includes ${\tilde{U}}_{4}$
and ${\tilde{\tilde{U}}}_{4}$, and the third group contains ten remaining
interactions: $U_{i}$, ${\bar{U}}_{i}$ with $i=1-5$. Below we analyse
these three groups of pRG equations separately.

\subsection{The pRG for the interactions $\tilde{U}_{5}$ and $\tilde{\tilde{U}}_{5}$}

\label{sec:Ut5} 
\begin{figure}[h]
\begin{centering}
\includegraphics[width=0.6\columnwidth]{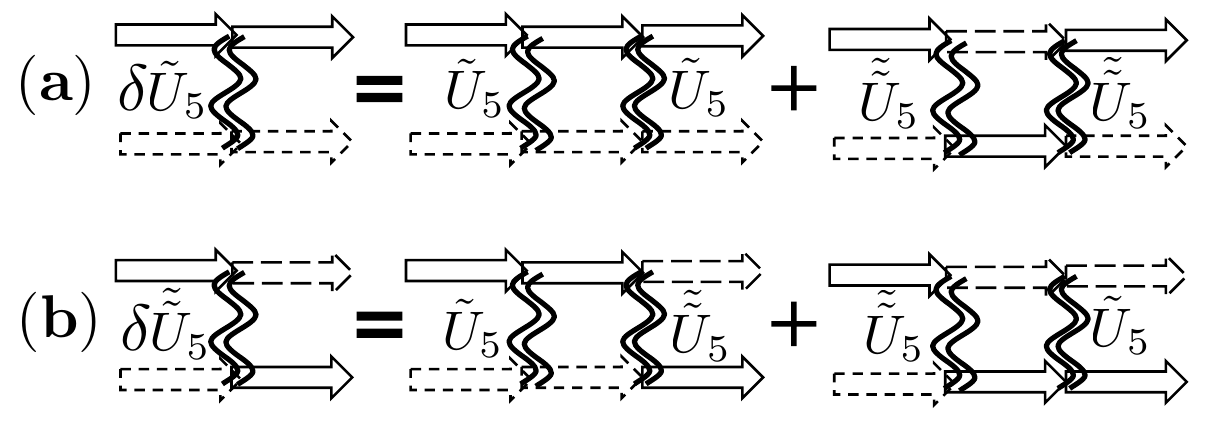} \protect\caption{Diagrammatic representation of the renormalizations of the interactions
$\tilde{U}_{5}$, (a) and $\tilde{\tilde{U}}_{5}$, (b), to second
order in the interactions. \label{fig:rgU5t}}

\par\end{centering}

\centering{}
\end{figure}

It is instructive to consider first the flow of $\tilde{U}_{5}$
and $\tilde{\tilde{U}}_{5}$ (see Fig.~\ref{fig:amp1}) because they
describe interactions between fermions from the two electron pockets
and get renormalized only in the particle-particle channel (see Fig.~\ref{fig:rgU5t}).

The corresponding terms in the four-fermion Hamiltonian\eqref{H_int13}
are
\begin{align}
H_{\tilde{U}_{5}}+H_{\tilde{\tilde{U}}_{5}} & =\tilde{U}_{5}\sum_{\sigma\sigma'}\sum'[f_{1,\sigma,\bm{k}_{1}}^{\dag}f_{2,\sigma',\bm{k}_{3}}^{\dag}][f_{2,\sigma',\bm{k}_{4}}f_{1,\sigma,\bm{k}_{2}}]
+\tilde{\tilde{U}}_{5}\sum_{\sigma\sigma'}\sum'[f_{1,\sigma,\bm{k}_{1}}^{\dag}f_{2,\sigma',\bm{k}_{3}}^{\dag}][f_{1,\sigma',\bm{k}_{4}}f_{2,\sigma,\bm{k}_{2}}]\,.\label{t5_13}
\end{align}
The logarithmic renormalization in the particle-particle channel is
obtained when either $\bm{k}_{4}\approx-\bm{k}_{2}\approx\bm{k}$
or $\bm{k}_{1}\approx-\bm{k}_{3}\approx\bm{k}$ in Eq.~\eqref{t5_13}
are the running (larger) momentum, with no kinematical constrains
on external (small) momenta $\bm{k}_{1(4)},\bm{k}_{3(2)}$, which
we assume to be of the same order and label as $k_{ext}$. 

Let the amplitudes at a running momentum $k$ be $\tilde{U}_{5}(k)$
and $\tilde{\tilde{U}}_{5}(k)$. Using the standard reasoning for
pRG, i.e., selecting the cross-section in the diagram in Fig.~\ref{fig:rgU5t}

with the smallest running momentum momentum $k$, integrating over
larger momenta (in logarithmical sense) on both sides of this cross-section
to get running $\tilde{U}_{5}(k)$ and $\tilde{\tilde{U}}_{5}(k)$,
and integrating over $k$ with $k_{ext}$ as the lower limit, we obtain
running couplings $\tilde{U}_{5}(k_{ext})$ and $\tilde{\tilde{U}}_{5}(k_{ext})$.
The equation for $\tilde{U}_{5}(k_{ext})$, obtained this way, reads
\begin{align}
\tilde{U}_{5}(k_{ext})=-\int_{k_{ext}}\frac{d^{2}\bm{k}}{4\pi^{2}}((\tilde{U}_{5}(k))^{2}+(\tilde{\tilde{U}}_{5}(k))^{2})\int\frac{d\epsilon}{2\pi}G_{f_{1}}(i\epsilon,\bm{k})G_{f_{2}}(-i\epsilon,-\bm{k})\,,\label{u5_RG_1}
\end{align}
where
\begin{align}
G_{f_{1,2}}(i\epsilon,k)=\frac{1}{i\epsilon-\epsilon_{1,2}(\bm{k})-\mu}\label{G_f}
\end{align}
are the Green functions for the $xz$ and $yz$ electrons with dispersions
\eqref{disp}. The integration over frequency and over directions
of ${\bf k}$ yield
\begin{eqnarray}
 &  & \int\frac{d\phi}{2\pi}\int\frac{d\epsilon}{2\pi}G_{f_{1}}(i\epsilon,\epsilon_{f_{1}}(k))G_{f_{2}}(-i\epsilon,\epsilon_{f_{2}}(-k))=\int\frac{d\phi}{2\pi}\frac{1}{\xi_{f_{1}}+\xi_{f_{2}}}\nonumber \\
 &  & =\int\frac{d\phi}{2\pi}\frac{1}{k_{x}^{2}/(2m_{x})+k_{y}^{2}/(2m_{y})+k_{x}^{2}/(2m_{y})+k_{y}^{2}/(2m_{x})}\nonumber \\
 &  & =\frac{2m_{x}m_{y}}{m_{x}+m_{y}}\frac{1}{k^{2}}\, .
\end{eqnarray}
Substituting this into (\ref{u5_RG_1}) we obtain
\begin{align}
\tilde{U}_{5}(k_{ext})=-\frac{2m_{x}m_{y}}{m_{x}+m_{y}}\int_{k_{ext}}\frac{dk^{2}}{4\pi k^{2}}((\tilde{U}_{5}(k))^{2}+(\tilde{\tilde{U}}_{5}(k))^{2})\label{U5_tt_2}\, .
\end{align}
Introducing the logarithmical variable $L=\log{\frac{Wm}{k_{ext}^{2}}}$
we obtain
\begin{align}
4\pi\frac{d\tilde{U}_{5}(L)}{dL}=-\frac{2m_{x}m_{y}}{m_{x}+m_{y}}((\tilde{U}_{5}(L))^{2}+(\tilde{\tilde{U}}_{5}(L))^{2})\,.\label{u5t_RG}
\end{align}
Similarly,
\begin{align}
4\pi\frac{d\tilde{\tilde{U}}_{5}}{dL}=-2\frac{2m_{x}m_{y}}{m_{x}+m_{y}}\tilde{U}_{5}\tilde{\tilde{U}}_{5}\,.\label{u5tt_RG}
\end{align}
Introducing dimensionless interactions as
\begin{align}
\tilde{u}_{5}=\frac{2m_{x}m_{y}}{m_{x}+m_{y}}\frac{\tilde{U}_{5}}{4\pi}\,,\quad\tilde{\tilde{u}}_{5}=\frac{2m_{x}m_{y}}{m_{x}+m_{y}}\frac{\tilde{\tilde{U}}_{5}}{4\pi}
\end{align}
we cast the pRG equations \eqref{u5t_RG} and \eqref{u5tt_RG} in
the following form,
\begin{align}
\frac{d\tilde{u}_{5}}{dL} & =-(\tilde{u}_{5}^{2}+\tilde{\tilde{u}}_{5}^{2})\,,\notag\label{u5_RG}\\
\frac{d\tilde{\tilde{u}}_{5}}{dL} & =-2\tilde{u}_{5}\tilde{\tilde{u}}_{5}\,.
\end{align}

Eqs. (\ref{u5_RG}) could be obtained also in the Wilsonian RG scheme,
in which one \textit{assumes} renormalizability (i.e assumes that
the coupings depend on the running rather than initial momenta) and
integrates in (\ref{u5_RG}) over momenta in the annulus $k-dk<k'<k$.
In this procedure
\begin{align}
d\tilde{U}_{5}(k) & =-((\tilde{U}_{5}(k))^{2}+(\tilde{\tilde{U}}_{5}(k))^{2})\int_{dk}\frac{d^{2}\bm{k}}{4\pi^{2}}\int\frac{d\epsilon}{2\pi}G_{f_{1}}(i\epsilon,\bm{k})G_{f_{2}}(-i\epsilon,-\bm{k})\nonumber \\
 & =-((\tilde{U}_{5}(k))^{2}+(\tilde{\tilde{U}}_{5}(k))^{2})\frac{2m_{x}m_{y}}{m_{x}+m_{y}}\frac{dL}{4\pi}\,.
\end{align}
Differentiating over $dL$ one obtains the same equation as (\ref{u5t_RG}).
The Wilsonian RG scheme is more common and we will use it for the
derivation of other RG equations.

It follows from the Eq.~\eqref{u5_RG} that the pRG flow moves the
repulsive interactions $\tilde{u}_{5},\tilde{\tilde{u}}_{5}>0$ towards
zero provided at the bare level (i.e. at energies comparable to $\Lambda$)
$\tilde{u}_{5}>\tilde{\tilde{u}}_{5}$. According according to Eq.~\eqref{Hubbard_relation}
this holds when $U'>J$. As this condition is supposed to be satisfied,
we may safely set $\tilde{u}_{5}$ and $\tilde{\tilde{u}}_{5}$ to
zero.

We note in passing that the logarithmical renormalization in the particle-particle
channel is not the Cooper effect because we integrate over momenta
well above $k_{F}$. Rather it is related to the fact that in 2D and
for $k^{2}$ dispersion of fermions, the scattering amplitude is logarithmically
singular, what physically implies that even a weak attraction between
two fermions gives rise to the development of a bound state.

\subsection{The pRG for the interactions $\tilde{U}_{4}$ and $\tilde{\tilde{U}}_{4}$}

\label{sec:Ut4} The interactions $\tilde{U}_{4}$ and $\tilde{\tilde{U}}_{4}$
(see Fig.~\ref{fig:amp2}) are also renormalized only in the particle-particle
channel. In this case, however the band basis differs from the orbital
basis and the transformation \eqref{band_G} is required in order
to find the right pRG equations.

\begin{figure}[h]
\begin{centering}
\includegraphics[width=1\columnwidth]{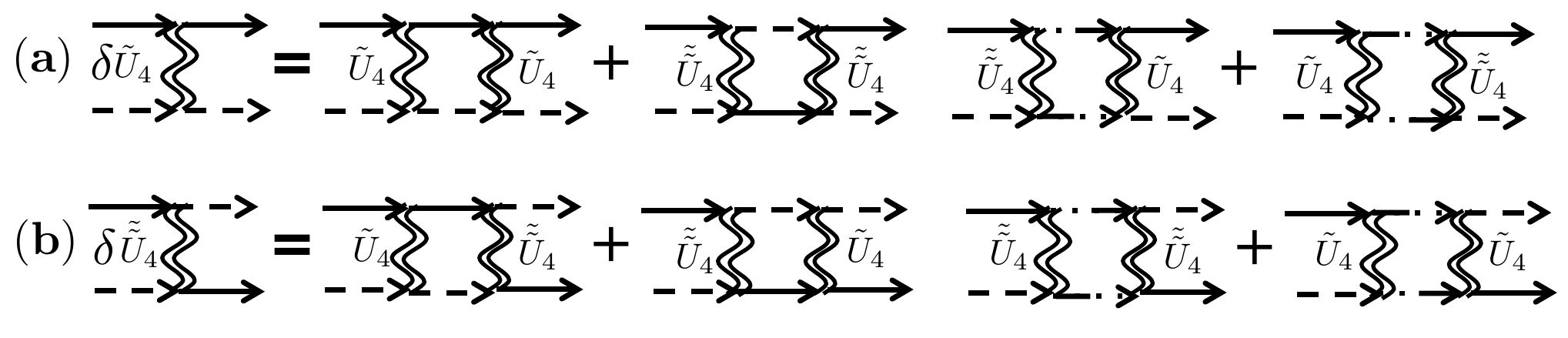} \protect\caption{Diagrammatic representation of the renormalizations of the interactions
$\tilde{U}_{4}$, (a) and $\tilde{\tilde{U}}_{4}$, (b), to second
order in the interactions. \label{fig:rgU4t}}

\par\end{centering}

\centering{}
\end{figure}

In the orbital representation the change of $\tilde{U}_{4}$ and $\tilde{\tilde{U}}_{4}$
due to integration over the ring $k-dk<k<k$ is
\begin{align}
d\tilde{U}_{4}= & -(\tilde{U}_{4}^{2}+\tilde{\tilde{U}}_{4}^{2})\int_{dk}\frac{d^{2}\bm{k}}{4\pi^{2}}\int\frac{d\epsilon}{2\pi}G_{d_{1};d_{1}}(i\epsilon,\bm{k})G_{d_{2};d_{2}}(-i\epsilon,-\bm{k})\notag\label{U4_t}\\
 & -2\tilde{U}_{4}\tilde{\tilde{U}}_{4}\int_{dk}\frac{d^{2}\bm{k}}{4\pi^{2}}\int\frac{d\epsilon}{2\pi}G_{d_{1};d_{2}}(i\epsilon,\bm{k})G_{d_{2};d_{1}}(-i\epsilon,-\bm{k})\notag\\
d\tilde{\tilde{U}}_{4}= & -2\tilde{U}_{4}\tilde{\tilde{U}}_{4}\int_{dk}\frac{d^{2}\bm{k}}{4\pi^{2}}\int\frac{d\epsilon}{2\pi}G_{d_{1};d_{1}}(i\epsilon,\bm{k})G_{d_{2};d_{2}}(-i\epsilon,-\bm{k})\notag\\
 & -(\tilde{U}_{4}^{2}+\tilde{\tilde{U}}_{4}^{2})\int_{dk}\frac{d^{2}\bm{k}}{4\pi^{2}}\int\frac{d\epsilon}{2\pi}G_{d_{1};d_{2}}(i\epsilon,\bm{k})G_{d_{2};d_{1}}(-i\epsilon,-\bm{k})\,,
\end{align}
The Greens functions in the orbital representation, $G_{d_{i};d_{j}}(i\epsilon,\bm{k})$
are expressed via the propagators of low-energy fermions in the band
representation,
\begin{align}
G_{c(d)}(i\epsilon,\bm{k})=\frac{1}{i\epsilon-\epsilon_{c(d)}(\bm{k})-\mu}\label{G_hb}
\end{align}
as
\begin{align}
G_{d_{1},d_{1}}(i\epsilon,\bm{k}) & =G_{c}(i\epsilon,\bm{k})\cos^{2}\theta_{\bm{k}}+G_{d}(i\epsilon,\bm{k})\sin^{2}\theta_{\bm{k}},\notag\label{G_orb}\\
G_{d_{2},d_{2}}(i\epsilon,\bm{k}) & =G_{c}(i\epsilon,\bm{k})\sin^{2}\theta_{\bm{k}}+G_{d}(i\epsilon,\bm{k})\cos^{2}\theta_{\bm{k}},\notag\\
G_{d_{1},d_{2}}(i\epsilon,\bm{k}) & =G_{d_{2},d_{1}}(i\epsilon,\bm{k})=\left[G_{d}(i\epsilon,\bm{k})-G_{c}(i\epsilon,\bm{k})\right]\sin\theta_{\bm{k}}\cos\theta_{\bm{k}}\,.
\end{align}
The band dispersions $\epsilon_{c(d)}(\bm{k})$ are given in Eq.~\eqref{disp}.
The Eq.~\eqref{U4_t} is illustrated in Fig.~\ref{fig:rgU4t}.

The energy and momentum integrations in (\ref{U4_t}) are performed
using the expressions for the angular averages,
\begin{align}
\langle\cos^{4}\theta\rangle=\langle\sin^{4}\theta\rangle=\frac{3}{8}\,,\quad\langle\cos^{2}\theta\sin^{2}\theta\rangle=\frac{1}{8}\,.\label{ang}
\end{align}
We further have
\begin{align}
\int_{dk} & \frac{d^{2}\bm{k}}{4\pi^{2}}\int\frac{d\epsilon}{2\pi}G_{d(c)}(i\epsilon,\bm{k})G_{d(c)}(-i\epsilon,-\bm{k})=\frac{dL}{4\pi}m_{c(d)}\,,\notag\label{G_dc}\\
\int_{dk} & \frac{d^{2}\bm{k}}{4\pi^{2}}\int\frac{d\epsilon}{2\pi}G_{d(c)}(i\epsilon,\bm{k})G_{c(d)}(-i\epsilon,-\bm{k})=\frac{dL}{4\pi}\frac{2m_{c}m_{d}}{m_{c}+m_{d}}\,.
\end{align}

Substituting Eqs. \eqref{disp}, \eqref{G_hb} and \eqref{G_orb}
into \eqref{U4_t} and using \eqref{ang} we obtain
\begin{align}
4\pi\frac{d\tilde{U}_{4}}{dL}= & -(\tilde{U}_{4}^{2}+\tilde{\tilde{U}}_{4}^{2})\left[\frac{1}{8}(m_{c}+m_{d})+\frac{3}{8}\frac{4m_{c}m_{d}}{m_{c}+m_{d}}\right]
-2\tilde{U}_{4}\tilde{\tilde{U}}_{4}\frac{1}{8}\frac{(m_{c}-m_{d})^{2}}{m_{c}+m_{d}}\notag\label{U4_t2}\, , \\
4\pi\frac{d\tilde{\tilde{U}}_{4}}{dL}= & -2\tilde{U}_{4}\tilde{\tilde{U}}_{4}\left[\frac{1}{8}(m_{c}+m_{d})+\frac{3}{8}\frac{4m_{c}m_{d}}{m_{c}+m_{d}}\right]
-(\tilde{U}_{4}^{2}+\tilde{\tilde{U}}_{4}^{2})\frac{1}{8}\frac{(m_{c}-m_{d})^{2}}{m_{c}+m_{d}}\, .
\end{align}
It follows that
\begin{align}
4\pi\frac{d(\tilde{U}_{4}\pm\tilde{\tilde{U}}_{4})}{dL}= & -(\tilde{U}_{4}\pm\tilde{\tilde{U}}_{4})^{2}\left[\frac{1}{8}(m_{c}+m_{d})+\frac{3}{8}\frac{4m_{c}m_{d}}{m_{c}+m_{d}}\pm\frac{1}{8}\frac{(m_{c}-m_{d})^{2}}{m_{c}+m_{d}}\right]\,.\label{U4_t3}
\end{align}
Solving Eq.~\eqref{U4_t3} we find that the interactions $\tilde{U}_{4}$
and $\tilde{\tilde{U}}_{4}$ flow to zero under pRG, provided at the
bare level $\tilde{U}_{4}>\tilde{\tilde{U}}_{4}$. Like before, this
holds when $U'>J$. As this condition is supposed to be satisfied,
we may safely set $\tilde{U}_{4}$ and $\tilde{\tilde{U}}_{4}$ to
zero.

\subsection{The third group of pRG equations}

Finally we derive and solve the pRG equations obeyed by the third
group of couplings shown in Fig.~\ref{fig:amp3}.

\begin{figure}[h]
\begin{centering}
\includegraphics[width=0.6\columnwidth]{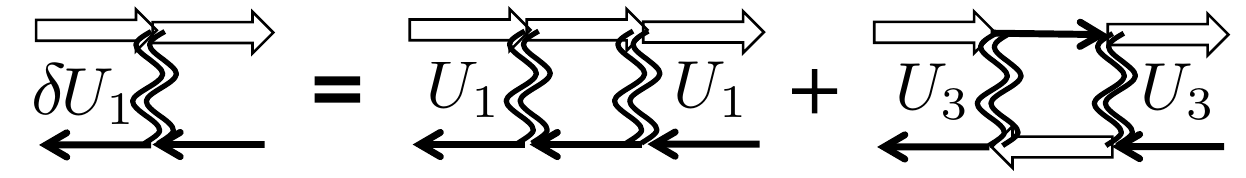} \protect\caption{Diagrammatic representation of the renormalizations of the interaction
$U_{1}$ to second order in the interactions. The diagrams for $\bar{U}_{1}$
have the same form in terms of $\bar{U}_{1}$ and $\bar{U}_{3}$. \label{fig:rgU1}}

\par\end{centering}

\centering{}
\end{figure}

The interaction $U_{1}$ flows due to renormalizations in the particle-hole
channel. The corresponding diagrams are shown in Fig.~\ref{fig:rgU1}.
In analytical form we have
\begin{align}
dU_{1}=-(U_{1}^{2}+U_{3}^{2})\int_{dk}\frac{d^{2}\bm{k}}{4\pi^{2}}\int\frac{d\epsilon}{2\pi}G_{d_{1};d_{1}}(i\epsilon,\bm{k})G_{f_{1}}(i\epsilon,\bm{k}).\label{dU1}
\end{align}
Using Eq.~\eqref{G_orb} we re-express Green's functions in the orbital
basis via the Green's functions in the band basis:
\begin{align}
\int_{dk}\frac{d^{2}\bm{k}}{4\pi^{2}}\int\frac{d\epsilon}{2\pi}G_{d_{1};d_{1}}(i\epsilon,\bm{k})G_{f_{1}}(i\epsilon,\bm{k})=\int_{dk}\frac{d^{2}\bm{k}}{4\pi^{2}}\int\frac{d\epsilon}{2\pi}[\cos^{2}\theta_{\bm{k}}G_{c}(i\epsilon,\bm{k})+\sin^{2}\theta_{\bm{k}}G_{d}(i\epsilon,\bm{k})]G_{f_{1}}(i\epsilon,\bm{k})\, .\label{GGdf_1}
\end{align}
We further write,
\begin{align}
\int_{dk}\frac{d^{2}\bm{k}}{(2\pi)^{2}}\int\frac{d\epsilon}{2\pi}\cos^{2}\theta G_{c}G_{f_{1}}=\int_{dk}\frac{kdk}{2\pi}\int\frac{d\theta}{2\pi}\cos^{2}\theta\int\frac{d\epsilon}{2\pi}G_{c}(i\epsilon,\xi_{c}(k))G_{f_{1}}(i\epsilon,\xi_{f_{1}}(k))\label{GGdf_2}
\end{align}
As $\xi_{c}(k)<0$ and $\xi_{f_{1}}>0$ integration over the energy
gives,
\begin{equation}
\int\frac{d\epsilon}{2\pi}\frac{1}{i\epsilon+|\xi_{c}(k)|}\frac{1}{i\epsilon-\xi_{f_{1}}(k)}=-\frac{1}{|\xi_{c}(k)|+\xi_{f_{1}}(k)}\,.\label{U1_e}
\end{equation}
We then obtain using Eq.~\eqref{disp}
\begin{align}
\int_{dk}\frac{d^{2}\bm{k}}{(2\pi)^{2}}\int\frac{d\epsilon}{2\pi}\cos^{2}\theta G_{c}G_{f_{1}}=-\frac{1}{4\pi}\frac{dk^{2}}{k^{2}}\int_{0}^{2\pi}\frac{d\theta}{\pi}\frac{\cos^{2}\theta}{\frac{1}{m_{c}}+\frac{\cos^{2}\theta}{m_{x}}+\frac{\sin^{2}\theta}{m_{y}}}=-\frac{dL}{4\pi}A_{1}\,,\label{U1_14}
\end{align}
where we have defined
\begin{equation}
A_{1}=\int_{0}^{2\pi}\frac{d\theta}{\pi}\frac{\cos^{2}\theta}{\frac{1}{m_{c}}+\frac{\cos^{2}\theta}{m_{x}}+\frac{\sin^{2}\theta}{m_{y}}}\,.\label{A1}
\end{equation}
Similarly
\begin{align}
\int_{dk}\frac{d^{2}\bm{k}}{(2\pi)^{2}}\int\frac{d\epsilon}{2\pi}\cos^{2}\theta G_{d}G_{f_{2}}=-\frac{dL}{4\pi}A_{2}\,,\label{U1_15}
\end{align}
where we have introduced,
\begin{align}
A_{2}=\int_{0}^{2\pi}\frac{d\theta}{\pi}\frac{\sin^{2}\theta}{\frac{1}{m_{d}}+\frac{\cos^{2}\theta}{m_{x}}+\frac{\sin^{2}\theta}{m_{y}}}\, .\label{A2}
\end{align}
We see that the momentum integral is still logarithmical $\int kd(k)/k^{2}$,
this time because hole and electronic excitations have opposite signs
of the dispersion. This does not require a true nesting, i.e. hole
and electron masses do not have to be equal and electron dispersion
does not have to be circular. Still, the logarithmical behavior in
the particle-hole channel for momenta $k\ll\Lambda$ holds only if
both pockets are tiny, i.e, both Fermi momenta are small. We also
note that Eq.~\eqref{U1_e} contains an additional minus sign compared
to the contribution from the particle-particle channel.

The pRG equation for $U_{1}$ is obtained by substituting Eqs.~\eqref{U1_14}
and \eqref{U1_15} into Eq.~\eqref{dU1}. This yields
\begin{align}
4\pi\frac{dU_{1}}{dL}=(U_{1}^{2}+U_{3}^{2})A\,,\label{dU1_A}
\end{align}
where
\begin{equation}
A=A_{1}+A_{2}\,.\label{A}
\end{equation}

The pRG equation for the interaction $\bar{U}_{1}$ is obtained in
a similar way and is
\begin{align}
4\pi\frac{d\bar{U}_{1}}{dL}=(\bar{U}_{1}^{2}+\bar{U}_{3}^{2})\bar{A}\,,\label{dU1bar_A}
\end{align}
where
\begin{equation}
\bar{A}=\bar{A}_{1}+\bar{A}_{2}\, ,\label{Ab}
\end{equation}
and
\begin{equation}
\bar{A}_{1}=\int_{0}^{2\pi}\frac{d\theta}{\pi}\frac{\sin^{2}\theta}{\frac{1}{m_{c}}+\frac{\cos^{2}\theta}{m_{x}}+\frac{\sin^{2}\theta}{m_{y}}}\,,\quad\bar{A}_{2}=\int_{0}^{2\pi}\frac{d\theta}{\pi}\frac{\cos^{2}\theta}{\frac{1}{m_{d}}+\frac{\cos^{2}\theta}{m_{x}}+\frac{\sin^{2}\theta}{m_{y}}}\,.\label{Abar}
\end{equation}

\begin{figure}[h]
\begin{centering}
\includegraphics[width=0.8\columnwidth]{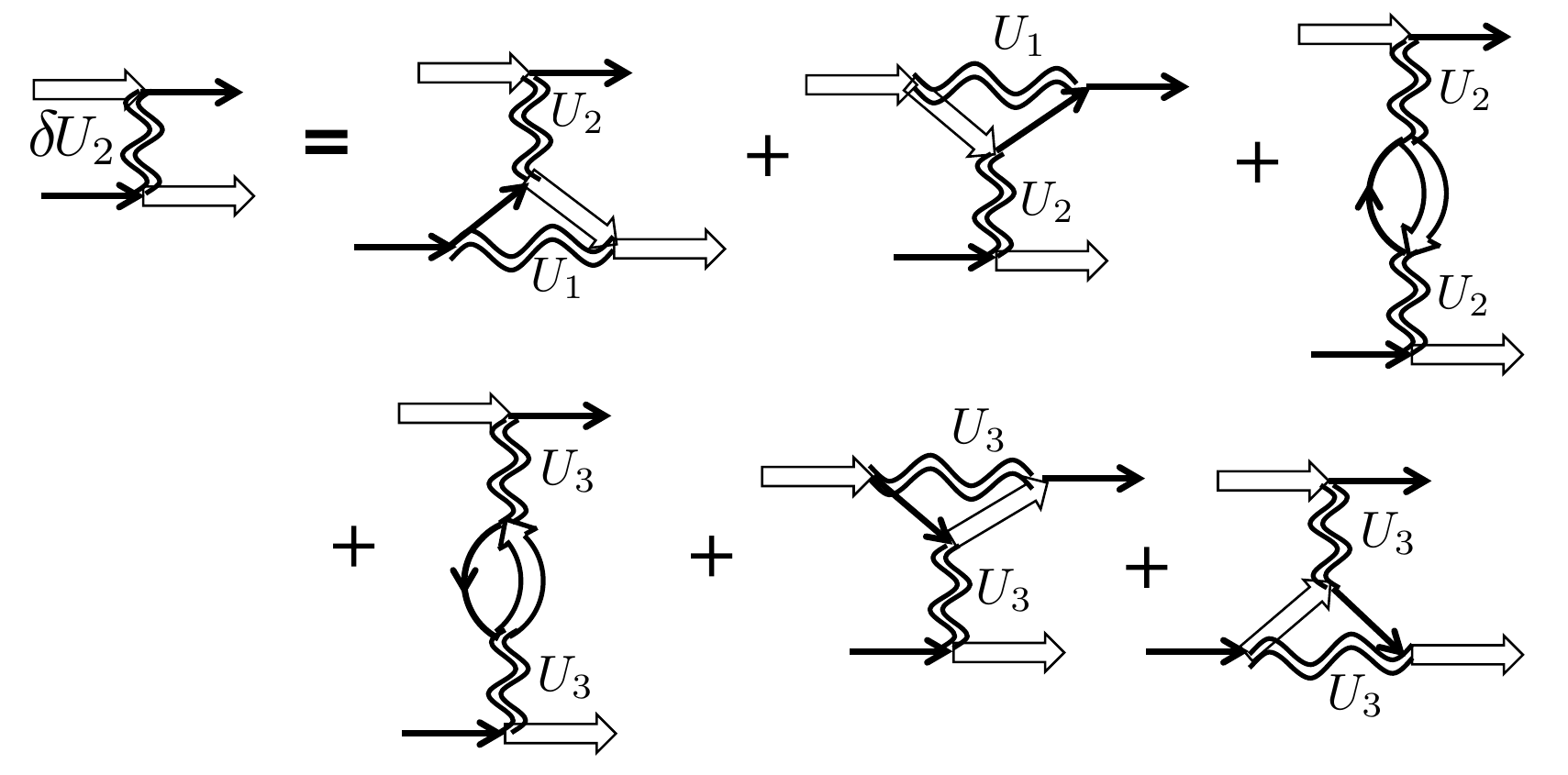} \protect\caption{Diagrammatic representation of the renormalizations of the interaction
$U_{2}$ to second order in the interactions. The two contributions
$\propto U_{3}^{2}$ in the second row cancel each other. The diagrams
for $\bar{U}_{2}$ have the same form in terms of $\bar{U}_{1}$ and
$\bar{U}_{2}$. \label{fig:rgU2}}

\par\end{centering}

\centering{}
\end{figure}

The interactions $U_{2}$ and $\bar{U}_{2}$ are also renormalized
in the particle-hole channel. The corresponding diagrams are shown
in Fig.~\ref{fig:rgU2}.

The corresponding pRG equations are:
\begin{align}
4\pi\frac{dU_{2}}{dL}=2(U_{1}U_{2}-U_{2}^{2})A\,,\label{dU2}
\end{align}
and
\begin{align}
4\pi\frac{d\bar{U}_{2}}{dL}=2(\bar{U}_{1}\bar{U}_{2}-\bar{U}_{2}^{2})A\,.\label{dU2_bar}
\end{align}
\begin{figure}[h]
\begin{centering}
\includegraphics[width=0.9\columnwidth]{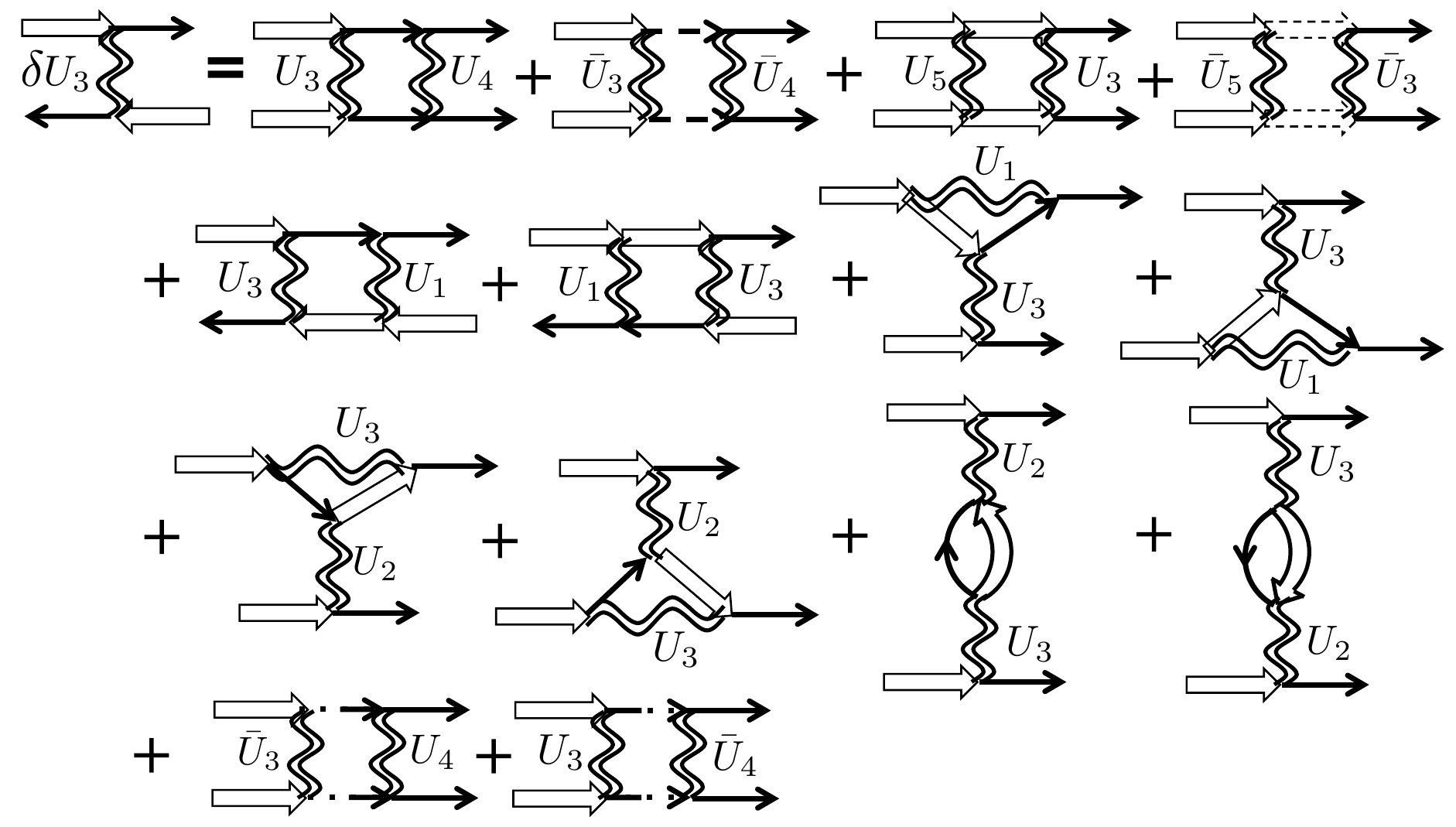} \protect\caption{Diagrammatic representation of the renormalizations of the interaction
$U_{3}$ to second order in the interactions. The last two diagrams
include the hole propagators, off-diagonal in the orbital index. There
two diagrams vanish in the limit $m_{c}=m_{d}$. The diagrams for
$\bar{U}_{3}$ are the same in terms of cross-products $\bar{U}_{i}U_{j}$. \label{fig:rgU3}}

\par\end{centering}

\centering{}
\end{figure}

The flow of $U_{3}$ is due to renormalizations in both particle-hole
and particle-particle channels. The corresponding diagrams are shown
in Fig.~\ref{fig:rgU3}. As is clear from the Fig.~\ref{fig:rgU3},
there are three contributions to the renormalization of $U_{3}$:
\begin{align}
dU_{3}=d_{h}U_{3}+d_{e}U_{3}+d_{eh}U_{3}.\label{dU_13}
\end{align}
Here $d_{h}U_{3}$ is the contribution to $dU_{3}$ from integration
over the hole momenta in the particle-particle channel, $d_{e}U_{3}$
is the contribution to $dU_{3}$ from integration over the electron
momenta in particle-particle channel, and $d_{h,e}U_{3}$ is the contribution
originating from the integration over both electron and hole momenta
in the particle-hole channel. We start with the electronic contribution:
\begin{align}
d_{e}U_{3}=-(U_{5}U_{3}+\bar{U}_{5}\bar{U}_{3})\int_{dk}\frac{d^{2}\bm{k}}{(2\pi)^{2}}\int\frac{d\epsilon}{2\pi}G_{f_{1,2}}(\bm{k},\epsilon)G_{f_{1,2}}(-\bm{k},-\epsilon)\, .\label{dU_e}
\end{align}
We have
\begin{align}
\int_{dL} & \frac{d^{2}k}{4\pi^{2}}\int\frac{d\epsilon}{2\pi}G_{f_{1}}(i\epsilon,\epsilon_{f_{1}}(k))G_{f_{1}}(-i\epsilon,\epsilon_{f_{1}}(-k))=\int_{k-dk}^{k}\frac{kdk}{2\pi}\int\frac{d\phi}{2\pi}\frac{1}{2\xi_{f_{1}}}\notag\label{G_f11}\\
 & =\int_{k-dk}^{k}\frac{kdk}{2\pi}\int\frac{d\phi}{2\pi}\frac{1}{k^{2}[\cos^{2}\theta/m_{x}+\sin^{2}\theta/m_{y}]}=\frac{dL}{4\pi}\sqrt{m_{x}m_{y}}\, .
\end{align}
As a result, Eq.~\eqref{dU_e} takes the form,
\begin{align}
4\pi\frac{d_{e}U_{3}}{dL}=-(U_{5}U_{3}+\bar{U}_{5}\bar{U}_{3})A_{e}\,,\label{dU_e1}
\end{align}
where
\begin{align}
A_{e}=\sqrt{m_{x}m_{y}}\,.\label{Ae}
\end{align}

The contribution from integrating over momenta of hole excitations
in the particle-particle channel is
\begin{align}
d_{h}U_{3} & =-U_{3}U_{4}\int_{dk}\frac{d^{2}\bm{k}}{4\pi^{2}}\int\frac{d\epsilon}{2\pi}G_{d_{1};d_{1}}(i\epsilon,\bm{k})G_{d_{1};d_{1}}(-i\epsilon,-\bm{k})\notag\\
 & -\bar{U}_{3}\bar{U}_{4}\int_{dk}\frac{d^{2}\bm{k}}{4\pi^{2}}\int\frac{d\epsilon}{2\pi}G_{d_{2};d_{2}}(i\epsilon,\bm{k})G_{d_{2};d_{2}}(-i\epsilon,-\bm{k})\notag\\
 & -U_{3}\bar{U}_{4}\int_{dk}\frac{d^{2}\bm{k}}{4\pi^{2}}\int\frac{d\epsilon}{2\pi}G_{d_{1};d_{2}}(i\epsilon,\bm{k})G_{d_{1};d_{2}}(-i\epsilon,-\bm{k})\notag\\
 & -\bar{U}_{3}U_{4}\int_{dk}\frac{d^{2}\bm{k}}{4\pi^{2}}\int\frac{d\epsilon}{2\pi}G_{d_{2};d_{1}}(i\epsilon,\bm{k})G_{d_{2};d_{1}}(-i\epsilon,-\bm{k})\,.
\end{align}
Here the last two terms are graphically presented by the last two
diagrams in the Fig.~\ref{fig:rgU3}. The integrations in the first
two terms are entirely analogous to those in Eq.~\eqref{U4_t} and
we just quote the result,
\begin{align}
\int_{dk}\frac{d^{2}\bm{k}}{4\pi^{2}}\int\frac{d\epsilon}{2\pi}G_{d_{1,2};d_{1,2}}(i\epsilon,\bm{k})G_{d_{1,2};d_{1,2}}(-i\epsilon,-\bm{k})=\frac{dL}{4\pi}A_{h},\label{Gdd_3}
\end{align}
where
\begin{align}
A_{h}=\frac{3}{8}(m_{c}+m_{d})+\frac{1}{8}\frac{4m_{c}m_{d}}{m_{c}+m_{d}}.\label{A_h}
\end{align}
The two remaining integrals yield
\begin{align}
\int_{dk}\frac{d^{2}\bm{k}}{4\pi^{2}}\int\frac{d\epsilon}{2\pi}G_{d_{1,2};d_{2,1}}(i\epsilon,\bm{k})G_{d_{1,2};d_{2,1}}(-i\epsilon,-\bm{k})=\frac{dL}{4\pi}A_{h}^{-}\, ,\label{Gdd_4}
\end{align}
where
\begin{align}
A_{h}^{-}=\frac{1}{8}\frac{(m_{c}-m_{d})^{2}}{m_{c}+m_{d}}\,.\label{A_hm}
\end{align}
We have therefore
\begin{align}
4\pi\frac{d_{h}U_{3}}{dL} & =-(U_{3}U_{4}+\bar{U}_{3}\bar{U}_{4})A_{h}-(U_{3}\bar{U}_{4}+\bar{U}_{3}U_{4})A_{h}^{-}\,.\label{dU_h13}
\end{align}
Finally, $d_{eh}U_{3}$ contains the same integrals as $dU_{2}$.
Borrowing the results we obtain
\begin{align}
4\pi\frac{d_{eh}U_{3}}{dL} & =(4U_{3}U_{1}-2U_{2}U_{3})A\,.\label{dU_eh13}
\end{align}
Adding the contributions \eqref{dU_e1}, \eqref{dU_eh13} and \eqref{dU_h13}
we obtain
\begin{align}
4\pi\frac{dU_{3}}{dL}=-(U_{5}U_{3}+\bar{U}_{5}\bar{U}_{3})A_{e}-(U_{3}U_{4}+\bar{U}_{3}\bar{U}_{4})A_{h}-(U_{3}\bar{U}_{4}+\bar{U}_{3}U_{4})A_{h}^{-}+(4U_{3}U_{1}-2U_{2}U_{3})A\,.\label{dU_15}
\end{align}

The equations for the remaining five amplitudes are obtained in a
similar fashion. We list these five equations below together with
the equations that we already obtained:
\begin{align}
4\pi\dot{U}_{1} & =AU_{1}^{2}+AU_{3}^{2}\notag\label{U_all_13}\\
4\pi\dot{\bar{U}}_{1} & =\bar{A}\bar{U}_{1}^{2}+\bar{A}\bar{U}_{3}^{2}\notag\\
4\pi\dot{U}_{2} & =2AU_{1}U_{2}-2AU_{2}^{2}\notag\\
4\pi\dot{\bar{U}}_{2} & =2\bar{A}\bar{U}_{1}\bar{U}_{2}-2\bar{A}\bar{U}_{2}^{2}\notag\\
4\pi\dot{U}_{3} & =-A_{h}U_{3}U_{4}-A_{h}\bar{U}_{3}\bar{U}_{4}-A_{h}^{-}U_{3}\bar{U}_{4}-A_{h}^{-}\bar{U}_{3}U_{4}+4AU_{3}U_{1}-2AU_{2}U_{3}-A_{e}U_{5}U_{3}-A_{e}\bar{U}_{5}\bar{U}_{3}\notag\\
4\pi\dot{\bar{U}}_{3} & =-A_{h}\bar{U}_{3}U_{4}-A_{h}U_{3}\bar{U}_{4}-A_{h}^{-}U_{3}U_{4}-A_{h}^{-}\bar{U}_{3}\bar{U}_{4}+4\bar{A}\bar{U}_{3}\bar{U}_{1}-2\bar{A}\bar{U}_{2}\bar{U}_{3}-A_{e}U_{5}\bar{U}_{3}-A_{e}\bar{U}_{5}U_{3}\notag\\
4\pi\dot{U}_{4} & =-A_{h}U_{4}^{2}-A_{h}\bar{U}_{4}^{2}-2A_{h}^{-}U_{4}\bar{U}_{4}-A_{e}U_{3}^{2}-A_{e}\bar{U}_{3}^{2}\notag\\
4\pi\dot{\bar{U}}_{4} & =-2A_{h}U_{4}\bar{U}_{4}-A_{h}^{-}U_{4}^{2}-A_{h}^{-}\bar{U}_{4}^{2}-2A_{e}U_{3}\bar{U}_{3}\notag\\
4\pi\dot{U}_{5} & =-A_{e}U_{5}^{2}-A_{e}\bar{U}_{5}^{2}-A_{h}U_{3}^{2}-A_{h}\bar{U}_{3}^{2}-2A_{h}^{-}U_{3}\bar{U}_{3}\notag\\
4\pi\dot{\bar{U}}_{5} & =-2A_{e}U_{5}\bar{U}_{5}-2A_{h}U_{3}\bar{U}_{3}-A_{h}^{-}U_{3}^{2}-A_{h}^{-}\bar{U}_{3}^{2}\,.
\end{align}
We now introduce the dimensionless couplings
\begin{align}
u_{1,2} & =\frac{A}{4\pi}U_{1,2}\,,\quad u_{3}=\frac{A}{4\pi}CU_{3}\,,\quad\bar{u}_{1,2}=\frac{\bar{A}}{4\pi}\bar{U}_{1,2}\,,\quad\bar{u}_{3}=\frac{\bar{A}}{4\pi}\bar{C}\bar{U}_{3}\,,\notag\label{u_dim}\\
u_{4} & =\frac{A_{h}}{4\pi}U_{4}\,,\quad\bar{u}_{4}=\frac{A_{h}}{4\pi}\bar{U}_{4}\,,\quad u_{5}=\frac{A_{e}}{4\pi}U_{5}\,,\quad\bar{u}_{5}=\frac{A_{e}}{4\pi}\bar{U}_{5}\,,
\end{align}
and the parameters
\begin{align}
C=\frac{\sqrt{A_{h}A_{e}}}{A}\,,\quad\bar{C}=\frac{\sqrt{A_{h}A_{e}}}{\bar{A}}\,.\label{C}
\end{align}
Re-expressing \eqref{U_all_13} in terms of dimensionless couplings
from Eq.~\eqref{u_dim} and the parameters $C$ and ${\bar{C}}$,
we obtain
\begin{align}
\dot{u}_{1} & =u_{1}^{2}+u_{3}^{2}/C^{2}\notag\label{U_all_15}\\
\dot{\bar{u}}_{1} & =\bar{u}_{1}^{2}+\bar{u}_{3}^{2}/\bar{C}^{2}\notag\\
\dot{u}_{2} & =2u_{1}u_{2}-2u_{2}^{2}\notag\\
\dot{\bar{u}}_{2} & =2\bar{u}_{1}\bar{u}_{2}-2\bar{u}_{2}^{2}\notag\\
\dot{u}_{3} & =-u_{3}u_{4}-(C/\bar{C})\bar{u}_{3}\bar{u}_{4}-(A_{h}^{-}/A_{h})u_{3}\bar{u}_{4}-(A_{h}^{-}/A_{h})(C/\bar{C})\bar{u}_{3}u_{4}\notag\\
 & +4u_{3}u_{1}-2u_{2}u_{3}-u_{5}u_{3}-(C/\bar{C})\bar{u}_{5}\bar{u}_{3}\notag\\
\dot{\bar{u}}_{3} & =-\bar{u}_{3}u_{4}-(\bar{C}/C)u_{3}\bar{u}_{4}-(A_{h}^{-}/A_{h})(\bar{C}/C)u_{3}u_{4}-(A_{h}^{-}/A_{h})\bar{u}_{3}\bar{u}_{4}\notag\\
 & +4\bar{u}_{3}\bar{u}_{1}-2\bar{u}_{2}\bar{u}_{3}-u_{5}\bar{u}_{3}-(\bar{C}/C)\bar{u}_{5}u_{3}\notag\\
\dot{u}_{4} & =-u_{4}^{2}-\bar{u}_{4}^{2}-2(A_{h}^{-}/A_{h})u_{4}\bar{u}_{4}-u_{3}^{2}-\bar{u}_{3}^{2}\notag\\
\dot{\bar{u}}_{4} & =-2u_{4}\bar{u}_{4}-(A_{h}^{-}/A_{h})u_{4}^{2}-(A_{h}^{-}/A_{h})\bar{u}_{4}^{2}-2u_{3}\bar{u}_{3}\notag\\
\dot{u}_{5} & =-u_{5}^{2}-\bar{u}_{5}^{2}-u_{3}^{2}-\bar{u}_{3}^{2}-2(A_{h}^{-}/A_{h})u_{3}\bar{u}_{3}\notag\\
\dot{\bar{u}}_{5} & =-2u_{5}\bar{U}_{5}-2u_{3}\bar{u}_{3}-(A_{h}^{-}/A_{h})u_{3}^{2}-(A_{h}^{-}/A_{h})\bar{u}_{3}^{2}\,.
\end{align}
We further notice that in 122 systems, the masses $m_{x}$ and $m_{y}$
get interchanged once $k_{z}$ changes to $\rightarrow k_{z}+\pi$.
Averaging over $k_{z}$ then makes the parameters $A_{1}$ and $\bar{A}_{1}$
equal. The pRG equations can be further simplified by setting $m_{c}=m_{d}=m_{h}$.
Then
\begin{align}
A=\bar{A}=\frac{2m_{e}m_{h}}{m_{e}+m_{h}}\,,\quad A_{e}=m_{e}\,,\quad A_{h}=m_{h}\,,\quad A_{h}^{-}=0\,,\quad C=\bar{C}=\frac{m_{e}+m_{h}}{2\sqrt{m_{e}m_{h}}}\,.\label{reduction}
\end{align}
Under this approximation, the set of pRG Eqs. \eqref{U_all_15} simplifies
to
\begin{align}
\dot{u}_{1} & =u_{1}^{2}+u_{3}^{2}/C^{2}\notag\label{RG_13_a}\\
\dot{\bar{u}}_{1} & =\bar{u}_{1}^{2}+\bar{u}_{3}^{2}/C^{2}\notag\\
\dot{u}_{2} & =2u_{1}u_{2}-2u_{2}^{2}\notag\\
\dot{\bar{u}}_{2} & =2\bar{u}_{1}\bar{u}_{2}-2\bar{u}_{2}^{2}\notag\\
\dot{u}_{3} & =-u_{3}u_{4}-\bar{u}_{3}\bar{u}_{4}+4u_{3}u_{1}-2u_{2}u_{3}-u_{5}u_{3}-\bar{u}_{5}\bar{u}_{3}\notag\\
\dot{\bar{u}}_{3} & =-\bar{u}_{3}u_{4}-u_{3}\bar{u}_{4}+4\bar{u}_{3}\bar{u}_{1}-2\bar{u}_{2}\bar{u}_{3}-u_{5}\bar{u}_{3}-\bar{u}_{5}u_{3}\notag\\
\dot{u}_{4} & =-u_{4}^{2}-\bar{u}_{4}^{2}-u_{3}^{2}-\bar{u}_{3}^{2}\notag\\
\dot{\bar{u}}_{4} & =-2u_{4}\bar{u}_{4}-2u_{3}\bar{u}_{3}\notag\\
\dot{u}_{5} & =-u_{5}^{2}-\bar{u}_{5}^{2}-u_{3}^{2}-\bar{u}_{3}^{2}\notag\\
\dot{\bar{u}}_{5} & =-2u_{5}\bar{u}_{5}-2u_{3}\bar{u}_{3}\,.
\end{align}
These are the equations which we presented in the main text. They
are more general than the ones obtained earlier~ (Refs. \cite{chubukov,maiti,cee}),
which neglected orbital content of low-energy excitations. The earlier
pRG equations are reproduced if we set ${\bar{u}}_{i}=0$ from the
beginning and also set $C=1$, i.e., assume that excitations near
hole and electron pockets have equal masses.

\section{Solution of RG equations}

In this section we analyse the pRG Eq.~\eqref{RG_13_a}. For completeness
and for comparison with earlier works we first set bare values of
all ${\bar{u}}_{i}=0$. Eqs. ~\eqref{RG_13_a} then show that all
$\bar{u}_{i}$ remain zero in the pRG flow. We consider the fixed
trajectory for arbitrary $C\geq1$ and show the earlier results are
recovered in the limit $C=1$.

Next, we show that the trajectories with ${\bar{u}}_{i}=0$, $i=1-5$
are unstable already for arbitrary small non-zero bare values of ${\bar{u}}_{i}$
and find the fixed trajectory for the full model. We show that the
only stable fixed trajectory is the one with $u_{i}={\bar{u}}_{i}$,
$i=1-5$.

\subsection{The fixed trajectories with ${\bar{u}}_{i}=0$, $i=1-5$.}

\label{sec:single} 
these amplitudes remain zero under the pRG flow, as follows from Eq.~\eqref{RG_13_a}.
The remaining pRG equations are
\begin{align}
\dot{u}_{1} & =u_{1}^{2}+u_{3}^{2}/C^{2}\notag\label{rg1}\\
\dot{u}_{2} & =-2u_{2}^{2}+2u_{1}u_{2}\notag\\
\dot{u}_{3} & =[4u_{1}-2u_{2}-(u_{4}+u_{5})]u_{3}\notag\\
\dot{u}_{4} & =-u_{4}^{2}-u_{3}^{2}\notag\\
\dot{u}_{5} & =-u_{5}^{2}-u_{3}^{2}
\end{align}
For $C=1$, the equations are the same as in Refs.~\cite{chubukov,maiti}.
Notice that, if $u_{4}=u_{5}$ at the bare level, they remain equal
under pRG. For simplicity we set $u_{4}=u_{5}$.

The fixed trajectories are the solutions of (\ref{rg1}) to which
the system flows at large $L$. One can easily verified that such
solutions satisfy
\begin{equation}
u_{2}=\gamma_{2}u_{1}\,,u_{3}=\gamma_{3}u_{1}\,,u_{4}=\gamma_{4}u_{1}\label{def_g}
\end{equation}
with constant $\gamma_{i}$, i.e., the ratios of the couplings tend
to finite values under pRG. To obtain $\gamma_{i}$ we substitute
(\ref{def_g}) into (\ref{rg1}). This gives
\begin{align}
\dot{u}_{1} & =u_{1}^{2}\left(1+\gamma_{3}^{2}/C^{2}\right)\label{r0_1}
\end{align}
and
\begin{align}
\gamma_{2}\left(1+\gamma_{3}^{2}/C^{2}\right) & =\gamma_{2}[2-2\gamma_{2}]\notag\label{r0}\\
\gamma_{3}\left(1+\gamma_{3}^{2}/C^{2}\right) & =\gamma_{3}[4-2\gamma_{2}-2\gamma_{4}]\notag\\
\gamma_{4}\left(1+\gamma_{3}^{2}/C^{2}\right) & =-(\gamma_{4}^{2}+\gamma_{3}^{2})\,.
\end{align}
The last equation implies that $\gamma_{4}<0$, and we write $\gamma_{4}=-|\gamma_{4}|$.
The solution with all $\gamma_{i}\neq0$ does not exist, as one can
easily verify. However, the solutions with either $\gamma_{2}=0$
and/or $\gamma_{3}=0$ do exist.

Consider first the case $\gamma_{3}=0$, $\gamma_{2}\neq0$. In this
case the second equation \eqref{r0} should be disregarded. The other
two equations give $\gamma_{2}=1/2$ and $\gamma_{4}=-1$. Hence,
along the fixed trajectory
\begin{equation}
u_{1}(L)=\frac{1}{L_{0}-L},~\gamma_{2}=1/2\,,\gamma_{3}=0,\gamma_{4}=-1\,.\label{fix_13}
\end{equation}
This fixed trajectory describes the $u_{3}=0$ separatrix. The solution
exists per se, but the fixed trajectory is unstable in the sense that
once bare $u_{3}$ is arbitrary small but finite, the pRG trajectory
runs out of Eq. (\ref{fix_13}) Indeed, the pRG equation for $u_{3}$
in (\ref{rg1}), linearized in the proximity of the solution \eqref{fix_13},
gives $\dot{u}_{3}=[4u_{1}-2u_{2}-(u_{4}+u_{5})]u_{3}\approx5u_{1}u_{3}$.
Because $u_{1}$ is positive, $u_{3}$ increases by magnitude, no
matter whether its bare value is positive or negative. The fixed trajectory
with $\gamma_{2}=0\,,\gamma_{3}=0,\gamma_{4}=-1$ is equally unstable.

Consider next the case $\gamma_{3}\neq0$, and $\gamma_{2}=0$. In
this case the first equation in \eqref{r0} should be disregarded.
The remaining three equations give
\begin{align}
4+2|\gamma_{4}| & =1+\gamma_{3}^{2}/C^{2}\notag\label{rg3}\\
1+\gamma_{3}^{2}/C^{2} & =|\gamma_{4}|+\gamma_{3}^{2}/|\gamma_{4}|\,.
\end{align}
Solving this set we obtain
\begin{align}
\gamma_{2}=0\,,\quad\gamma_{3}=\pm C\left[-1+2C^{2}+2\sqrt{(2-C^{2})^{2}+3C^{2}}\right]^{1/2}\,,\quad\gamma_{4}=(2-C^{2})-\sqrt{(2-C^{2})^{2}+3C^{2}}\,.\label{fixed2}
\end{align}
For $C=1$ we recover the earlier results, $\gamma_{2}=0$, $\gamma_{3}=\sqrt{5}$,
$\gamma_{4}=-1$ \cite{chubukov}. The initial conditions in our model
are such that bare $u_{3}>0$, hence we choose the plus sign in the
second equation in (\ref{fixed2}). Along the fixed trajectory \eqref{fixed2},
\begin{equation}
{\dot{u}}_{1}(L)=(u_{1}(L))^{2}\left(1+\left(\frac{\gamma_{3}}{C}\right)^{2}\right).\label{chu_9}
\end{equation}
Solving this equation, we obtain
\begin{align}
u_{1}(L)=\frac{1}{1+\gamma_{3}^{2}/C^{2}}\frac{1}{L_{0}-L}\,,\label{u1L}
\end{align}
The scale $L_{0}$ cannot be explicity obtained by solving pRG equations
only along the fixed trajectory. Roughly,
\begin{align}
L_{0}=1/[u_{1}(0)(1+\gamma_{3}^{2}/C^{2})].\label{L0}
\end{align}
We remark that the condition $\gamma_{2}=0$ implies that $u_{2}/u_{1}$
tends to zero under pRG, but does not necessary imply that $u_{2}$
itself tends to zero under pRG. In fact, by going beyond the leading
approximation, one finds that $u_{2}$ also increases as $L$ approaches
$L_{0}$, but scales as 

\subsection{The fixed trajectory in the full model with non-zero bare values
of $u_{i}$ and ${\bar{u}}_{i}$}

\label{sec:full} We first show the trajectory with $\bar{u}_{i}=0$,
$i=1-5$, found in Sec.~\ref{sec:single}, is unstable. To see this
we perform a linear stability analysis around the fixed trajectory
with $\bar{u}_{i}=0$, Eq.~\eqref{def_g}, \eqref{fixed2}, assuming
that the bare values of $\bar{u}_{i}$ are small but finite. From
the second and fourth equations in the set \eqref{RG_13_a} we see
that, to the linear order, we still have $\dot{\bar{u}}_{1}=\dot{\bar{u}}_{2}=0$
For simplicity we also set have $\bar{u}_{4}=\bar{u}_{5}$. The remaining
two equations on $\bar{u}_{3}$ and $\bar{u}_{4}$ are
\begin{align}
\dot{\bar{u}}_{3} & \approx-2u_{4}\bar{u}_{3}-2u_{3}\bar{u}_{4}\notag\label{RG_lin}\\
\dot{\bar{u}}_{4} & \approx-2u_{3}\bar{u}_{3}-2u_{4}\bar{u}_{4}\,.
\end{align}
Along the fixed trajectory, Eq.~\eqref{RG_lin} can be written in
the matrix form as
\begin{align}
\begin{bmatrix}\dot{\bar{u}}_{3}\\
\dot{\bar{u}}_{4}
\end{bmatrix}=2u_{1}\hat{M}\begin{bmatrix}{\bar{u}}_{3}\\
{\bar{u}}_{4}
\end{bmatrix}\, ,\label{RG_lin1}
\end{align}
where
\begin{align}
\hat{M}=\begin{bmatrix}-\gamma_{4} & -\gamma_{3}\\
-\gamma_{3} & -\gamma_{4}
\end{bmatrix}\, .
\end{align}
This matrix is guaranteed to have at least one positive eigenvalue
because $\gamma_{4}<0$. Because $u_{1}$ is positive, this means
that the trajectory with $\bar{u}_{i}$ $i=1-5$ is unstable.

We next conjecture that the only stable fixed trajectory of the full
set of pRG equations is the one with
\begin{align}
u_{i}=\bar{u}_{i}\,,i=1-5\,,\,\,\, u_{4}=u_{5}\,.\label{p2}
\end{align}
Note that if Eq.~\eqref{p2} is satisfied at the bare level, it holds
under pRG. Along the phase trajectory of Eq.~\eqref{p2} the couplings
$u_{i}$, $i=1-4$ satisfy
\begin{align}
\dot{u}_{1} & =u_{1}^{2}+u_{3}^{2}/C^{2}\notag\label{p3}\\
\dot{u}_{2} & =2u_{1}u_{2}-2u_{2}^{2}\notag\\
\dot{u}_{3} & =4u_{3}u_{1}-2u_{2}u_{3}-4u_{3}u_{4}\notag\\
\dot{u}_{4} & =-2u_{4}^{2}-2u_{3}^{2}
\end{align}

We again assume that the ratios of the couplings tend to finite values
as the system approaches the fixed trajectory and write
\begin{align}
u_{2}=\gamma_{2}u_{1},u_{3}=\gamma_{3}u_{1},u_{4}=\gamma_{4}u_{1}\,.\label{def_g1}
\end{align}
Substituting this into \eqref{p3} we obtain
\begin{align}
\dot{u}_{1} & =u_{1}^{2}\left(1+\gamma_{3}^{2}/C^{2}\right)\label{p4_1}
\end{align}
and
\begin{align}
\gamma_{2}\left(1+\gamma_{3}^{2}/C^{2}\right) & =\gamma_{2}(2-2\gamma_{2})\notag\label{p4}\\
\gamma_{3}\left(1+\gamma_{3}^{2}/C^{2}\right) & =\gamma_{3}[4(1-\gamma_{4}-2\gamma_{2}]\notag\\
\gamma_{4}\left(1+\gamma_{3}^{2}/C^{2}\right) & =-2\left(\gamma_{4}^{2}+\gamma_{3}^{2}\right)\,.
\end{align}
We again see that (i) $\gamma_{4}$ must be negative, i.e. $\gamma_{4}=-|\gamma_{4}|$,
and (ii) that the solution with both $\gamma_{2}\neq0$ and $\gamma_{3}\neq0$
does not exist. as in this case the first and the second equations
in \eqref{p4} give $\gamma_{4}=1/2$, inconsistent with the third
equation. Hence either $\gamma_{2}$ or $\gamma_{3}$, or both, must
vanish.

For $\gamma_{2}=0$ and $\gamma_{3}\neq0$ we obtain from (\ref{p4})
\begin{align}
\gamma_{3}=\pm C\sqrt{8C^{2}-1+4\sqrt{1-C^{2}+4C^{4}}},\gamma_{4}=1-2C^{2}-\sqrt{1-C^{2}+4C^{4}}\,.\label{p8}
\end{align}
For $C=1$ this gives $\gamma_{2}=0$, $\gamma_{3}=\sqrt{15}$, $\gamma_{4}=-3$.
We verified, both numerically and analytically that this fixed trajectory
is stable. 
the condition $\gamma_{2}=0$ actually means that $u_{2}$ and ${\bar{u}}_{2}$
scale as $u_{2}=u_{20}/(L_{0}-L)^{p}$, ${\bar{u}}_{2}={\bar{u}}_{20}/(L_{0}-L)^{p}$,
with $p=2/(1+(\gamma_{3}/C)^{2})<1$, such that $u_{2}/u_{1}$ and
${\bar{u}}_{2}/u_{1}$ both tend to zero. At the same time, the prefactors
$u_{20}$ and ${\bar{u}}_{20}$ depend on initial conditions and in
general are not equal, i.e., the ratio $u_{2}/{\bar{u}}_{2}$ does
not become equal to one along the fixed trajectory.

For $\gamma_{2}\neq0$ and $\gamma_{3}=0$ we obtain from \eqref{p4}
\begin{align}
\gamma_{2}=1/2,\gamma_{3}=0,\gamma_{4}=-1/2\,.\label{p10}
\end{align}
This fixed trajectory exists pr se but is unstable because at small
deviations from $\gamma_{3}=0$ (and hence $u_{3}=0$) we have $\dot{u}_{3}\approx u_{1}u_{3}$,
hence if $u_{3}$ is initially non-zero, it grows, i.e., the system
moves away from the trajectory specified by (\ref{p10}).

Finally, if we set $\gamma_{2}=\gamma_{3}=0$, we obtain fixed trajectory
with
\begin{align}
\gamma_{2}=0,\gamma_{3}=0,\gamma_{4}=-1/2\,.\label{p11}
\end{align}
This trajectory is also unstable because once we make $\gamma_{2}$
(and, hence, $u_{2}$) small but non-zero, $u_{2}$ will flow according
to $\dot{u}_{2}\approx2u_{2}u_{1}$ and keep increasing.

We see therefore that the only stable fixed trajectory is the one
specified by Eq.~\eqref{p8}. The running coupling $u_{1}$ satisfies
the same equation ${\dot{u}}_{1}=u_{1}^{2}(1+(\gamma_{3}/C)^{2})$
as for the case when ${\bar{u}}_{i}=0$, and its flow is given by
Eq. \eqref{u1L}.

\section{SDW, SC, and orbital channels: vertices and relevant interactions}

\subsection{Interaction channels}

The tetragonal symmetry further allows us to decompose the running
interactions in Eq.~\eqref{H_int13} into different channels. To
achieve this goal we construct bilinear fermion operators that transform
irreducibly under the symmetry group of the lattice. We consider separately
the bilinear combinations in the particle-hole channel at zero momentum
and at momenta $\bm{Q}_{1,2}$, and and the bilinear combinations
in the particle-particle channel at zero total momentum.

\subsubsection{Bilinear fermion combinations in the charge and spin particle-hole
channels at large momentum transfer }

The two possible order parameters which describe charge-density-wave
(CDW) order with momenta $(\pi,0)$ and $(0,\pi)$ are
\begin{align}
\delta_{1,2}^{r}=f_{1,2}^{\dag}d_{1,2}+d_{1,2}^{\dag}f_{1,2},\,\,\,\,
\delta_{1,2}^{i}=i(f_{1,2}^{\dag}d_{1,2}- d_{1,2}^{\dag}f_{1,2})
\, .\label{del_r}
\end{align}
Another two possible charge order with large momentum transfer describe
anti-ferro-orbital order. The corresponding order parameters are
\begin{align}
\bar{\delta}_{1,2}^{r}=f_{1,2}^{\dag}d_{2,1}+ d_{2,1}^{\dag}f_{1,2}, \,\,\,\,
\bar{\delta}_{1,2}^{r}=i(f_{1,2}^{\dag}d_{2,1}- d_{2,1}^{\dag}f_{1,2})
\,.\label{bar_del}
\end{align}
These order parameters differ from the ones in Eqs.~\eqref{del_r}
because they are off-diagonal in the orbital index.

The four possible SDW order parameters with the same momenta are
\begin{align}
\bm{s}_{1,2}^{r}=f_{1,2}^{\dag}\bm{\sigma}d_{1,2}+ d_{1,2}^{\dag}\bm{\sigma}f_{1,2}, \,\,\,\,
\bm{s}_{1,2}^{i}=i (f_{1,2}^{\dag}\bm{\sigma}d_{1,2}- d_{1,2}^{\dag}\bm{\sigma}f_{1,2})\, ,
\label{s1}
\end{align}
\begin{align}
\bar{\bm{s}}_{1,2}^{r}=f_{1,2}^{\dag}\bm{\sigma}d_{2,1} +d_{1,2}^{\dag}\bm{\sigma}f_{2,1}, \,\,\,\,
\bar{\bm{s}}_{1,2}^{i}=i(f_{1,2}^{\dag}\bm{\sigma}d_{2,1}- d_{1,2}^{\dag}\bm{\sigma}f_{2,1})\,.\label{s1_b}
\end{align}

The components of Eq.~\eqref{H_int13} which describe the interactions
in CDW and SDW channels are
\begin{align}
H_{\delta,\pi}= & \frac{1}{8}(-U_{1}+2U_{2}+U_{3})\left[\delta_{1}^{r}\delta_{1}^{r}+\delta_{2}^{r}\delta_{2}^{r}\right]+\frac{1}{8}(-U_{1}+2U_{2}-U_{3})\left[\delta_{1}^{i}\delta_{1}^{i}+\delta_{2}^{i}\delta_{2}^{i}\right]\notag\label{H_del}\\
 & +\frac{1}{8}(-\bar{U}_{1}+2\bar{U}_{2}+\bar{U}_{3})\left[\bar{\delta}_{1}^{r}\bar{\delta}_{1}^{r}+\bar{\delta}_{2}^{r}\bar{\delta}_{2}^{r}\right]+\frac{1}{8}(-\bar{U}_{1}+2\bar{U}_{2}-\bar{U}_{3})\left[\bar{\delta}_{1}^{i}\bar{\delta}_{1}^{i}+\bar{\delta}_{2}^{i}\bar{\delta}_{2}^{i}\right]
\end{align}
and
\begin{align}
H_{\bm{s},\pi}= & \frac{1}{8}(-U_{1}-U_{3})\left[\bm{s}_{1}^{r}\bm{s}_{1}^{r}+\bm{s}_{2}^{r}\bm{s}_{2}^{r}\right]+\frac{1}{8}(-U_{1}+U_{3})\left[\bm{s}_{1}^{i}\bm{s}_{1}^{i}+\bm{s}_{2}^{i}\bm{s}_{2}^{i}\right]\notag\label{H_s}\\
 & +\frac{1}{8}(-\bar{U}_{1}-\bar{U}_{3})\left[\bar{\bm{s}}_{1}^{r}\bar{\bm{s}}_{1}^{r}+\bar{\bm{s}}_{2}^{r}\bar{\bm{s}}_{2}^{r}\right]+\frac{1}{8}(-\bar{U}_{1}+\bar{U}_{3})\left[\bar{\bm{s}}_{1}^{i}\bar{\bm{s}}_{1}^{i}+\bar{\bm{s}}_{2}^{i}\bar{\bm{s}}_{2}^{i}\right]\,.
\end{align}

\subsubsection{Bilinear fermion combinations in the particle-particle channel}

We focus on the singlet pairing with zero total momentum. We introduce
the notations
\begin{align}
\kappa_{\mu\mu'}^{f}=f_{\mu\uparrow}f_{\mu'\downarrow}
\,,\quad\kappa_{\mu\mu'}^{d}=d_{\mu\uparrow}d_{\mu'\downarrow}\,.
\label{Co}
\end{align}
The fermion bilinear combinations are classified as follows,
\begin{align}
\kappa_{A_{1}}^{f(d)} & =\kappa_{11}^{f(d)}+\kappa_{22}^{f(d)}\notag\label{C_channels}\\
\kappa_{B_{1}}^{f(d)} & =\kappa_{11}^{f(d)}-\kappa_{22}^{f(d)}\notag\\
\kappa_{B_{2}}^{f(d)}&=\kappa_{12}^{f(d)}+\kappa_{21}^{f(d)}\, .
\end{align}
Note that the $A_{2g}$ combination, $\kappa_{A_{2}}^{f(d)}=\kappa_{12}^{f(d)}-\kappa_{21}^{f(d)}$
vanishes as it is odd in the orbital index. The interaction component
in the Cooper channel is obtained by setting $\bm{k}_{1}=-\bm{k}_{2}$
in Eq.~\eqref{H_int13}. Expressing Eq.~\eqref{H_int13} in terms
of the combinations \eqref{C_channels} we obtain
\begin{align}
H_{\kappa}=H_{\kappa_{A_{1}}}+H_{\kappa_{B_{1}}}+H_{\kappa_{B_{2}}}\,,
\end{align}
\begin{align}
H_{\kappa_{A_{1}}}=\frac{1}{2}(U_{5}+\bar{U}_{5})[\kappa_{A_{1}}^{f}]^{\dag}\kappa_{A_{1}}^{f}+\frac{1}{2}(U_{4}+\bar{U}_{4})[\kappa_{A_{1}}^{d}]^{\dag}\kappa_{A_{1}}^{d}+\frac{1}{2}(U_{3}+\bar{U}_{3})([\kappa_{A_{1}}^{f}]^{\dag}\kappa_{A_{1}}^{d}+h.c.)
\end{align}
\begin{align}
H_{\kappa_{B_{1}}}=\frac{1}{2}(U_{5}-\bar{U}_{5})[\kappa_{B_{1}}^{f}]^{\dag}\kappa_{B_{1}}^{f}+\frac{1}{2}(U_{4}-\bar{U}_{4})[\kappa_{B_{1}}^{d}]^{\dag}\kappa_{B_{1}}^{d}+\frac{1}{2}(U_{3}-\bar{U}_{3})([\kappa_{B_{1}}^{f}]^{\dag}\kappa_{B_{1}}^{d}+h.c.)
\end{align}
\begin{align}
H_{\kappa_{B_{2}}}=\frac{1}{2}(\tilde{U}_{5}+\tilde{\tilde{U}}_{5})[\kappa_{B_{2}}^{f}]^{\dag}\kappa_{B_{2}}^{f}+\frac{1}{2}(\tilde{U}_{4}+\tilde{\tilde{U}}_{4})[\kappa_{B_{2}}^{f}]^{\dag}\kappa_{B_{2}}^{f}
\end{align}

\subsubsection{Bilinear fermion combinations in particle-hole charge channel with
zero momentum transfer}

The bilinear combinations of fermions with zero momentum transfer
in the particle-hole charge channel are
\begin{align}
\rho_{\mu\mu'}^{f}=\sum_{\sigma}f_{\mu\sigma}^{\dag}f_{\mu'\sigma}\,,\quad\rho_{\mu\mu'}^{d}=\sum_{\sigma}d_{\mu\sigma}^{\dag}d_{\mu'\sigma}\label{rho}
\end{align}
These combinations form reducible representations of the $D_{4h}$
group, separately for electrons, $f$ and holes, $d$ (Ref. \cite{Cvetkovic2013}).
All bilinear combinations are even under inversion, hence we only
consider one-dimensional (even) irreducible presentations of the $D_{4h}$
group: $A_{1g}$, $A_{2g}$, $B_{1g}$ and $B_{2g}$.

The combinations
\begin{align}
\rho_{A_{1}}^{f(d)} & =\rho_{11}^{f(d)}+\rho_{22}^{f(d)}\notag\label{rho_channels}\\
\rho_{B_{1}}^{f(d)} & =\rho_{11}^{f(d)}-\rho_{22}^{f(d)}\notag\\
\rho_{A_{2}}^{f(d)} & =\rho_{12}^{f(d)}-\rho_{21}^{f(d)}\notag\\
\rho_{B_{2}}^{f(d)} & =\rho_{12}^{f(d)}+\rho_{21}^{f(d)}
\end{align}
transform as $A_{1g}$, $B_{1g}$, $A_{2g}$, $B_{2g}$ respectively.

To obtain the interactions in the particle-hole charge channel at
zero momentum transfer (the ones which renormalize bilinear combinations
in (\ref{rho_channels}) we set $\bm{k}_{1}=\bm{k}_{2}$ or $\bm{k}_{1}=\bm{k}_{4}$
in Eq.~\eqref{H_int13}. Expressing Eq.~\eqref{H_int13} in terms
of the combinations \eqref{rho_channels} we obtain
\begin{align}
H_{\rho}=H_{\rho_{A_{1}}}+H_{\rho_{A_{2}}}+H_{\rho_{B_{1}}}+H_{\rho_{B_{2}}}\,,
\end{align}
where
\begin{align}
H_{\rho_{A_{1}}}=\frac{1}{8}(U_{5}+2\tilde{U}_{5}-\tilde{\tilde{U}}_{5})[\rho_{A_{1}}^{f}]^{2}+\frac{1}{8}(U_{4}+2\tilde{U}_{4}-\tilde{\tilde{U}}_{4})[\rho_{A_{1}}^{d}]^{2}+\frac{1}{4}\rho_{A_{1}}^{f}\rho_{A_{1}}^{d}(2U_{1}-U_{2}+2\bar{U}_{1}-\bar{U}_{2})\label{H_r_A1}
\end{align}
\begin{align}
H_{\rho_{B_{1}}}=\frac{1}{8}(U_{5}-2\tilde{U}_{5}+\tilde{\tilde{U}}_{5})[\rho_{B_{1}}^{f}]^{2}+\frac{1}{8}(U_{4}-2\tilde{U}_{4}+\tilde{\tilde{U}}_{4})[\rho_{B_{1}}^{d}]^{2}+\frac{1}{4}\rho_{B_{1}}^{f}\rho_{B_{1}}^{d}(2U_{1}-U_{2}-2\bar{U}_{1}+\bar{U}_{2})\label{chu_4}
\end{align}
\begin{align}
H_{\rho_{A_{2}}}=\frac{1}{8}(\bar{U}_{5}-2\tilde{\tilde{U}}_{5}+\tilde{U}_{5})[\rho_{A_{2}}^{f}]^{2}+\frac{1}{8}(\bar{U}_{4}-2\tilde{\tilde{U}}_{4}+\tilde{U}_{4})[\rho_{A_{2}}^{d}]^{2}\label{H_r_A2}
\end{align}
\begin{align}
H_{\rho_{B_{2}}}=\frac{1}{8}(\bar{U}_{5}+2\tilde{\tilde{U}}_{5}-\tilde{U}_{5})[\rho_{B_{2}}^{f}]^{2}+\frac{1}{8}(\bar{U}_{4}+2\tilde{\tilde{U}}_{4}-\tilde{U}_{4})[\rho_{B_{2}}^{d}]^{2}\label{H_r_B1}
\end{align}

\subsection{pRG equations for the flow of vertex functions in different channels}

We now use the solutions of pRG equations for the running couplings
as inputs and obtain pRG equations for bilinear vertices which describe
coupling of fermions to various order parameters introduced in the
previous subsection. We will not consider all vertices with large/small
momentum transfer in particle-hole and particle-particle channels
and focus only on the ones which show the strongest divergencies.
Once we obtain vertices, it will be straightforward to obtain susceptibilities.
For convenience we summarize the interaction amplitudes in different
channels in the Table~\ref{tab:couplings} at the end of the section.

\subsubsection{SDW channel}

\label{sec:SDW} The SDW verices describe the coupling of fermions
to SDW order parameters $\bm{s}_{1,2}^{r,i}$ and $\bm{{\bar{s}}}_{1,2}^{r,i}$,
defined by Eqs.~\eqref{s1} and ~\eqref{s1_b}. The subscript $1,2$
refers to the orbital, while the superscript $r,i$ refers to a true
SDW (real, $r$) or spin-current (imaginary, $i$) magnetic order.
The order parameter $\bm{s}_{1,2}^{r}$, given by Eq.~\eqref{s1},
is diagonal in orbital index and in real space describes the SDW magnetism
of Fe atoms. The the order parameter $\bm{{\bar{s}}}_{1,2}^{r,i}$,
given by Eq.~\eqref{s1_b}, is off-diagonal in orbital index, and,
when converted to real space, describes magnetism on pnictogen/chalcogen
atoms rather than on Fe (Ref. \cite{Cvetkovic2013}). We will refer
to this order as off-diagonal spin polarization.

The order parameters in the orbital basis are $\bm{s}_{1}^{r,i}=d_{xz,\alpha}^{\dagger}\bm{\sigma}_{\alpha\beta}f_{xz,\beta}\pm f_{xz,\alpha}^{\dagger}\bm{\sigma}_{\alpha\beta}d_{xz,\beta}$
and $\bm{{\bar{s}}}_{1}^{r,i}=d_{xz,\alpha}^{\dagger}\bm{\sigma}_{\alpha\beta}f_{yz,\beta}\pm f_{xz,\alpha}^{\dagger}\bm{\sigma}_{\alpha\beta}d_{yz,\beta}$.
In the band basis,
$\bm{s}_{1}^{r,i}=\langle c_{\alpha}^{\dagger}\bm{\sigma}_{\alpha\beta}f_{1,\beta}\rangle \cos{\theta}+ \langle d_{\alpha}^{\dagger}\bm{\sigma}_{\alpha\beta}f_{1,\beta} \rangle \sin{\theta}\pm\left(\langle f_{1,\alpha}^{\dagger}\bm{\sigma}_{\alpha\beta}c_{\beta}\rangle \cos{\theta}+\langle f_{1,\alpha}^{\dagger}\bm{\sigma}_{\alpha\beta}d_{\beta} \rangle \sin{\theta}\right)$,
$\bm{{\bar{s}}}_{1}^{r,i}=\langle f_{1,\alpha}^{\dagger}\bm{\sigma}_{\alpha\beta}d_{\beta}\rangle \cos{\theta}-\langle f_{1,\alpha}^{\dagger}\bm{\sigma}_{\alpha\beta}c_{\beta}\rangle \sin{\theta}\pm\left(\langle c_{\alpha}^{\dagger}\bm{\sigma}_{\alpha\beta}f_{2,\beta} \rangle \cos{\theta}+\langle d_{\alpha}^{\dagger}\bm{\sigma}_{\alpha\beta}f_{2,\beta} \rangle \sin{\theta}\right)$,
and analogous expressions for $\bm{s}_{2}^{r,i}$ and $\bm{{\bar{s}}}_{2}^{r,i}$.
In all formulas the summation over small momenta near the corresponding
$E_{F}$ is implied with transferred momentum $(\pi,0)$ for $\bm{s}_{1}$
and $(0,\pi)$ for $\bm{s}_{2}$.

\begin{figure}[h]
\begin{centering}
\includegraphics[width=0.9\columnwidth]{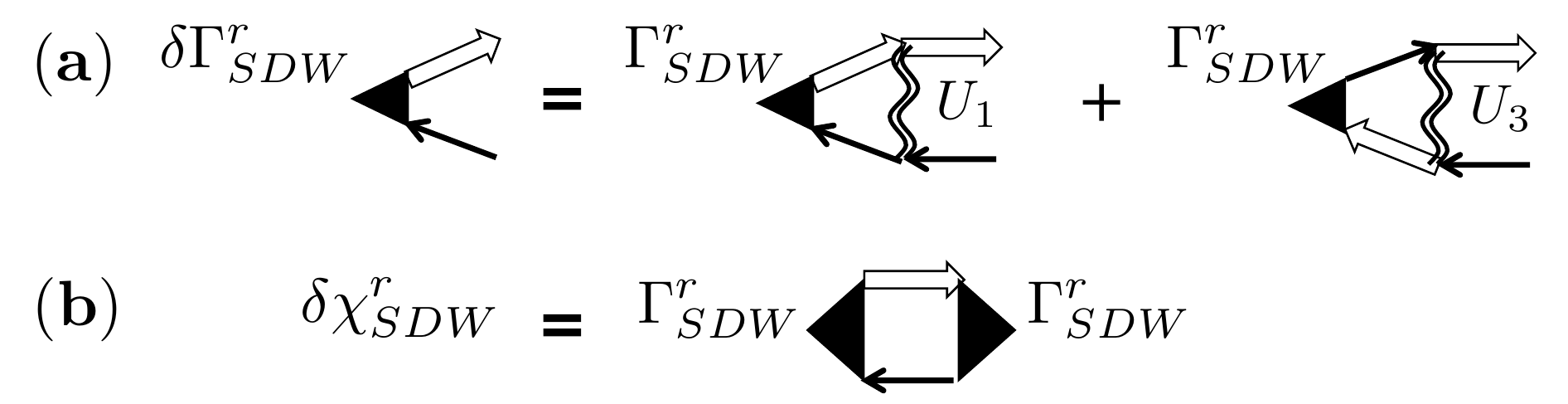} \protect\caption{(a) The diagrammatic representation of the renormalization of the
vertex $\Gamma_{SDW}^{r}$. The effective interaction in this channel
is $U_{1}+U_{3}$. (b) The diagrammatic representation of the flow
equation for the spin susceptibility $\chi_{SDW}^{r}$. \label{fig:G_chi}}

\par\end{centering}

\centering{}
\end{figure}

The pRG flow of the vertices is derived following the same procedure
which we used to derive the pRG equations for the interactions {[}see
Fig.~\ref{fig:G_chi}(a){]}

The equations decouple between diagonal and non-diagonal SDW vertices
and between even real and imaginary order parameters. The interaction
in the SDW channel with real order parameter, diagonal in the orbital
index, is $-(U_{1}+U_{3})$ (see Eq. (\ref{H_s})), where $U_{1}$
and $U_{3}$ should be understood as running variables. We label the
corresponding vertex as $\Gamma_{SDW}^{r}$ In the Wilsonian computational
scheme, the change in the SDW vertex due to the integration over the
momenta $k$ in the annulus between $k$ and $k-\delta k$ is
\begin{align}
d\Gamma_{SDW}^{r}=\frac{1}{2}\Gamma_{SDW}^{r}(U_{1}+U_{3})\int_{dk}\frac{d^{2}\bm{k}}{4\pi^{2}}\int\frac{d\epsilon}{2\pi}\left[G_{d_{1};d_{1}}(i\epsilon,\bm{k})G_{f_{1}}(i\epsilon,\bm{k})+G_{d_{2};d_{2}}(i\epsilon,\bm{k})G_{f_{2}}(i\epsilon,\bm{k})\right]\, .\label{G_SDW}
\end{align}
The prefactor of $1/2$ in Eq.~\eqref{G_SDW} includes $1/8$ in
Eq.~\eqref{H_s}, the factor of $2$ due to the summation over two
spin components, and the combinatorial factor of $2$ obtained from
two possible contractions with the two spin operators appearing in
Eq.~\eqref{H_s}. Evaluating the momentum and frequency integrals
we obtain 
\begin{align}
d\Gamma_{SDW}^{r}=\Gamma_{SDW}^{r}(U_{1}+U_{3})\frac{A}{4\pi}dL\,.\label{G_SDW_1}
\end{align}
Expressing the interactions via dimensionless couplings, we re-write
Eq. (\ref{G_SDW_1}) as
\begin{align}
\frac{d\Gamma_{SDW}^{r}}{dL}=\Gamma_{SDW}^{r}(u_{1}+u_{3}/C)\,.\label{G_SDW1}
\end{align}
On the fixed trajectory, defined by Eqs.~\eqref{def_g1} and \eqref{p8},
Eq.~\eqref{G_SDW1} becomes
\begin{align}
\frac{d\Gamma_{SDW}^{r}}{dL}=\Gamma_{SDW}^{r}u_{1}(1+\gamma_{3}/C)\,.\label{G_SDW2}
\end{align}
where $u_{1}(L)$ is given by Eq.~\eqref{u1L}. Solving the differential
equation, we obtain
\begin{align}
\Gamma_{SDW}^{r}(L)=\Gamma_{SDW,0}^{r}\left(\frac{L_{0}}{L_{0}-L}\right)^{\beta_{SDW}^{r}}\,,\label{G_SDW3}
\end{align}
where $\Gamma_{SDW,0}^{r}$ is of the same order as the bare SDW vertex,
and
\begin{align}
\beta_{SDW}^{r}=\frac{1+\gamma_{3}/C}{1+\gamma_{3}^{2}/C^{2}}\,.\label{G_SDW4}
\end{align}

The pRG flow of the spin susceptibility is shown in Fig.~\ref{fig:G_chi}(b).
It flows under pRG according to
\begin{align}
\frac{d\chi_{SDW}^{r}}{dL}=\left(\Gamma_{SDW}^{r}\right)^{2}\,.\label{G_SDW5}
\end{align}
Solving this equation we obtain
\begin{align}
\chi_{SDW}^{r}(L)\propto\frac{1}{(L_{0}-L)^{\alpha_{SDW}^{r}}}\,,\label{G_SDW6}
\end{align}
where
\begin{align}
\alpha_{SDW}^{r}=2\beta_{SDW}^{r}-1\,.\label{a_8_SM}
\end{align}

We see that the SDW spin susceptibility diverges at $L=L_{0}$, indicating
the instability towards SDW magnetism, but only when $\alpha_{SDW}^{r}>0$,
i.e., when $\beta_{SDW}^{r}>1/2$.

We emphasize that the present analysis does not resolve the orbital
degeneracy of the SDW magnetism. Indeed the two observables, $\bm{s}_{1}^{r}$
and $\bm{s}_{2}^{r}$ defined in Eq.~\eqref{s1}, describe the spin
polarization of states made of $xz$ and $yz$ atomic orbitals respectively
and the susceptibilities for the two order parameters are identical.
Below the SDW instability, the coupling between these two order parameters
determines whether they appear together or separately, i.e., whether
SDW order is a stripe or a checkerboard.

\begin{figure}[h]
\begin{centering}
\includegraphics[width=1\columnwidth]{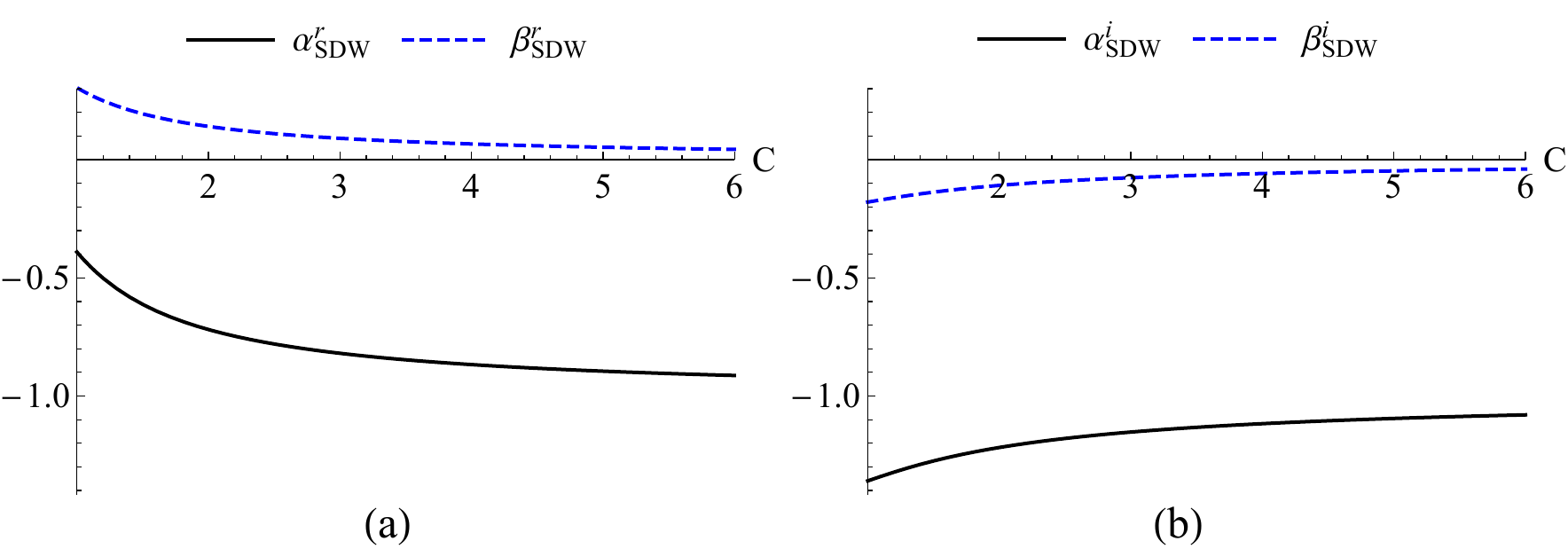} \protect\caption{ The exponents $\alpha_{SDW}^{r}$, $\beta_{SDW}^{r}$ (a) and $\alpha_{SDW}^{i}$,
$\beta_{SDW}^{i}$ (b) as functions of the parameter $C=(m_{e}+m_{h})/(2\sqrt{m_{e}m_{h}})\geq1$. \label{fig:exp_SDW}}

\par\end{centering}

\centering{}
\end{figure}

The computation of the vertex and the susceptibility for the diagonal
imaginary SDW order parameter (spin-current) proceeds in the same
way, and the result is
\begin{align}
\chi_{SDW}^{i}(L)\propto\frac{1}{(L_{0}-L)^{\alpha_{SDW}^{i}}}\,,\label{G_SDW6_1}
\end{align}
where
\begin{align}
\alpha_{SDW}^{i}=2\beta_{SDW}^{i}-1\,.\label{a_8_SM_1}
\end{align}
and
\begin{align}
\beta_{SDW}^{i}=\frac{1-\gamma_{3}/C}{1+\gamma_{3}^{2}/C^{2}}\,.\label{G_SDW4_1}
\end{align}
This exponent is smaller than $\beta_{SDW}^{r}$, hence spin-current
order is subleading to the real SDW order. The exponents, Eqs.~\eqref{G_SDW4},
\eqref{a_8_SM}, \eqref{a_8_SM_1}, \eqref{G_SDW4_1} are plotted
as functions of the parameter $C=(m_{e}+m_{h})/(2\sqrt{m_{e}m_{h}})\geq1$
in Fig.~\ref{fig:exp_SDW}

Next, we consider the vertices for the coupling to other two order
parameters $\bar{\bm{s}}_{1}^{r}$ and $\bar{\bm{s}}_{2}^{r}$, each
with mixed orbital content. Using Eq.~\eqref{H_s} and performing
the same calculations as above, we find that the equations for the
vertices with different orbital index decouple, and each $\bar{\Gamma}_{SDW}^{r}$
obeys
\begin{align}
\frac{d\bar{\Gamma}_{SDW}^{r}}{dL}={\bar{\Gamma}}_{SDW}^{r}(\bar{u}_{1}+\bar{u}_{3}/C)\,.\label{G_SDW1_b}
\end{align}
Because $u_{1,3}=\bar{u}_{1,3}$ on the fixed trajectory, see Eq.~\eqref{p2},
the corresponding susceptibilities scale as
\begin{align}
\bar{\chi}_{SDW}^{r}(L)\propto\frac{1}{(L_{0}-L)^{\bar{\alpha}_{SDW}^{r}}}\label{G_SDW6_b}
\end{align}
has the same exponent as $\alpha_{SDW}^{r}$, i.e., $\bar{\alpha}_{SDW}^{r}=\alpha_{SDW}^{r}$.
This indicates that the spatial spin arrangement below the magnetic
transition must include all four types of SDW order.

Similarly, the susceptibility $\bar{\chi}_{SDW}^{r}(L)$ behaves as
\begin{align}
\bar{\chi}_{SDW}^{i}(L)\propto\frac{1}{(L_{0}-L)^{\bar{\alpha}_{SDW}^{i}}}\, , \label{G_SDW6_c}
\end{align}
where along fixed trajectory $\bar{\alpha}_{SDW}^{i}=\alpha_{SDW}^{i}$.

\subsubsection{CDW channel}

The CDW order parameters are defined in Eq.~\eqref{del_r} and Eq.
\eqref{bar_del}. The corresponding interaction components are presented
in Eq.~\eqref{H_del}.

The order parameters in the orbital basis are $\delta_{1}^{r,i}=d_{xz,\alpha}^{\dagger}f_{xz,\alpha}\pm f_{xz,\alpha}^{\dagger}d_{xz,\alpha}$
and
${\bar{\delta}}_{1}^{r,i}=d_{xz,\alpha}^{\dagger}f_{yz,\alpha}\pm f_{xz,\alpha}^{\dagger}d_{yz,\alpha}$.
In the band basis,
$\delta_{1}^{r,i}=\langle c_{\alpha}^{\dagger}f_{1,\alpha}\rangle \cos{\theta}+ \langle d_{\alpha}^{\dagger}f_{1,\alpha}\rangle \sin{\theta}\pm\left(\langle f_{1,\alpha}^{\dagger}c_{\alpha}\rangle \cos{\theta}+\langle f_{1,\alpha}^{\dagger}d_{\alpha}\rangle \sin{\theta}\right)$,
${\bar{\delta}}_{1}^{r,i}=\langle f_{1,\alpha}^{\dagger}d_{\alpha}\rangle \cos{\theta}-\langle f_{1,\alpha}^{\dagger}c_{\alpha}\rangle \sin{\theta}\pm\left(\langle c_{\alpha}^{\dagger}f_{2,\alpha}\rangle \cos{\theta}+\langle d_{\alpha}^{\dagger}f_{2,\alpha}\rangle \sin{\theta}\right)$,
and analogous expressions for $\delta_{2}^{r,i}$ and ${\bar{\delta}}_{2}^{r,i}$.
Again, the summation over momentum is implied, the transferred momentum
is $\pi,0)$ for $\delta_{1}$ and $(0,\pi)$ for $\delta_{2}$.

\begin{figure}[h]
\begin{centering}
\includegraphics[width=1\columnwidth]{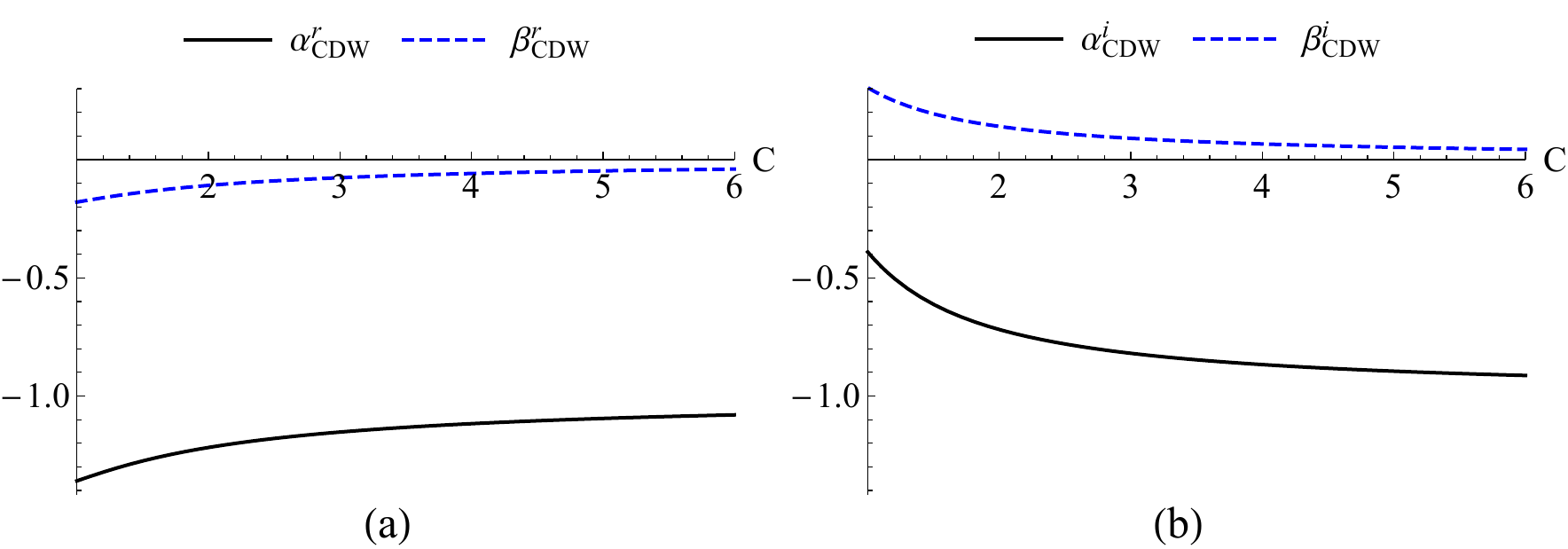} \protect\caption{(a) The exponents $\alpha_{CDW}^{r}$, $\beta_{CDW}^{r}$ and (b)
$\alpha_{CDW}^{i}$, $\beta_{CDW}^{i}$ as functions of the parameter
$C$. \label{fig:exp_CDW}}
\par\end{centering}
\centering{}
\end{figure}

The analysis of susceptibilities in the CDW channel is analogous to
what we just did for the SDW channel and we skip intermediate steps.
Along the fixed pRG trajectory the CDW susceptibilities for DCDW order
parameters, diagonal in orbital index, scale as
\begin{eqnarray}
 &  & \chi_{CDW}^{r}(L)\propto\frac{1}{(L_{0}-L)^{\alpha_{CDW}^{r}}}\nonumber \\
 &  & \chi_{CDW}^{i}(L)\propto\frac{1}{(L_{0}-L)^{\alpha_{CDW}^{i}}}\, ,
\end{eqnarray}
where the exponents are
\begin{align}
\alpha_{CDW}^{r}=2\beta_{CDW}^{r}-1\,,\quad\alpha_{CDW}^{i}=2\beta_{CDW}^{i}-1\label{a_CDW_1}
\end{align}
with
\begin{eqnarray}\label{G_CDW2}
 &  & \beta_{CDW}^{r}=\frac{1-2\gamma_{2}-\gamma_{3}/C}{1+\gamma_{3}^{2}/C^{2}}\,\nonumber \\
 &  & \beta_{CDW}^{i}=-\frac{1-2\gamma_{2}+\gamma_{3}/C}{1+\gamma_{3}^{2}/C^{2}}\, .
\end{eqnarray}
The results Eq.~\eqref{a_CDW_1} and \eqref{G_CDW2} are presented
graphically in Fig.~\ref{fig:exp_CDW}.

For order parameters which are odd in the orbital index (anti-ferro-orbital
order parameters) the susceptibilities along the fixed trajectory
are
\begin{eqnarray}
 &  & {\bar{\chi}}_{CDW}^{r}(L)\propto\frac{1}{(L_{0}-L)^{{\bar{\alpha}}_{CDW}}}\, ,\nonumber \\
 &  & {\bar{\chi}}_{CDW}^{i}(L)\propto\frac{1}{(L_{0}-L)^{{\bar{\beta}}_{CDW}}}\, .
\end{eqnarray}
The exponents ${\bar{\alpha}}_{CDW}^{r}$ and ${\bar{\alpha}}_{CDW}^{i}$
are the same as in the CDW channel, diagonal in the orbital index.
One can verify this using Eq.~\eqref{H_del} and the relation Eq.~\eqref{p2},
i.e
\begin{align}
\bar{\alpha}_{CDW}^{r}=\alpha_{CDW}^{r}\, ,~~~~\bar{\alpha}_{CDW}^{i}=\alpha_{CDW}^{i}\, .\label{G_CDW4}
\end{align}

\subsubsection{Particle-particle channel}
\label{sec:SC}

For simplicity, we will refer to the instability in the particle-particle
channel as SC instability and to particle-particle channel as Cooper
channel, although within our pRG this instability involves fermions
with energies away from $E_{F}$ and is towards the formation of a
bound state of two fermions with zero total momentum.

We remind that there are three pairing channels with non-zero order
parameters: $A_{1g}$, $B_{1g}$, and $B_{2g}$. The corresponding
order parameters in the orbital basis are $\kappa_{A_{1}}^{f}=f_{xz,\uparrow}f_{xz,\downarrow}+f_{yz,\uparrow}f_{yz,\downarrow}$,
$\kappa_{A_{1}}^{d}=d_{xz,\uparrow}d_{xz,\downarrow}+d_{yz,\uparrow}d_{yz,\downarrow}$,
$\kappa_{B_{1}}^{f}=f_{xz,\uparrow}f_{xz,\downarrow}-f_{yz,\uparrow}f_{yz,\downarrow}$,
$\kappa_{B_{1}}^{d}=d_{xz,\uparrow}d_{xz,\downarrow}-d_{yz,\uparrow}d_{yz,\downarrow}$,
$\kappa_{B_{2}}^{f}=f_{xz,\uparrow}f_{yz,\downarrow}+f_{yz,\uparrow}f_{xz,\downarrow}$,
$\kappa_{B_{2}}^{d}=d_{xz,\uparrow}d_{yz,\downarrow}+d_{yz,\uparrow}d_{xz\downarrow}$.
In the band basis, $\kappa_{A_{1}}^{f}=f_{1,\uparrow}f_{1,\downarrow}+f_{2,\uparrow}f_{2,\downarrow}$,
$\kappa_{A_{1}}^{d}= c_{\uparrow}c_{\downarrow}+d_{\uparrow}d_{\downarrow}$,
$\kappa_{B_{1}}^{f}=f_{1,\uparrow}f_{1,\downarrow}-f_{2,\uparrow}f_{2,\downarrow}$,
$\kappa_{B_{1}}^{d}=\left( c_{\uparrow}c_{\downarrow}   -  d_{\uparrow}d_{\downarrow}   \right)\cos{2\theta}+\left( c_{\uparrow}d_{\downarrow}   + d_{\uparrow}c_{\downarrow}  \right)\sin{2\theta}$,
$\kappa_{B_{2}}^{f}=f_{1,\uparrow}f_{2,\downarrow}+f_{2,\uparrow}f_{1,\downarrow}$,
$\kappa_{B_{2}}^{d}=\left( c_{\uparrow}c_{\downarrow}   + d_{\uparrow}d_{\downarrow}  \right)\sin{2\theta}+\left(c_{\uparrow}d_{\downarrow}   - d_{\uparrow}c_{\downarrow}  \right)\cos{2\theta}$.
The summation over momenta is implied, the total momentum in each
term is zero.

The running interactions in the $B_{2g}$ channel, $\tilde{U}_{4}$,
$\tilde{\tilde{U}}_{4}$ $\tilde{U}_{5}$, and $\tilde{\tilde{U}}_{5}$,
all scale to zero, see Secs.~\ref{sec:Ut5} and \ref{sec:Ut4}. In
the other two channels, we introduce the vertices $\Gamma_{C,A_{1}}^{f}$
and $\Gamma_{C,A_{1}}^{d}$ for the coupling of fermions from electron
and hole pockets to SC order parameter with $A_{1g}$ symmetry, and
$\Gamma_{C,B_{1}}^{f}$ and $\Gamma_{C,B_{1}}^{d}$ for the same in
$B_{1g}$ symmetry channel.

\begin{figure}[h]
\begin{centering}
\includegraphics[width=0.5\columnwidth]{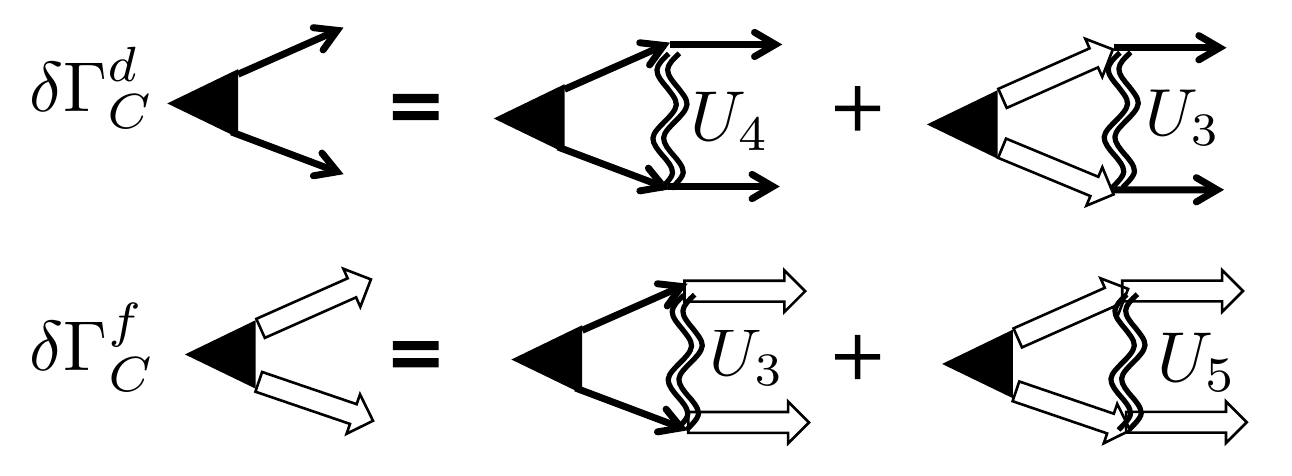} \protect\caption{The diagrammatic representation of the equations for the interaction
vertices in the particle-particle channel. The contributions from
$\bar{U}_{3}$, $\bar{U}_{4}$ and $\bar{U}_{5}$ are not shown. They
have the same structure as the ones we kept in the figure. \label{fig:G_SC}}

\par\end{centering}

\centering{}
\end{figure}

The pRG equations for these vertices are obtained using the same computational
procedure as before:
\begin{align}
d\Gamma_{C,A_{1}(B_{1})}^{f} & =-\Gamma_{C,A_{1}(B_{1})}^{f}\frac{1}{2}(U_{5}\pm\bar{U}_{5})d\Pi_{C,A_{1}(B_{1})}^{f}-\Gamma_{C,A_{1}(B_{1})}^{d}\frac{1}{2}(U_{3}\pm\bar{U}_{3})d\Pi_{C,A_{1}(B_{1})}^{d}\notag\label{dG_SC1}\\
d\Gamma_{C,A_{1}(B_{1})}^{d} & =-\Gamma_{C,A_{1}(B_{1})}^{f}\frac{1}{2}(U_{3}\pm\bar{U}_{3})d\Pi_{C,A_{1}(B_{1})}^{f}-\Gamma_{C,A_{1}(B_{1})}^{d}\frac{1}{2}(U_{4}\pm\bar{U}_{4})d\Pi_{C,A_{1}(B_{1})}^{d}\,,
\end{align}
where the upper (lower) sign are for $A_{1g}$ and $B_{1g}$ channels,
respectively. The quantities $d\Pi_{C,A_{1}(B_{1})}^{f}$ and $d\Pi_{C,A_{1}(B_{1})}^{d}$
are given by
\begin{align}
d\Pi_{C,A_{1}(B_{1})}^{f}= & \int_{dL}\frac{d^{2}k}{4\pi^{2}}\int\frac{d\epsilon}{2\pi}\left[G_{f_{1}}(i\epsilon,\epsilon_{f_{1}}(k))G_{f_{1}}(-i\epsilon,-k)+G_{f_{2}}(i\epsilon,\epsilon_{f_{1}}(k))G_{f_{2}}(-i\epsilon,-k)\right]\notag\label{dG_SC3}\\
 & =\frac{dL}{2\pi}A_{e}\,,
\end{align}
and
\begin{align}
d\Pi_{C,A_{1}(B_{1})}^{d}= & \int_{dL}\frac{d^{2}k}{4\pi^{2}}\int\frac{d\epsilon}{2\pi}\big[G_{d_{1},d_{1}}(i\epsilon,k)G_{d_{1},d_{1}}(-i\epsilon,-k)+G_{d_{2},d_{2}}(i\epsilon,k)G_{d_{1},d_{1}}(-i\epsilon,-k)\notag\label{dG_SC5}\\
 & \pm G_{d_{1},d_{2}}(i\epsilon,k)G_{d_{1},d_{2}}(-i\epsilon,-k)\pm G_{d_{2},d_{1}}(i\epsilon,k)G_{d_{2},d_{1}}(-i\epsilon,-k)\big]\notag\\
 & =\frac{dL}{2\pi}(A_{h}\pm A_{h}^{-})\,,
\end{align}
In obtaining these expressions we used Eqs.~\eqref{G_f11}, \eqref{Gdd_3},
\eqref{Gdd_4},~\eqref{Ae}, ~\eqref{A_h}, and \eqref{A_hm}.
We show the equations for the interaction vertices graphically in
Fig.~\ref{fig:G_SC}.

Substituting Eqs.~\eqref{dG_SC3} and \eqref{dG_SC5} into Eq.~\eqref{dG_SC1},
introducing dimensionless couplings, and approximating hole masses
$m_{c}$ and $m_{d}$ as $m_{h}$, we obtain
\begin{align}
\frac{d\Gamma_{C,A_{1}(B_{1})}^{f}}{dL} & =-\Gamma_{C,A_{1}(B_{1})}^{f}(u_{5}\pm\bar{u}_{5})-\Gamma_{C,A_{1}(B_{1})}^{d}\frac{A_{h}}{AC}(u_{3}\pm\bar{u}_{3})\notag\label{dG_SC7}\\
\frac{d\Gamma_{C,A_{1}(B_{1})}^{d}}{dL} & =-\Gamma_{C,A_{1}(B_{1})}^{f}\frac{A_{e}}{AC}(u_{3}\pm\bar{u}_{3})-\Gamma_{C,A_{1}(B_{1})}^{d}(u_{4}\pm\bar{u}_{4})\,.
\end{align}

\begin{figure}[h]
\begin{centering}
\includegraphics[width=0.6\columnwidth]{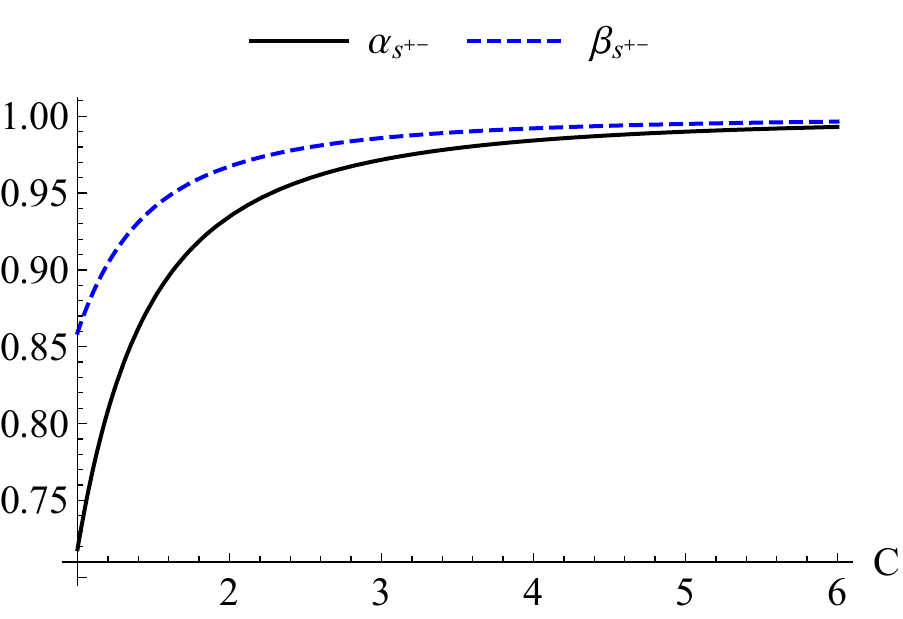} \protect\caption{The exponents $\alpha_{s^{+-}}$ and $\beta_{s^{+-}}$ as functions
of the parameter $C$. Both exponents approach $1$ at large $C$. \label{fig:al_sc}}

\par\end{centering}

\centering{}
\end{figure}

Along the stable fixed trajectory $u_{i}={\bar{u}}_{i}$, hence the
vertex in $B_{1g}$ channel does not renormalize, while the one in
$A_{1g}$ channel obeys
\begin{align}
\frac{d}{dL}\begin{bmatrix}\Gamma_{C,A_{1}}^{f}\\
\Gamma_{C,A_{1}}^{d}
\end{bmatrix}=2u_{1}M_{C}\begin{bmatrix}\Gamma_{C,A_{1}}^{f}\\
\Gamma_{C,A_{1}}^{d}
\end{bmatrix}\,,\,\,\, M_{C}=\begin{bmatrix}|\gamma_{4}| & -\gamma_{3}\frac{A_{h}}{AC}\\
-\gamma_{3}\frac{A_{e}}{AC} & |\gamma_{4}|
\end{bmatrix}\,.\label{dG_SC9}
\end{align}
Combining $\Gamma_{C,A_{1}}^{f}$ and $\Gamma_{C,A_{1}}^{d}$ into
symmetric, $s^{++}$, and anti-symmetric, $s^{+-}$ channels as
\begin{align}
\begin{bmatrix}\Gamma_{C,A_{1}}^{f}\\
\Gamma_{C,A_{1}}^{d}
\end{bmatrix}=\Gamma_{s^{+-},s^{++}}\begin{bmatrix}\sqrt{A_{h}/A_{e}}\\
\mp1
\end{bmatrix},\label{Eig1}
\end{align}
where the upper and lower signs are for $s^{+-}$ $s^{++}$ channels,
and using Eq.~\eqref{C}, we obtain from Eq.~\eqref{dG_SC9},
\begin{align}
\frac{d\Gamma_{s^{+-},s^{++}}}{dL}=2u_{1}(|\gamma_{4}|\pm\gamma_{3})\Gamma_{s^{+-},s^{++}}\,.\label{Eig2}
\end{align}
The exponent for the $s^{+-}$ channel is obviously larger, and focusing
on this channel only we obtain from (\ref{Eig2})
\begin{align}
\Gamma_{s^{+-}}(L)=\Gamma_{s^{+-},0}\left(\frac{L_{0}}{L_{0}-L}\right)^{\beta_{s^{+-}}}\,,\label{dG_SC11}
\end{align}
where
\begin{align}
\beta_{s^{+-}}=2\frac{|\gamma_{4}|+\gamma_{3}}{1+\gamma_{3}^{2}/C^{2}}\,.\label{dG_SC13}
\end{align}
The pRG equation for the pairing susceptibility has the same form
as in the SDW channel:
\begin{align}
\frac{d\chi_{s^{+-}}}{dL}=\Gamma_{s^{+-}}^{2}.\label{dG_SC15}
\end{align}
Using Eq.~\eqref{dG_SC13} we obtain
\begin{align}
\chi_{s^{+-}}(L)\propto\frac{1}{(L_{0}-L)^{\alpha_{s^{+-}}}}\,,\label{dG_SC17}
\end{align}
where the exponent is
\begin{align}
\alpha_{s^{+-}}=2\beta_{s^{+-}}-1=4\frac{|\gamma_{4}|+\gamma_{3}}{1+\gamma_{3}^{2}/C^{2}}-1\,.\label{dG_SC19}
\end{align}
The exponents $\alpha_{s^{+-}}$ and $\beta_{s^{+-}}$ controlling
the Cooper channel susceptibility and the vertex respectively are
shown as functions of the parameter $C$ in Fig.~\ref{fig:al_sc}.


As a side note, we remark that for the unstable fixed trajectory,
Eq.~\eqref{fixed2}, with $\bar{u}_{i}=0$ for $i=1-5$, the exponents
in the $A_{1g}$ and $B_{1g}$ Cooper channels are identical. In both
channels the superconducting susceptibilities scale as $\chi_{s^{+-}}\approx\chi_{d^{+-}}\propto1/(L_{0}-L)^{\alpha_{sc}}$
with $\alpha_{sc}=2(|\gamma_{4}|+\gamma_{3})/(1+\gamma_{3}^{2}/C^{2})-1$.

We next compute susceptibilities in Pomeranchuk channels. There are
four channels, even under inversion - $A_{1}$, $A_{2}$, $B_{1}$
and $B_{2}$. The order parameters in these channels are presented
in Eq. \eqref{rho_channels}. We consider each channel separately.

\subsubsection{$B_{1}$ Pomeranchuk channel}

We first compute the vertex and the susceptibility in the $B_{1}$
charge Pomeranchuk channel. The corresponding order parameter in the
orbital basis is $\rho_{B_{1}}=n_{xz}-n_{yz}$ with contributions
from states near hole and electron pockets, $\rho_{B_{1}}^{d}$ and
$\rho_{B_{1}}^{f}$, see Eq. \eqref{rho_channels}. We label corresponding
vertices as $\Gamma_{ph,B_{1}}^{d}$ and $\Gamma_{ph,B_{1}}^{f}$.
In the band basis, $\rho_{B_{1}}^{d}$ is the combination of $\langle c^{\dagger}c-d^{\dagger}d \rangle  \cos{2\theta}$
and $\langle c^{\dagger}d+d^{\dagger}c \rangle  \sin{2\theta}$ with equal amplitudes,
while $\rho_{B_{1}}^{f}$ is just $\langle f_{1}^{\dagger}f_{1}-f_{2}^{\dagger}f_{2} \rangle  $.
As before, the summation over momentum is implied, the transferred
momentum in all terms is equal to zero.

\begin{figure}[h]
\begin{centering}
\includegraphics[width=0.9\columnwidth]{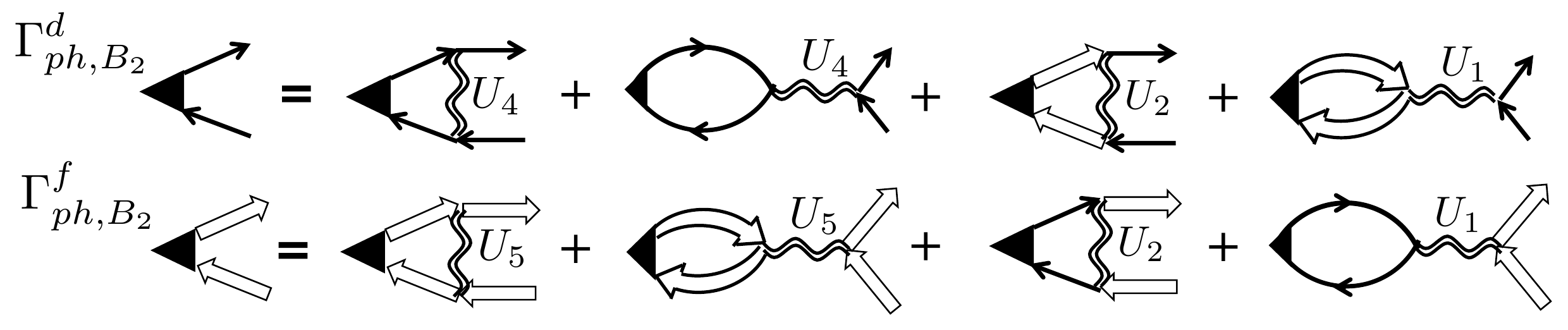} \protect\caption{The diagrammatic representation of the equations for the interaction
vertices in $B_{1}$ Pomeranchuk channel. The contributions from $\tilde{U}_{4}$,
$\tilde{\tilde{U}}_{4}$, $\bar{U}_{1}$ and $\bar{U}_{2}$ are not
shown. They have the same structure as the ones which we kept in the
figure. \label{fig:Pom_d}}

\par\end{centering}

\centering{}
\end{figure}

The polarization operator in the Pomeranchuk channel is not logarithmical
and, moreover, internal and external energies are of the same order,
i.e., if one probes the vertices $\Gamma_{ph,B_{1}}^{d}$ and $\Gamma_{ph,B_{1}}^{f}$
at a scale $L$, typical internal scale in the diagram for vertex
renormalization is also of order $L$. The vertex still flows logarithmically
because its renormalization involves running interactions. However,
because the running interaction in the $B_{2g}$ Pomeranchuk channel
is the only source of logarithmical flow, the ladder series of vertex
renormalizations reduce to algebraic rather than differential equations
for $\Gamma_{ph,B_{1}}^{d}$ and $\Gamma_{ph,B_{1}}^{f}$. Using Eq.
(\ref{chu_4}) for the vertices and evaluating ladder series of vertex
renormalizations we obtain
\begin{align}
\begin{bmatrix}\Gamma_{ph,B_{1}}^{d}\\
\Gamma_{ph,B_{1}}^{f}
\end{bmatrix}=M_{B_{1},ph}\begin{bmatrix}\Gamma_{ph,B_{1}}^{d}\\
\Gamma_{ph,B_{1}}^{f}
\end{bmatrix}+\begin{bmatrix}\Gamma_{ph,B_{1}}^{d(0)}\\
\Gamma_{ph,B_{1}}^{f(0)}
\end{bmatrix}\,,\label{chu_8}
\end{align}
where $\Gamma_{ph,B_{1}}^{d(0)}$ and $\Gamma_{ph,B_{1}}^{f(0)}$
are the bare vertices, and
\begin{align}
M_{B_{1},ph}=\begin{bmatrix}-2(u_{4}-2\tilde{u}_{4}+\tilde{\tilde{u}}_{4}) & -2\frac{A_{e}}{A}(2u_{1}-2\bar{u}_{1}-u_{2}+\bar{u}_{2})\\
-2\frac{A_{h}}{A}(2u_{1}-2\bar{u}_{1}-u_{2}+\bar{u}_{2}) & -2(u_{5}-2\tilde{u}_{5}+\tilde{\tilde{u}}_{5})
\end{bmatrix}\,.\label{M_ph}
\end{align}
The relations Eq.~\eqref{chu_8} and \eqref{M_ph} are illustrated
in Fig.~\ref{fig:Pom_d}. 

The two eigenvalues of the matrix $M_{B_{1},ph}$ are
\begin{align}
\lambda_{\pm}^{B_{1}} & =-(u_{4}-2\tilde{u}_{4}+\tilde{\tilde{u}}_{4}+u_{5}-2\tilde{u}_{5}+\tilde{\tilde{u}}_{5})\pm\sqrt{D_{M}}\,,\notag\label{lambda_P}\\
D_{M} & =(u_{4}-2\tilde{u}_{4}+\tilde{\tilde{u}}_{4}-u_{5}+2\tilde{u}_{5}-\tilde{\tilde{u}}_{5})^{2}+4C^{2}(2u_{1}-2\bar{u}_{1}-u_{2}+\bar{u}_{2})^{2}
\end{align}
Along the stable fixed trajectory the two eigenvalues are degenerate:
\begin{align}
\lambda_{+}^{B_{1}}=\lambda_{-}^{B_{1}}=\lambda^{B_{1}}=2|\gamma_{4}|u_{1}=\frac{2|\gamma_{4}|}{1+\gamma_{3}^{2}C^{2}}\frac{1}{L_{0}-L}\, .\label{pom_c}
\end{align}
Accordingly,
\begin{equation}
\Gamma_{ph,B_{1}}^{d(f)}=\frac{\Gamma_{ph,B_{1}}^{d(0)(f(0))}}{1-\lambda^{B_{1}}}\propto\frac{1}{L_{P_{d}}-L}\, ,
\end{equation}
where
\begin{align}
L_{P_{d}}=L_{0}-\frac{2|\gamma_{4}|}{1+\gamma_{3}^{2}C^{2}}\, \label{L_Pd}
\end{align}
implies that $B_{1}$ Pomeranchuk vertices for states near hole and
electron pockets grow independent on each other, each is proportional
to its bare value. At $L\approx L_{P_{d}}$, when the running $\Gamma_{ph,B_{1}}^{d(f)}$
are much larger than their bare values, the ratio $\Gamma_{ph,B_{1}}^{d}/\Gamma_{ph,B_{1}}^{f}$
actually tends to a universal number, determined by the way how the
system approaches the fixed trajectory. Solving for the ratio $\Gamma_{ph,B_{1}}^{d}/\Gamma_{ph,B_{1}}^{f}$
by setting $\Gamma_{ph,B_{1}}^{d(0)(f(0))}$ to zero but keeping $u_{i}-{\bar{u}}_{i}$
small but finite, we obtain from (\ref{chu_8}) and (\ref{lambda_P})
that $\Gamma_{ph,B_{1}}^{d}/\Gamma_{ph,B_{1}}^{f}$ approaches 4 at
$L=L_{0}$. We cited this result in the main text.

The $B_{1}$ Pomeranchuk susceptibility is given by a series of diagrams
which consist of a particle-hole bubble with ladder series of vertex
renormalizations. Because integration over internal momenta in each
cross-section does not give rise to logarithms, ladder renormalizations
can be absorbed into the renormalization of just one of side vertices.
As a result,
\begin{equation}
\chi_{ph,B_{1}}\sim\Gamma_{ph,B_{1}}^{d(f)}\propto\frac{1}{L_{P_{d}}-L}\, .
\end{equation}

We see that Pomeranchuk susceptibility diverges with the exponent
$\alpha_{P_{d}}=1$ and, moreover, $L_{P_{d}}<L_{0}$, i.e., Pomeranchuk
susceptibility diverges at a smaller $L$ (i.e., at a larger temperature)
than the susceptibilities in SDW and SC channels. We discuss the consequences
in the main text.

We emphasize the role of the flow of the couplings plays the major
role in this analysis. If we did the same calculation as above but
with the bare couplings related to Hund and Hubbard interaction terms,
we would obtain
\begin{align}
M_{ph,B_{1}}=-(U-2U'+J)\frac{m}{2\pi}\begin{bmatrix}1 & 1\\
1 & 1
\end{bmatrix}\,.\label{M_ph1}
\end{align}
For simplicity we set $m_{e}=m_{h}=m$. The matrix $M_{ph,B_{1}}$
has one zero eigenvalue $\lambda_{-}=0$ due to the particle-hole
symmetry at the bare level, and the other one is $\lambda_{+}=-(U-2U'+J)\frac{m}{2\pi}$.
This eigenvalue is positive only when $U'>(U+J)/2$, and, even if
itr is positive, $2U'-(U+J)$ has to exceed the critical value, otherwise
the Pomeranchuk instability does not develop. In the full theory,
which incorporates the flow of the couplings, the Pomeranchuk instability
develops at arbitrary Hubbard and Hund repulsive interactions and
for arbitrary ratios of $U,U'$ and $J$.

\subsubsection{$A_{1}$ channel}

The Pomeranchuk suceptibility in the $A_{1}$ channel is analysed
in a similar way. The $A_{1}$ order parameter in the orbital basis
is $\rho_{A1}=n_{xz}+n_{yz}$ with contributions from states near
hole and electron pockets, $\rho_{A_{1}}^{d}$ and $\rho_{A_{1}}^{f}$,
see Eq. \eqref{rho_channels}. We label corresponding vertices as
$\Gamma_{ph,A_{1}}^{d}$ and $\Gamma_{ph,A_{1}}^{f}$. In the band
basis, $\rho_{A_{1}}^{d}$ is $\langle c^{\dagger}c+d^{\dagger}d \rangle$ and
$\rho_{A_{1}}^{f}$
is $\langle f_{1}^{\dagger}f_{1}+f_{2}^{\dagger}f_{2} \rangle $. By analogy with
superconductivity, we label the state with the same sign of $\rho_{A_{1}}^{d}$
and $\rho_{A_{1}}^{f}$ as $s^{++}$ and the state with opposite signs
as $s^{+-}$.

The interaction in the $A_{1}$ channel is presented in Eq. (\ref{H_r_A1}).
Using this equation and performing the same analysis as in the $B_{2}$
channel, we find the set of self-consistent equations for $A_{1}$
vertices in the form
\begin{align}
\begin{bmatrix}\Gamma_{ph,A_{1}}^{d}\\
\Gamma_{ph,A_{1}}^{f}
\end{bmatrix}=M_{A_{1},ph}\begin{bmatrix}\Gamma_{ph,A_{1}}^{d}\\
\Gamma_{ph,A_{1}}^{f}
\end{bmatrix}+\begin{bmatrix}\Gamma_{ph,A_{1}}^{d(0)}\\
\Gamma_{ph,A_{1}}^{f(0)}
\end{bmatrix}\,,\label{M_Aph}
\end{align}
where $\Gamma_{ph,A_{1}}^{d(0)}$ and $\Gamma_{ph,A_{1}}^{f(0)}$
are the bare vertices, and
\begin{align}
M_{A_{1},ph}=\begin{bmatrix}-2(u_{4}+2\tilde{u}_{4}-\tilde{\tilde{u}}_{4}) & -2\frac{A_{e}}{A}(2u_{1}+2\bar{u}_{1}-u_{2}-\bar{u}_{2})\\
-2\frac{A_{h}}{A}(2u_{1}+2\bar{u}_{1}-u_{2}-\bar{u}_{2}) & -2(u_{5}+2\tilde{u}_{5}-\tilde{\tilde{u}}_{5})
\end{bmatrix}\,.\label{M_ph_A}
\end{align}
The two eigenvalues of the matrix $M_{A_{1},ph}$ are
\begin{align}
\lambda_{\pm}^{A_{1}} & =-(u_{4}+2\tilde{u}_{4}-\tilde{\tilde{u}}_{4}+u_{5}+2\tilde{u}_{5}-\tilde{\tilde{u}}_{5})\pm\sqrt{D_{M}}\,,\notag\label{lambda_P_A}\\
D_{M} & =(u_{4}+2\tilde{u}_{4}-\tilde{\tilde{u}}_{4}-u_{5}-2\tilde{u}_{5}+\tilde{\tilde{u}}_{5})^{2}+4C^{2}(2u_{1}+2\bar{u}_{1}-u_{2}-\bar{u}_{2})^{2}
\end{align}
These two eigenvalues are not degenerate along the stable fixed trajectory,
Eq.~\eqref{p2}, and are given by
\begin{align}
\lambda_{\pm}^{A_{1}}=2(|\gamma_{4}|\pm4C)u_{1}\,.\label{lambda_P_A1}
\end{align}
It follows from Eq.~\eqref{M_Aph} that that $\lambda_{+}^{A_{1}}$
is the effective coupling in $s^{+-}$ channel and $\lambda_{-}^{A_{1}}$
is the effective coupling in $s^{++}$ channel. The coupling in the
$s^{+-}$ channel is obviously larger and below we focus only on this
channel. The corresponding susceptibility scales as
\begin{equation}
\chi_{A_{1}}^{s^{+-}}\propto\frac{1}{L_{P_{s}}-L}\,,\label{chu_5}
\end{equation}
where
\begin{align}
L_{P_{s}}=L_{0}-\frac{2(|\gamma_{4}|+4C)}{1+\gamma_{3}^{2}C^{2}}\,.\label{L_Ps}
\end{align}
In our model the instability in $A_{1}$ channel occurs prior to the
instability in $B_{2}$ channel, but in a generic three-orbital low-energy
model $B_{2}$ instability well may come first.

In physical terms, $A_{1}$, $s^{+-}$ order leads to opposite shifts
in the chemical potentials for electrons and holes. This does not
break any symmetry, and the opposite shift of $\mu_{h}$ and $\mu_{e}$
can be obtained from fermionic self-energy (which is neglected in
RG analysis). Because no symmetry is broken, there will be no true
instability in $A_{1}$ channel once effects beyond pRG are included.
Still, at some distance from $L_{P_{s}}$ the $A_{1}$ susceptibility
obeys Eq. (\ref{chu_5}), what in practice mean that the separation
between $\mu_{e}$ and $\mu_{h}$ grows as the temperature approaches
the one which corresponds to $L=L_{P_{s}}$ (we recall that $L$ can
be interpreted as $\log{W/T}$, where $W$ is of order bandwidth).

We again emphasize the role of the flow of the couplings. If we used
the bare interactions instead of the running ones, we would obtain
the matrix $M_{A_{1},ph}$ in the form,
\begin{align}
M_{A_{1},ph}=-(U+2U'-J)\frac{m}{2\pi}\begin{bmatrix}1 & 1\\
1 & 1
\end{bmatrix}\,.\label{M_ph2}
\end{align}
As a result, $\lambda_{+}^{A_{1}}\propto J-(U+2U')$ and it would
be negative for realistic $U,U'$, and $J$. This implies that the
increase of the susceptibility in $A_{1}$ $s^{+-}$ channel is entirely
due to the flow of the couplings.

\subsubsection{$A_{2}$ and $B_{2}$ channels}

The $A_{2}$ order parameter in the orbital basis is $\rho_{A_{2}}^{d}=d_{xz}^{\dagger}d_{yz}-d_{yz}^{\dagger}d_{xz}$,
$\rho_{A_{2}}^{f}=f_{xz}^{\dagger}f_{yz}-f_{yz}^{\dagger}f_{xz}$,
see Eq. \eqref{rho_channels}. In the band basis, $\rho_{A_{2}}^{d}=<c^{\dagger}d-d^{\dagger}c>$
and $\rho_{B_{2}}^{f}$ is $<f_{1}^{\dagger}f_{2}-f_{2}^{\dagger}f_{1}>$.
The $B_{2}$ order parameter in the orbital basis is $\rho_{B_{2}}^{d}=d_{xz}^{\dagger}d_{yz}+d_{yz}^{\dagger}d_{xz}$
and $\rho_{B_{2}}^{f}=f_{xz}^{\dagger}f_{yz}+f_{yz}^{\dagger}f_{xz}$
In the band basis, $\rho_{B_{2}}^{d}=<c^{\dagger}d+d^{\dagger}c>\cos{2\theta}+<d^{\dagger}d-c^{\dagger}c>\sin{2\theta}$
and $\rho_{B_{2}}^{f}$ is $<f_{1}^{\dagger}f_{2}+f_{2}^{\dagger}f_{1}>$.

\begin{figure}[h]
\begin{centering}
\includegraphics[width=0.5\columnwidth]{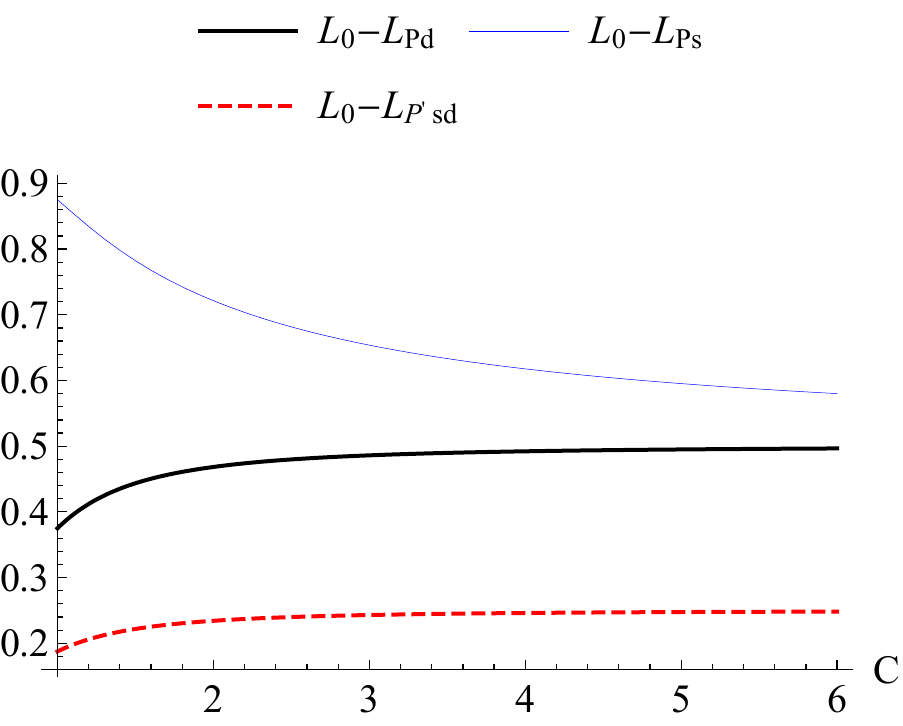} \protect\caption{The differences $L_{0}-L_{Pd}$, $L_{0}-L_{Ps}$, and $L_{0}-L_{P'sd}$
as functions of the parameter $C$. $L_{Ps},L_{Pd}$ and $L_{P'sd}$
are the values of $L$ at which Pomeranchuk susceptibilities in $A_{1}$,
$B_{1}$, and $A_{2}/B_{2}$ channels diverge within RG. The larger
in the difference, the larger is the temperature at which the instability
occurs. The divergence of the susceptibility in $A_{1}$ channel is
an artefact of RG approximation as it does not give rise to a symmetry
breaking. This divergence is cut by terms not included into one-loop
RG. The divergence in the $B_{1}$ channel is the real one, and leads
to $d-$wave orbital order. \label{fig:L_diff}}

\par\end{centering}

\centering{}
\end{figure}

The computation of susceptibilities in these two channels proceeds
in the same way as for $A_{1}$ and $B_{1}$ channels. 
The vertices $\Gamma_{A_{2},B_{2}}^{f}$ do not renormalize because
one of energies in the bubble made out of $f_{1}$ and $f_{2}$ fermions
is necessary large. The vertices $\Gamma_{A_{2},B_{2}}^{d}$ do renormalize,
and along the fixed trajectory we obtained $\Gamma_{A_{2},B_{2}}^{d}=\Gamma_{A_{2},B_{2}}^{d,0}/(1-\lambda_{A_{2},B_{2}})$,
with
\begin{equation}
\lambda_{A_{2},B_{2}}=|\gamma_{4}|u_{1}\, .
\end{equation}
Accordingly, the susceptibilities in these two channels scale as
\begin{equation}
\chi_{A_{2},B_{2}}\propto\frac{1}{L_{P'_{s,d}}-L}\, ,
\end{equation}
where
\begin{equation}
L_{P'_{s,d}}=L_{0}-\frac{|\gamma_{4}|}{1+\gamma_{3}^{2}C^{2}}\,.\label{L_P'sd}
\end{equation}
Comparing this form with critical $L$ in $A_{1}$ and $B_{1}$ channels,
we see that $L_{P'_{s,d}}>L_{P_{d}}$. As the result, the susceptibilities
in $A_{2}$ and $B_{2}$ channels diverge at a lower $T$ than the
one in $B_{1}$ channel, hence these channels are subleading to $B_{1}$
channel. 
plot $L_{0}-L_{Pd}$, $L_{0}-L_{Ps}$, and $L_{0}-L_{P'sd}$ as functions
of $C$ in Fig. ~\ref{fig:L_diff}.

\begin{figure}[h]
\begin{centering}
\includegraphics[width=0.5\columnwidth]{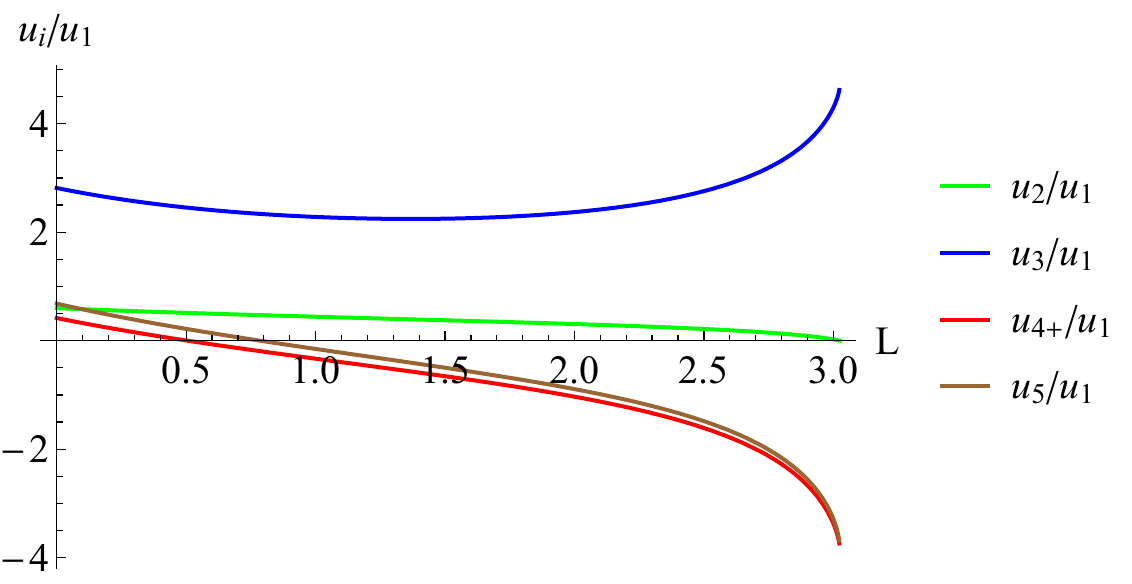} \protect\caption{The solution of the pRG equations, Eqs.~\protect\eqref{RG_13_a_2}, for the model where the electron pockets have $d_{xy}$ orbital content.
The pRG equations and the fixed trajectory are the same as in the model with $d_{xz}/d_{yz}$ electron pockets, however initial values of the couplings are different.
 The convergence towards the fixed trajectory is much better for $d_{xy}$  electron pockets. \label{fig:RG_xy}}

\par\end{centering}

\centering{}
\end{figure}

\section{The model with $d_{xy}$ electron pockets.}
\label{sec:xy}

For completeness, we also analyzed the model in which we approximated the two electron pockets as purely $d_{xy}$.
In this approximation, inter-orbital Hubbard interaction acts within the subset of the two hole pockets, which, like before, are made out of $d_{xz}$ and $d_{yz}$ orbitals, and within the subset of the two electron pockets.  The  corresponding interaction terms are $U_4$,  $U_5$, ${\bar U}_5$, ${\tilde U}_5$,
 ${\tilde{\tilde U}}_5$  terms  in Eq. (\ref{H_T}). The  bare values of all these couplings are Hubbard $U$, i.e.,
 \beq
U_{4,0} =U_{5,0} ={\bar U}_{5,0} ={\tilde U}_{5,0} = {\tilde{\tilde U}}_{5,0} = U\, .
\label{sun_1}
\eeq
Inter-orbital Hubbard terms include density-density interactions $U_1$ and ${\bar U}_1$ between hole and electron pockets and interaction ${\tilde U}4$ within $d_{xz}$ and $d_{yz}$ components  of hole pockets.  Because $d_{xy}$ orbital interacts equally with $d_{xz}$ and $d_{yz}$ orbitals, the bare value of $U_1$ and ${\bar U}_1$ are equal, i.e.,
 \beq
U_{1,0} ={\bar U}_{1,0} ={\tilde U}_{4,0} = U'\, .
\label{sun_2}
\eeq
The exchange Hund interaction $J$ acts in the subspace of $d_{xz}$ and $d_{yz}$ orbitals (${\tilde {\tilde U}}_4$ term) and  between $d_{xy}$ and $d_{xz}/d_{yz}$ orbitals ($U_2$ and ${\bar U}_2$ terms). Again,  $d_{xy}$ orbital interacts equally with $d_{xz}$ and $d_{yz}$ orbitals, hence the bare values  of $U_2$ and ${\bar U}_2$ are equal:
 \beq
U_{2,0} ={\bar U}_{2,0} ={\tilde {\tilde U}}_{4,0} = J\, .
\label{sun_3}
\eeq
Finally, pair-hopping interaction $J'$ also acts in the acts in the subspace of $d_{xz}$ and $d_{yz}$ orbitals (${\bar U}_4$ term) and  between $d_{xy}$ and $d_{xz}/d_{yz}$ orbitals ($U_3$ and ${\bar U}_3$ terms).  Like for other interactions, bare values of $U_3$ and ${\bar U}_3$ are equal:
 \beq
U_{3,0} ={\bar U}_{3,0} ={\bar U}_{4,0} = J'\, .
\label{sun_4}
\eeq
The structure of low-energy electronic states is the same as in the model which we considered in the main text, hence pRG equations are the same as in (\ref{RG_13_a}).
 The couplings ${\tilde U}_4$ and ${\tilde {\tilde U}}_4$  still flow to zero if the bare ${\tilde U}_4$ exceeds the bare ${\tilde {\tilde U}}_4$, which is the case when  $U'>J$.  The couplings ${\tilde U}_5$ and ${\tilde {\tilde U}}_5$   reman equal under pRG, and both tens to zero when $U >0$.
  One can further make sure that the couplings $u_1$ and ${\bar u}_1$,  $u_2$ and ${\bar u}_2$, $u_3$ and ${\bar u}_3$,  $u_5$ and ${\bar u}_5$, which are equal at the bare level, remain equal under pRG.  This reduces the set of pRG equations to
 \begin{align}\label{RG_13_a_1}
\dot{u}_1 & =  u_1^2 +  u_3^2/C^2
\notag \\
\dot{u}_2 & = 2  u_1 u_2  - 2   u_2^2
\notag \\
\dot{u}_3 & = - u_3 \left(u_4  + \bar{u}_4\right) + 4  u_3 u_1- 2 u_2 u_3
- 2 u_5 u_3
\notag \\
\dot{u}_4 & = - u_4^2 -  \bar{u}_4^2 - 2 u^2_3
\notag \\
\dot{\bar{u}}_4 & = - 2 u_4 \bar{u}_4 - 2 u^2_3
\notag \\
\dot{u}_5 &= - 2 u_5^2 - 2 u_3^2
\end{align}
The transformation from $U_i$ to dimensionless $u_i$ is the same as before, and we remind that $C = (m_e+m_h)/2\sqrt{m_e m_h}$.

Introducing $u_{4+} = (u_4 + {\bar u}_4)/2$ and $u_{4-} = (u_4 - {\bar u}_4)/2$
 we immediately find that  the equation for $u_{4-}$ decouples from the rest:
\beq
\dot{u}_{4-}  = - u^2_{4-}\, ,
\label{sun_5}
\eeq
i.e.,
\beq
{u}_{4-} = \frac{{u}_{4-,0}}{1 + {u}_{4-,0}L}\, .
\label{sun_6}
\eeq
At the bare level, $ u_{4-,0} > 0$.  Eq. (\ref{sun_6}) then shows that $u_{4-}$ tends to zero under pRG.
The other equations become
 \begin{align}\label{RG_13_a_2}
\dot{u}_1 & =  u_1^2 +  u_3^2/C^2
\notag \\
\dot{u}_2 & = 2  u_1 u_2  - 2   u_2^2
\notag \\
\dot{u}_3 & = - 2u_3 u_{4+} + 4  u_3 u_1- 2 u_2 u_3 - 2 u_5 u_3
\notag \\
\dot{u}_{4+} & = - 2 u^2_{4+}  - 2 u^2_3
\notag \\
\dot{u}_5 &= - 2 u_5^2 - 2 u_3^2\, .
\end{align}
The fixed trajectory for these equations is the same as for the model with $d_{xz}/d_{yz}$ electron pockets, namely
$u_i = {\bar u}_i$,
 $u_2/u_1 =0$,  $u_3 = \gamma_3 u_1$, $u_4 = u_5 = \gamma_4 u_1$, where $\gamma_3 = C\sqrt{8C^2 -1 + 4 \sqrt{1-C^2+4C^4}}$,
$\gamma_4 =1-2 C^2 -\sqrt{1-C^2 + 4 C^4}$, and
\beq
u_1 = \frac{1}{1 + \left(\frac{\gamma_3}{C}\right)^2} ~\frac{1}{L_0 -L}\, .
\label{sun_7}
\eeq
However, because $u_i = {\bar u}_i$, $i = 1,2,3$ already at the bare level, the system approaches the fixed trajectory faster than in the model which
 we studied in the main text.  We show pRG flow of the ratios of the couplings  in Fig.~\ref{fig:RG_xy}.


\begin{thebibliography}{10}
\bibitem{comm} When electron pockets are approximated as pure $d_{xy}$,
some of $u_{i}$ and ${\bar{u}}_{i}$ turn out to be equal already
at a bare level, see SM.
\end{thebibliography}
\end{document}